\documentclass[a4paper,11pt]{article}
\usepackage{fullpage}
\usepackage{color}
\usepackage{latexsym}
\usepackage{mathrsfs}
\usepackage{enumitem}
\usepackage{lscape}
\usepackage{lmodern}
\usepackage{amsmath,amssymb,amsthm}
\usepackage{bbm} 
\usepackage{tikz-cd} 
\usepackage{xspace,bbold}
\usepackage[margin=1in]{geometry} 
\usepackage[utf8]{inputenc}
\usepackage{stmaryrd}
\usepackage{comment}
\usepackage{multirow}
\usepackage{diagbox}

\usepackage[toc,page]{appendix}
\usepackage{etoolbox}
\appto\appendix{\addtocontents{toc}{\protect\setcounter{tocdepth}{1}}}

\appto\listoffigures{\addtocontents{lof}{\protect\setcounter{tocdepth}{1}}}
\appto\listoftables{\addtocontents{lot}{\protect\setcounter{tocdepth}{1}}}

\usepackage{float}

\numberwithin{equation}{section}       
\allowdisplaybreaks                    

\usepackage{jheppub}                   

\makeatletter
\gdef\@fpheader{\ }                    
\makeatother

\newcommand{\be}{\begin{equation}}
\newcommand{\ee}{\end{equation}}

\newcommand{\SU}[1]{\mathrm{SU}( #1 )}
\newcommand{\Suni}[1]{\mathrm{S}( #1 )}
\newcommand{\SUs}[1]{\mathrm{SU}^*\!(#1)}
\newcommand{\SOs}[1]{\mathrm{SO}^*\!(#1)}
\newcommand{\SL}[1]{\mathrm{SL}( #1 )}

\newcommand{\SO}[1]{\mathrm{SO}( #1 )}

\newcommand{\Spin}[1]{\mathrm{Spin}(#1)}

\newcommand{\Sp}[1]{\mathrm{Sp}(#1)}

\newcommand{\USp}[1]{\mathrm{USp}( #1 )}
\newcommand{\U}[1]{\mathrm{U}(#1)}

\DeclareMathOperator{\GL}{\mathit{GL}}
\newcommand{\Ex}[1]{\mathrm{E}_{#1(#1)}}

\DeclareMathOperator{\su}{\mathfrak{su}}

\newcommand{\ex}[1]{\mathfrak{e}_{#1(#1)}}

\newcommand{\GDiff}{\mathrm{GDiff}}

\newcommand{\tr}{\mathrm{tr}}

\numberwithin{equation}{section}

\newcommand{\bpm}{\begin{pmatrix}}
\newcommand{\epm}{\end{pmatrix}}

\newcommand{\calM}{\mathcal{M}}
\newcommand{\calN}{\mathcal{N}}

\newcommand{\dd}{\mathrm{d}}

\newcommand{\id}{\mathbb{1}}

\DeclareMathOperator{\adj}{ad}

\DeclareMathOperator{\Comm}{C}
\newcommand{\Com}[2]{\Comm_{#2}(#1)}

\newcommand{\Gst}{G_S}

\newcommand{\bbR}{\mathbb{R}}

\newcommand{\rep}[1]{\mathbf{#1}}

\usepackage{booktabs}

\newcommand{\sast}{{\star}}

\newcommand{\ad}{{\text{ad}}}

\title{Consistent Truncations and Generalised Geometry: Scanning through Dimensions and Supersymmetry}

\author[a]{Gr\'egoire Josse,}
\emailAdd{gregoire.josse@physik.hu-berlin.de}
\author[b]{Michela Petrini,}
\emailAdd{petrini@lpthe.jussieu.fr}
\author[c]{and Mart\'in Pico}
\emailAdd{martin.pico@uam.es}

\affiliation[a]{Institut für Physik, Humboldt-Universität zu Berlin, IRIS Gebäude, Zum Großen Windkanal 2, Berlin, 12489, Germany}
\affiliation[b]{Sorbonne Universit\'e, UPMC Paris 05, UMR 7589, LPTHE, 75005 Paris, France}
\affiliation[c]{Universidad Autónoma de Madrid and Instituto de Física Teórica UAM/CSIC, Madrid 28049, Spain}

\subheader{\hfill\textrm{IFT-UAM/CSIC-25-155}}

\abstract{We study consistent truncations  in the framework of Exceptional Generalised Geometry. 
We classify the 4-dimensional gauged supergravities
that can be obtained as a consistent truncation of 10/11-dimensional supergravity. 
Any truncation is associated to a (generalised) $\Gst$-structure with singlet intrinsic torsion. 
We give the full classification for all truncations associated to continuous structure groups and we discuss a few examples with discrete ones.  We recover gauged supergravities corresponding to known truncations as well as others for which explicit truncations are still to be constructed. We also summarise similar results obtained in the literature for truncations to $d=5,6,7$ dimensions and we complete them, when needed.}

\begin{document}

\maketitle

\section{Introduction}

A question that naturally arises in string theory is how to construct low-dimensional effective actions. This is of course crucial if we want to construct string models to be confronted with phenomenological observations, but it has also important applications in other contexts such as understanding the web of supergravity theories.

Supergravity theories have been constructed in any dimension $D=2, \ldots  11$. 
While the theories in $11$ and $10$ dimensions are very constrained and all have an  interpretation as low-energy effective theories of M-theory and string theory, respectively, when going down in dimension there is a multitude of supergravities, with or without gauge symmetries,  and it is an open question whether they have a string theory origin or not. 

A way to address this question is in terms of consistent truncations.  11/10-dimensional supergravity on a background of the form
\begin{equation}
    X_{10/11} = X_{D} \times M \, ,
\end{equation}
with $M$ a compact manifold of dimension $d=11/10-D$, can be seen 
as a $D$-dimensional theory with an infinite number of fields organised into representations of the symmetry group of the internal manifold. A consitent truncation is a procedure to truncate the theory to a finite set of modes, in such a way that all truncated modes decouple from the lower-dimensional equations of motion and that no dependence on the internal manifold is left. If the truncation is consistent any solution of the truncated theory can be uplifted to a solution of the 11/10-dimensional theory.

The main difficulty in constructing consistent truncations is to find a principle for selecting  the  modes to be kept in the truncated theory.  Starting from a series of explicit constructions (see for instance \cite{Cassani:2010uw, Gauntlett:2010vu, Gauntlett:2009zw, KashaniPoor:2007tr, Cassani:2009ck,  Cassani:2011fu, Cassani:2012pj}) it has become clear that the formalism of $G$-structures provides a powerful tool to construct consistent truncations in a systematic way. 
The notion of $G$-structure can be extended to the framework of Exceptional Geneneralised Geometry and Exceptional Field Theory. These are reformulations of 11/10-dimensional supergravity  that treat in a geometric way all the symmetries of the theory.  
Generalised $G_S$-structures provide a systematic and general derivation of consistent truncations \cite{Cassani:2019vcl}: 
 any generalised $G_S$-structure on $M$ with constant, singlet intrinsic torsion defines a consistent truncation of 10/11-dimensional supergravity.  

This approach allows to give a unified description of truncations in
different dimensions and with a different  amount of supersymmetry. 
For instance, 
all maximally supersymmetric truncations are associated to generalised identity structures and therefore can be seen as generalised Scherk-Schwarz reductions \cite{Lee:2014mla, Hohm:2014qga, Baguet:2015sma}. 
In particular,  all maximally supersymmetric truncations on spheres are unified in this class: truncations of 11-dimensional supergravity on S$^7$ and S$^4$,  IIB supergravity on S$^5$  \cite{Lee:2014mla,Baguet:2015sma} and massive IIA on spheres \cite{Ciceri:2016dmd, Cassani:2016ncu}.
Similar classifications can be given for half-maximal and quarter-maximal truncations by considering larger generalised structure groups \cite{Cassani:2019vcl, Malek:2017njj, Malek:2018zcz, Malek:2019ucd,Cassani:2020cod, Josse:2021put}.

The generalised $\Gst$-structure fully determines the lower-dimensional truncated theory:  field content, bosonic symmetries and  supersymmetry. 

In \cite{Josse:2021put} these features were exploited to classify the 5-dimensional supergravity theories with $\mathcal{N}=2$ supersymmetry that can be obtained as consistent truncations of 11/10-dimensional supergravity.

In this paper we will pursue this approach and apply it to truncations of 10/11-dimensional supergravity to 4 dimensions using the framework of $\Ex{7}$ generalised geometry. 
The main idea is that this reduces to the study of all possible generalised $\Gst$-structure compatible with a given amount of supersymmetry. 
We first solve the algebraic problem of finding all possible subgroups $\Gst$ of the generalised structure group $\Ex{7}$. Then, for any $\Gst$, from its embedding in $\Ex{7}$, we derive the  field content and symmetries of the 4-dimensional theory. 
In particular, the $\Gst$-singlet components of the generalised intrinsic torsion will give in a straightford way all the components of the  embedding tensor of the reduced theory and, hence, all the possible gaugings.

We obtain a classification of 4-dimensional theories  with $\mathcal{N} \geq 2$. As in  \cite{Josse:2021put}, we find that these form a very reduced set of the supergravity theories that can be constructed directly in 4 dimensions.  

Some of the theories we present in our classification  have already been obtained as explicit truncations on specific manifolds, while some others have not. A priori, nothing guaranties that such truncations exist at all.  This is because our algebraic analysis only gives the theories that could be a priori obtained. In this paper, we make  hypothesis that the only non-zero components of the intrinsic torsion are $\Gst$-singlets. However, 
in order to actually construct them, one has to find a compactification manifold with the appropriate $\Gst$-structure and the geometrical features to give a constant, singlet intrinsic torsion.

\medskip

Consistent truncations have many important applications in the context of the AdS/CFT correspondence, where most of the solutions dual to CFT's or deformations thereof have been first constructed in a lower-dimensional supergravity, which is a consistent truncation of 11/10-dimensional supergravity containing the  fields relevant for the solution, and 
then uplifted to the full theory.
The same is true for many black-hole solutions in string or M-theory. 
Of particular relevance are gauged supergravities in dimensions $  4 \leq D \leq 7$. For $D > 4$ a  systematic study of consistent truncations in the formalism of Exceptional Generalised Geometry and/or Exceptional field Theory can already be found in the literature   (see for instance in  \cite{Malek:2016bpu, Cassani:2019vcl, Malek:2017njj, Josse:2021put}).  
For completeness,  we summarise  such results and we present them in the language of Exceptional Generalised Geometry. In some cases we fill a few missing points in the classification. 

\medskip

Understanding lower-dimensional supergravity theories
as  consistent truncations of 11/10-dimensional supergravity has also a more fundamental meaning. 
Since supergravity theories are non-renormalisable, they  make only sense as low-energy effective theories.  11/10-dimensional supergravities are special as their 
ultraviolet completion is provided by string or M-theory. 
Thus, understanding whether a lower-dimensional supergravity theory is a consistent truncation of 11/10-dimensional supergravity gives it a proper embedding in a consistent theory of quantum gravity.

\medskip

The paper is organised as follows. In Section \ref{sec:GS} we briefly describe the formalism of $\Ex{7}$ generalised geometry, which is the relevant one for truncations to 4 dimensions, the notions of generalised $\Gst$-structure and how this gives the data of the truncated theory.
Section \ref{sec:scan} contains a summary of our results. We organise them according to the number of supersymmetries of the truncated theory and we also give a brief summary of the main features of the relevant 4$d$ supergravities. 
In Section \ref{Section_Algorithm} we discuss in more detail the methods we used to achieve the classification
and we illustrate them explicitly in the case of truncations with $\mathcal{N}=4$ supersymmetry in Appendix  \ref{explicit_ex_subsec}. In Section 
\ref{sec:higherD} 
 we summarise the results in the literature about  truncations to supergravities in other dimensions,  and complete them with a few details, when needed.
Our conventions for $\Ex{7}$ and $\SU{8}$ are given in  
 Appendix \ref{PreliminariesE77_Mth}.

\section{Generalised G-structures and consistent truncations}
\label{sec:GS}

In this Section we recall the main notions of Exceptional Generalised Geometry (EGG) we will use in the rest of the paper. We follow the conventions of \cite{Ashmore:2015joa} and \cite{Josse:2021put}.

\medskip

Exceptional Generalised Geometry (EGG)  is a  reformulation of supergravity which gives a unified geometrical description of the bosonic sector of 11/10-dimensional supergravity compactified on a $d$ dimensional manifold $M$. This is achieved by replacing the tangent bundle $TM$ with a generalised tangent bundle, 
defined as the extension of $TM$ by appropriate exterior powers of the cotangent bundle, and whose structure group is  the exceptional group 
$\Ex{d}$.

In this paper we will mainly be interested in compactifications of 11/10-dimensional supergravity to 4 dimensions
on backgrounds of the form
\begin{equation}
\label{eq:warpedsp}
     X_{10/11} = X_{4} \times M \, ,
\end{equation}
where the  internal manifold  $M$ has dimension $d=7$ or $d=6$, respectively. 
In this case the relevant exceptional group is $\Ex{7} \times \mathbb{R}^+$. 
The fibres of the generalised tangent bundle transform in the $\rep{56}_1$ of $\Ex{7}\times \mathbb{R}^+$, with the subscript denoting the 
$\mathbb{R}$ weight, and are called generalised vectors. 
One can  also introduce the dual generalised tangent bundle $E^*$  transforming in the $\rep{56}_{-1}$ of  $\Ex{7}\times \mathbb{R}^+$. 

\medskip

The ordinary notions of tensors, Lie derivative and covariant derivatives can be generalised to $E$ \cite{Hull:2007zu, Pacheco:2008ps, Coimbra:2011ky}.
Generalised tensors correspond to bundles whose fibres transform in given representations of the exceptional group and can  be  decomposed as local  sums of powers of $TM$ and $T^*M$.\footnote{
Consider, for instance, 11-dimensional supergravity. In this case $M$ has dimension
$d=7$ and the generalised tangent bundle can be written as 
\begin{equation}
\label{7dEb1}
    E  \simeq T M  \oplus \Lambda^2 T^* M  \oplus \Lambda^5 T^* M  \oplus
 ( T^* M \otimes \Lambda^7 T^* M ) \, . 
   \end{equation}
Its sections $V \in  \Gamma(E)$ are locally sums of a vector $v$, a real two-form $\omega$,  a real five-form $\sigma$ and the (1,7) tensor $\tau$ on $M$: 
$V  \sim v + \omega + \sigma + \tau$. 
The dual generalised vectors in $E^*$ are obtained by raising indices with the inverse metric on $M$. Similar decompositions are given for the other tensor bundles.  } 
There are four bundles of particular relevance for consistent truncations:
 the adjoint bundle, ${\rm ad}F$, the bundles $N$ and $K$, and  the generalised metric $G$. 
 
 The adjoint bundle has sections  transforming in the adjoint representation, $\rep{133}_0 \oplus \rep{1}_0$, of $\Ex{7}\times \mathbb{R}^+$, where $\rep{1}_0$ denotes the generator of $\mathbb{R}^+$. They are constructed via  the projection $E \times_{\rm ad} E^*$. 

The generalised bundle $N$ is a sub-bundle of the symmetrised product of two copies of the generalised tangent bundle
\begin{equation}
    N = \det T^*M \otimes E^* \subset S^2(E) \, ,
\end{equation} 
and the explicit form of the fibres in terms of ${\rm GL}(d)$ tensors can be found in \cite{Coimbra:2011ky, Ashmore:2015joa}.
The bundle $K$ has fibres transforming in the $\rep{912}_{-1}$ (see again \cite{Coimbra:2011ky} for more details). 

Finally,  the  generalised
metric, $G$, is defined  as  the symmetric tensor of rank 2 
\begin{equation}
\begin{aligned}
    G \,: &  \, E \otimes E \to \bbR^+ \\
 &   (V , W) \to G(V, W)  = G_{MN} V^M W^N \, , 
    \end{aligned}
\end{equation}
where $V,W$ are generalised vectors. At each point $p\in M$ the generalised metric parameterise the coset
\begin{equation}
    G \big|_p \in \frac{\Ex{7} \times \bbR^+}{\SU{8}/\mathbb{Z}_2 } \, , 
\end{equation}
where $\SU{8}/\mathbb{Z}_2$ is the maximally compact subgroup of $\Ex{7}$. 

\medskip

All bosonic transformations of the theory (diffeomorphisms and $p$-form gauge transformations) are treated in a geometrical way as generalised diffeomorphisms, $\GDiff$,  \cite{Hull:2007zu, Pacheco:2008ps}.  The 
 infinitesimal  $\GDiff$  are generated by the Generalised Lie derivative or  Dorfman derivative
\be 
\label{eq:Liedefgapp}
(L_V V')^M \,=\,  V^N \partial_N  V'^M - (\partial \times_{\adj} V)^M{}_N V'^N \, , 
\ee
where  $V, V^\prime \in \Gamma(E)$ are two generalised vectors, $\partial_M = (\partial_m, 0, \dots, 0)$  gives the embedding of the ordinary partial derivative $\partial_m$ on $M$ in $E^*$, and $\times_{\adj}$ denotes again the projection into the adjoint of $\Ex{7}$.

\medskip

As in ordinary geometry, one can define a generalised $\Gst$-structure
as the reduction of the exceptional structure group to a subgroup $\Gst \subset \Ex{7}$. 
More precisely, an exceptional $\Gst$-structure  defines a $\Gst$-principal subbundle of the $\Ex{7}$ frame bundle. 
In all the cases we are interested in, this is equivalent to the existence of globally defined $\Gst$-invariant generalised tensors.  
As an example, the generalised metric $G$ defines a $\SU{8}$ generalised structure. In  what follows we will be interested in generalised structures $\Gst$  that are subgroups of  $\SU{8}$.\footnote{Strictly speaking the $\Gst$-structure is a subgroup of $\SU{8}/\mathbb{Z}_2$. However, as discussed below, we are interested in its double cover acting on the fermions of the theory. For simplicity of notation, in the rest of the paper, we will not distinguish between the $\Gst$-structure and its double cover.} These are defined by generalised vectors, $K_I$, and/or elements of the adjoint bundle, $J_A$, 
\begin{equation}
    \Xi_i = \{K_I, J_A \}  \, .
\end{equation}
Starting from  $ \Xi_i$ it is always possible to construct the generalised metric as $G=G( \Xi_i)$  \cite{ Cassani:2019vcl}. 

A generalised $\Gst$-structure is characterised by its intrinsic torsion. Given a $\Gst$- compatible connection 
$D \Xi_i =0$,  the intrinsic torsion is the part of the torsion of the connection $D$ that cannot be eliminated by redefining the connection. It 
measures the obstruction to finding a torsion-free connection, compatible with the structure \cite{Coimbra:2014uxa}. 
The intrinsic torsion of a $\Gst$-structure is contained in the torsion bundle $W$
\begin{equation}
    W \simeq K \oplus E^*
\end{equation}
which transforms in the $\rep{912}_{-1} \oplus \rep{56}_{-1}$ of $\Ex{7} \times \mathbb{R}^+$ and can be decomposed into $\Gst$-representations. 
The component $\rep{56}$ provides the gauging of the scaling symmetry called trombone symmetry.  Theories with a trombone symmetry are not lagrangian, but can be studied by looking at the equations of motion. In this paper we will focus only on the $\rep{912}$ component, even if 
our results are trivially extended to include the trombone symmetry.\footnote{Any connection compatible  with $\Gst\subset \SU{8}$ does not mix with the trombone, since $D$ is $\Gst$-valued. Adding the trombone simply amounts to  taking into account the extra $\Gst$-singlets  coming from the $\rep{56}$.}

\medskip

Generalised structures  are the key ingredient for constructing  consistent truncations  \cite{Lee:2014mla, Hohm:2014qga, Cassani:2019vcl}. Consider 
11 or 10-dimensional supergravity on a manifold of the form \eqref{eq:warpedsp}. Since $M$ is compact, the theory can be seen as  an effective 4-dimensional theory with an infinite number of fields organised into representations of 
\begin{equation}
    \GL(4,\bbR) \times \Ex{7} \, . 
\end{equation}
The 4-dimensional  metric is a singlet of $\Ex{7}$, the scalars are arranged into the generalised metric, the  vectors are sections of the generalised tangent bundle $E$, while the two-form tensors are sections of the generalised tensor bundle  $N$
\begin{equation}
    \begin{aligned}
        \mbox{scalars} &  \qquad  G_{MN}(x,y) \in \Gamma(S^2 E^*)  \, , \\
        \mbox{vectors} &  \qquad   A_{\mu}^{M}(x,y) \in \Gamma(T^*X \otimes  E)  \, , \\
 \mbox{2-forms} &   \qquad   B_{\mu \nu  M N}(x,y) \in \Gamma(\Lambda^2 T^*X \otimes  N) \, .
    \end{aligned}
\end{equation}

\medskip
A consistent truncation is a procedure to truncate away the infinite towers  and  construct a 4-dimensional theory 
with only a finite set of fields.  The truncation is called consistent because the modes 
 that have been truncated away decouple from the equation of motion. In doing so, all dependence on the internal coordinates disappears from the 4-dimensional equations of motion and any given solution of the 4-dimensional theory can be uplifted to a full solution of the higher-dimensional one.

If the manifold $M$ admits a generalised $\Gst$-structure with constant singlet intrinsic torsion or zero torsion, then a consistent truncation is guaranteed to exist \cite{Cassani:2019vcl}. 

The consistent truncation is derived by expanding all  bosonic  10/11-dimensional fields  in terms of the generalised invariant tensors  $\{\Xi_i \}$ defining the 
$\Gst$-structure.  The coefficients in the expansions  only depend on the external  coordinates $x$ while the dependence on the internal space is in the tensors $\{\Xi_i \}$.

The 4-dimensional scalars are given by the $G_S$-singlets in the generalised metric $G_{MN}$. These are singlet deformations of the structure modulo the singlet deformations that do not deform the metric
\begin{equation}
\label{scalarman}
   \text{scalars:} \qquad h^{I}(x) \in
   \mathcal{M} = \frac{\Com{\Gst}{\Ex{7}}}{\Com{\Gst}{\SU{8}}}=\frac{G}{H}\, , 
\end{equation}
where $G$ and $H$ denote the groups 
that remain in the quotient after the common factors in the numerator and denominator  cancel out. The group  $G = :  \mathrm{G}_{\rm iso}$  gives  the isometry group of the scalar manifold. 

The vectors are determined by the number of $\Gst$-invariant generalised vectors
$\{K_I\}$
\begin{equation} 
\label{vectors}
   \text{vectors:} \qquad  \mathcal{A}_\mu^M (x, y)  = A_\mu^I(x)\, K_I^M \,\in\, \Gamma({T^*M}\otimes \mathcal{V}) \, , 
\end{equation}
where   $\mathcal{V}\subset \Gamma(E)$ is the vector space  spanned by the  $\{K_I\}$. 
Notice that the singlet generalised vectors determine all the vectors of the reduced theory, coming both from the reduction of the metric and the higher-rank potentials. Thus  the vectors $K_I$ generate all symmetries of the reduced theories.
This is an important difference with respect to the reductions based on conventional $G_S$-structures.

The fermionic sector of the truncated theory is constructed in a similar way.  The spinors are organised in representations of $\SU{8}$.  The structure group  $\Gst$ lifts to its double cover  $\SU{8}$ and the truncation is obtained by expanding the fermionic fields on the $\SU{8}$ singlets in the relevant representations. 

The supersymmetry parameters are embedded in the generalised  spinor bundle $\mathcal{S}$, which transforms in the $\rep{8} \oplus \rep{\bar{8}}$ of $\SU{8}$. For the truncated theory to have $\mathcal{N}$ supersymmetries the spinor bundle must contain  $\mathcal{N}$ $\Gst$-singlets transforming in the fundamental of the relevant  R-symmetry group. Thus 
$\Gst$  must be a subgroup of the commutant of the  R-symmetry group in $ \SU{8}$  
\begin{equation}
\Gst \subseteq \Com{G_R}{\SU{8}} 
\end{equation}
that allows for exactly $\mathcal{N}$ singlets in the spinorial representation of $\SU{8}$.

The truncation is consistent thanks to the fact that  the intrinsic  torsion  only consists of constant singlets of the $\Gst$-structure or is zero. 
Indeed, if there are only singlet representations in the intrinsic torsion,  the  generalised Levi--Civita connection acts on the  invariant generalised tensors $\Xi_i$ as 
\begin{equation}
   \label{eq:DQ}
   D_M \Xi_i = \Sigma_M \cdot \Xi_i \, ,
\end{equation}
where $\Sigma_M$ is completely determined in terms of the torsion.\footnote{$\Sigma_M$ is a section of $E^*\otimes{\rm adj}(\SU{8})$ and $\cdot$ denotes the adjoint action.} Thus, when plugging the truncated fields in the equations of motion, their derivatives only have expansions in terms of singlets.
Since products of singlet representations can never source the non-singlet ones that were truncated away,  the truncation is consistent. 
If the torsion is zero, the invariant tensors are covariantly constant and again no non-singlet terms can be generated in the equations of motion.

\medskip

As discussed above, the field content and the supersymmetry of the truncated theory are completely determined by the  $\Gst$-invariant tensors.  As we will now show,  the intrinsic torsion of the $\Gst$-structure also  encodes the information about the possible gaugings of the truncated theory. 

A gauged supergravity is obtained gauging a  subgroup of the rigid isometries of the scalar manifold, $\mathrm{G}_{\rm iso}$. The way the gauge group
$\mathrm{G}_{\rm gauge}$  is embedded in $\mathrm{G}_{\rm iso}$ is given by the embedding tensor 
(see~\cite{Samtleben:2008pe,Trigiante:2016mnt} for a review of this formalism)
\begin{equation}
   \label{eq:Theta}
   \Theta : \mathcal{V} \to {\rm Lie}\mathrm{G}_{\rm iso} \, , 
\end{equation}
which is a map from the set of vectors to the Lie algebra of the gauge group $\mathrm{G}_{\rm gauge}$.

In EGG the embedding tensor $\Theta$  is identified with the singlet intrinsic torsion. The generalised Lie derivative 
of the invariant  tensors  $\Xi_i$ along  any  invariant generalised vector  $K_I$  reduces to 
\begin{equation}
   \label{eq:LQ}
   L_{K_I}\Xi_i = - T_{\rm int}(K_I) \cdot \Xi_i  \, ,
\end{equation}
where  $T_{\rm int} \, : \, \mathcal{V} \to {\rm ad} F$ gives the map from the set of $G_S$-invariant vectors to the adjoint bundle. Since $T_{\rm int}(K_I)$ is a $\Gst$-singlet, the map select the $\Gst$-singlets in the adjoint, namely 
the elements of the Lie algebra  of the commutant group $G = \Com{\Gst}{\Ex{7}}$.  Since $G = {\rm G}_{\rm iso}$ gives the isometries of the scalar manifold, we 
can identify $-T_{\rm int}$  with the embedding tensor of the truncated theory. 
The Leibniz property of the generalised Lie derivative~\cite{Coimbra:2011ky,Lee2014mla}  translates into the quadratic condition on the embedding tensor. 
The generalised  Lie derivative \eqref{eq:LQ} defines the gauge Lie algebra of the 
truncated theory with structure constants  given by $T_{\rm int}$.  Since  the image of  the map $\Theta$ may not be the whole of ${\rm Lie}G_{\rm iso}$, the gauge group generated by the vectors can be a subgroup of  $G_{\rm iso}$ 
\begin{equation}
   \label{eq:gauge}
   \text{gauge group:} \qquad G_{\rm gauge} \subseteq G_{\rm iso} \, .
\end{equation}
Clearly, when the intrinsic torsion is zero, the truncated theory is ungauged. 

\medskip 

In summary we see that a $\Gst$-structure  completely fixes the data of the truncated theory:
 the topological properties of the  $\Gst$-structure, namely the existence of $\Gst$-invariant non-vanishing generalised tensors, determine
the field content and supersymmetry of the theory, while the differential conditions of having only constant, singlet intrinsic torsion (or zero intrinsic torsion),  beyond  guaranteeing  the consistency of the truncation,  determine its possible gaugings.


\section{Consistent truncations to four dimensions}
\label{sec:scan}

There is a  huge variety of 4-dimensional supergravities with different amount of supersymmetry and different gauge groups.
The aim of this paper is to use the formalism of generalised $\Gst$-structures  to classify those that can be obtained as consistent truncations of 11/10-dimensional supergravity.

\medskip

As discussed in the previous section, the algebraic properties of a generalised $\Gst$-structure are enough to fix 
the field content and  supersymmetries of the reduced theory, as well as the truncation ansatz.
Then imposing the differential constraint of having singlet, constant intrinsic torsion determines the embedding tensor and the possible gaugings.

We will follow the logic of \cite{Josse:2021put}: we assume that the differential constraints are satisfied and we classify  the  possible continous subgroup 
$\Gst$ of  $\Ex{7}$. We also briefly discuss some cases where the structure group is discrete.  
Differently from  \cite{Josse:2021put} we will not focus on a fixed amount of supersymmetry but we scan through all supersymetries: $\mathcal{N} = 2, \ldots 8$.\footnote{We also considered truncations with $\mathcal{N}=1$ and $\mathcal{N}=0$ supersymmetry. However,  the number of truncations is too big  to list  them in a synthetic and interesting way. We leave their discussion to a future publication. }

\medskip

In this section  we summarise our results, while 
the details of the analysis can be found in Section \ref{Section_Algorithm} and Appendix \ref{explicit_ex_subsec}. 
We organise the presentation by amount of supersymmetry. For any number  $\mathcal{N}$  of supercharges we give a brief summary of the corresponding 4-dimensional supergravity and we discuss the main details of the $\Gst$-structure giving rise to the truncation.

For  $\mathcal{N} \geq 3$ the generalised structures  are always defined only in terms of  invariant generalised vectors. In this cases, the  $\Gst$-singlets in the adjoint representation  are not independent and are obtained as 
products of the invariant vectors: ($J \sim K_I \times_{\rm ad} K_J$).

For $\mathcal{N} = 2$  this is no longer the case. The
generalised structure is defined by invariant vectors and invariant adjoint elements, which correspond to the presence of vector and hyper-multiplets in the truncated theory.

For any amount of supersymmetry $\mathcal{N}$ there is a maximal generalised structure, $\Gst^{max}$,  corresponding to the largest commutant in $\SU{8}$ of the R-symmetry group that admits  exactly $\mathcal{N}$ singlets in the $\rep{8}$ of  $\SU{8}$. 
In the table below we list the R-symmetry groups and the maximal generalised $\Gst$-structure for any amount of supersymmetry.  We also give the corresponding invariant generalised tensors. 
\begin{table}[h!]
    \centering
    \begin{tabular}{c|c|c|c}
 $\mathcal{N}$  & $G_R$  &  $\Gst^{max}$  & inv. tensors  \\
   \hline 
        8   & $\SU{8}$ &  $ \id$  &  $ \{ K_I \}_{I=1, \dots 56} $  \\ 
        6  &  $\SU{6} \times \U{1}$  & $\SU{2}$ &  $ \{ K_I \}_{I=1, \dots 32} $  \\ 
        5  &  $\SU{5}\times \U{1}$    & $\SU{3}$ &  $ \{ K_I \}_{I=1, \dots 20} $  \\ 
        4  &  $\SU{4}\times \U{1}$    & $\SU{4}$  &  $ \{ K_I \}_{I=1, \dots 12} $  \\  
        3   &  $\SU{3}\times \U{1}$   & $\SU{5}$   &  $ \{ K_I \}_{I=1, \dots 6} $ \\ 
        2    &  $\SU{2 }\times \U{1}$     & $\SU{6}$  &  $\{K, \hat{K} , J_\alpha \}_{\alpha=1,2,3}$  
    \end{tabular}
    \caption{R-symmetry groups and the maximal generalised $\Gst$-structure for $\mathcal{N}$ supersymmetries.}
    \label{tab:my_label}
\end{table}

As discussed in \cite{Cassani:2019vcl}, the truncated theories obtained from maximal structure groups correspond to pure supergravities. For  $\mathcal{N}= 5,6,8$ this is all one can obtain. For $\mathcal{N} \leq 4$ truncations with extra matter fields can be constructed considering 
subgroups $\Gst \subset \Gst^{max}$  that still give
exactly $\mathcal{N}$ singlets in the $\rep{8}$ of  $\SU{8}$. In what follows we only list the group $\Gst$ that give inequivalent field contents.

\medskip

Our results provide a complete classification for   
$\Gst$-structures where $\Gst$ is a Lie group, i.e. a continuous group. There  could also exist truncations corresponding to $\Gst$-structures defined by discrete groups.  We comment on them in Section \ref{Section_Algorithm}.

\medskip

The structure of the theories that can be obtained as consistent truncations is very restricted. For instance, the scalar manifolds must necessarily be symmetric and can all be written as cosets
\begin{equation}
    \calM = \frac{G}{H} \, ,
\end{equation}
where $G$ is the semisimple non-compact Lie group of global isometries and  $H$ is its maximal compact subgroup. 

For theories with $\mathcal{N} >2$  supersymmetry this property is a consequence of supersymmetry. The allowed manifolds are listed in Table \ref{hmnb3}. 

\begin{table}[h!]
\begin{center}    \hspace*{-0.3cm}
\def\arraystretch{1.8}
\begin{tabular}{|c|c|c|c|c|c|}
\hline
$\mathcal{N}$ &  8 &  6 & 5 & 4 & 3 \\
\hline
$\mathcal{M}$ & $\frac{\Ex{7}}{\SU{8}}$ & 
 $ \frac{\SOs{12}}{\SU{6}\times\U{1}}$ &  $\frac{\SU{5,1}}{\SU{5}\times\U{1}}$   &  $\frac{\SL{2,\mathbb{R}}}{\SO{2}} \times \frac{\SO{6,n_{\rm V}}}{\SO{6} \times\SO{n_{\rm V}}}$ & $\frac{\SU{3,n_{\rm V}}}{S[\U{3}\times \U{n_{\rm V}} ]}$   \\ 
\hline
\end{tabular} 
\caption{Homogeneous symmetric manifolds for $\mathcal{N} \geq3$}
\label{hmnb3}
\end{center}
\end{table}

For $\mathcal{N} =2$ the scalar manifold factorises into vector  and hypermultiplet spaces, which by supersymmetry, are special K\"aler (SK) and quaternionic K\"ahler (QK) manifolds, respectively.
The additional condition of being homogeneous and symmetric  is a prediction of EGG. The list of  homogeneous, symmetric spaces for theories with $\mathcal{N}=2$ supersymmetry is given in Table \ref{hmn2}.

\begin{table}[h!]
\begin{center}    \hspace*{-0.3cm}
\def\arraystretch{1.8}
\begin{tabular}{|c|c|c|c|c|}
\hline
 &  SK  &  $n_{\rm V}$ &  QK  & $n_{\rm H}$ \\
\hline
\multirow{8}{*}{$\mathcal{M}$} & $\frac{\SU{1,n_{\rm V}}}{\U{n_{\rm V}}}$   &  $n_{\rm V}$ 
& $\frac{\SU{2,n_{\rm H}}}{S[\U{2} \times \U{n_{\rm H}}]}$ & $n_{\rm H}$  \\
& $\frac{\SL{2,\mathbb{R}}}{\SO{2}} \times \frac{\SO{2,n_{\rm V}-1}}{\SO{2} \times\SO{n_{\rm V}-1}}$ &  $n_{\rm V}$ & $\frac{\SO{4, n_{\rm H}}}{\SO{4} \times\SO{n_{\rm H}}}$  & $n_{\rm H}$ \\
& $\frac{\SU{1,1}}{\U{1}}$   &  1 &  $\frac{{\rm G}_{2(2)}}{\SU{2} \times \SU{2}}$ & 2  \\
& $\frac{\Sp{6}}{\U{3}}$   &  6 &  $\frac{{\rm F}_{4(4)}}{\SU{2} \times \USp{6}}$ & 7 \\
& $\frac{\SU{3,3}}{S[\U{3}\times \U{3} ]}$  &  9&  $\frac{{\rm E}_{6(2)}}{\SU{2} \times \SU{6}}$& 10  \\
&  $ \frac{\SOs{12}}{\SU{6}\times\U{1}}$ & 15 &   $\frac{{\rm E}_{7(-5)}}{\SU{2} \times \SO{12}}$  & 16  \\
& $\frac{{\rm E}_{7(-25)}}{\U{1}\times\Ex{6}}$   &  27 &  $\frac{{\rm E}_{8(-24)}}{\SU{2} \times \Ex{7}}$ & 28   \\ 
& &  & $\frac{\USp{2,2 n_{\rm H}}}{\USp{2} \times \USp{2n_{\rm H}}}$ & $n_{\rm H}$ \\
\hline 
\end{tabular} 
\caption{Homogeneous symmetric manifolds for $\mathcal{N} =2 $ and the corresponding number of vector ($n_{\rm V}$) and hypermultiplets ($n_{\rm H}$). }
\label{hmn2}
\end{center}
\end{table}

In our classification, 
for $\mathcal{N} \geq 3$ we recover, as expected, all the classes of manifolds listed in Table \ref{hmnb3}, while  for $\mathcal{N} =2$, some manifolds are missing (see  Section 
\ref{sec:N2}). 

For theories with $\mathcal{N} \leq 4$ supersymmetry we find that the possible matter content is very constrained: there is a maximum number of vector and/or hypermultiplets and only a few possibilities are allowed. 

\medskip

For any amount of supersymmetry, the embedding tensor of the truncated theory is given by the $\Gst$-singlets components of the intrinsic torsion. From its analysis one can derive all the gaugings of the global isometries of the theory. In this article we are not interested in performing a detailed study of the gaugings. However,  for completeness, for any amount of supersymmetry we discuss in which  representations of the global isometry group the singlet intrinsic torsion transforms. 

\medskip

Finally, let us recall that the ungauged supergravity 
in 4 dimensions is invariant under electromagnetic duality, which is realised on-shell. Together, electric and dual magnetic vectors form a linear representation of the global symmetry group ${\rm G}_{\rm iso}$. In EGG they are all associated to the $\Gst$-invariant generalised vectors in the fundamental of $\Ex{7}$.

\subsection{\texorpdfstring{$\mathcal{N} = 8$}{} supergravity}

The field content of  $\mathcal{N} = 8$ supergravity in four dimensions \cite{Cremmer:1978ds} consists of the graviton, 28 electric vectors, 70 scalars, 
8 gravitini and  56 gaugini,  organised into a single (gravity) multiplet. The R-symmetry is $\SU{8}$ and the scalars parameterise the manifold 
\begin{equation}
\label{N8-sm}
\mathcal{M} = \frac{\Ex{7} \times \mathbb{R}^+}{\SU{8}/\mathbb{Z}_2} \, , 
\end{equation}
with $G_{\rm iso} = \Ex{7} \times \mathbb{R}^+$ the rigid isometry group. 
The subgroups of $G_{\rm iso}$ that can be gauged are determined by the embedding tensor $\Theta_I{}^\alpha$, with $I=1, \dots, 56$ and $\alpha = 1, \dots, 133$, which transforms in the $\rep{912}_{-1}$ of $ \Ex{7}$ \cite{deWit:2002vt,deWit:2007kvg}
\begin{equation}
     D_\mu  = \nabla_\mu - g A^I_\mu \Theta_I{}^\alpha t_\alpha  \, 
\end{equation}
where $t_\alpha$ are the $ \Ex{7}$ generators. 
\medskip

In generalised geometry maximally supersymmetric truncations correspond to a generalised identity structure  $\Gst = \id$.  The structure is defined by 56 generalised vectors $K_I$, $I=1, \dots, 56$, that give a Leibniz parallelisation of the generalised tangent bundle  \cite{Lee:2014mla, Baron:2014yua, Hohm:2014qga, Inverso:2017lrz}
\begin{equation}
\label{N8alg}
    L_{K_I} K_J = X_{IJ}{}^K K_K \,  , 
\end{equation}
where  $X_{IJ}{}^K$ are constant and $G(K_I, K_J) = \delta_{IJ}$ with $G$ the generalised metric.

The generalised vectors $K_I$
transform in the $\rep{28}_c$  of the $\SU{8}$ R-symmetry, and in the  $\rep{28}$ and $\rep{28}^\prime$ of $\SL{8, \mathbb{R}}$ (see Appendix \ref{PreliminariesE77_Mth})
\begin{equation}
\label{N8vec}
K_I  =  \{ K_{ij} , K^{ij} \} \qquad I=1, \ldots 56  \qquad i,j = 1, \ldots 8 \, , 
\end{equation} 
where $K_{ij}$ are  the 28 electric vectors of the truncated theory and  $K^{ij}$ their magnetic duals. 
The scalar manifold \eqref{N8-sm} is trivially obtained from  \eqref{scalarman}.

The tensor $X_{IJ}{}^K$ gives the intrinsic torsion of the identity structure: it transforms in the $\rep{912}_{-1}$ and 
is related to the embedding tensor of the truncated theory
\begin{equation}
    X_{IJ}{}^K = \Theta_I{}^\alpha (t_\alpha)_J{}^K \, ,  
\end{equation}
where $t_\alpha$ are the generators of the scalar isometry group ${\rm G}_{\rm iso} = \Ex{7}$.  The Leibniz property of the generalised Lie derivative~\cite{Coimbra:2011ky,Lee2014mla} 
\begin{equation}
\label{qconst}
[X_I, X_J] 
       =  - X_{I J}{}^K X_K \, , 
\end{equation}
with  $(X_I)_J{}^K=X_{I J}{}^K$ a matrix,  translates into the quadratic constraint  on the embedding tensor. 

\medskip

The theory with maximal gauge group $\SO{8}$ was constructed in 
 \cite{deWit:1982bul} and was  shown to be the consistent truncation of 11-dimensional supergravity on $S^7$ in \cite{deWit:1986oxb}. It was re-interpreted as a generalised Scherck-Schwarz reduction  in \cite{Lee:2014mla}. 
As shown in \cite{Lee:2014mla},  the intrinsic torsion only belongs to the component $\rep{36}$ in the decomposition
\begin{equation}
\rep{912} = \rep{36} \oplus \rep{36^\prime} \oplus \rep{420} \oplus
\rep{420^\prime}
\end{equation}
under $\SL{8, \mathbb{R}} \subset \Ex{7}$. The generalised Lie derivative \eqref{N8alg} among the generalised vectors \eqref{N8vec} gives the $\SO{8}$ algebra 
\begin{equation}
\label{So8}
    X_{[i i^\prime]  [j  j^\prime]}{}^{[k k^\prime]}=  - X_{[i i^\prime]}{}^{[k k^\prime]}{}_{ [j  j^\prime ]}  = R^{-1} (\delta_{ij} \delta_{i^\prime j^\prime}^{k k^\prime} - \delta_{i^\prime j} \delta_{i j^\prime}^{k k^\prime}-
    \delta_{ij^\prime} \delta_{i^\prime j}^{k k^\prime} + \delta_{i^\prime j^\prime } \delta_{ij}^{k k^\prime} ) \, ,
\end{equation}
where $[i,i^\prime]$, with $i, i^\prime= 1, \dots 8$, are antisymmetrised $ \SL{8,\mathbb{R}}$ indices,
$\delta_{i j }^{k k^\prime} =
 \delta_i^{[k} \delta_j^{k^\prime]}$,   and $R$ the radius of $S^7$. The tensor \eqref{So8}
reproduces the 4-dimensional embedding tensor for the $\SO{8}$ electric gauging.

In \cite{Guarino:2015vca} $\mathcal{N} = 8$ supergravity with a dyonic ISO(7) gauging has been obtained as a consistent truncation of massive type IIA supergravity on $S^6$. In this case it is natural to branch the various representations under $\SL{7,\mathbb{R}} \subset \SL{8,\mathbb{R}} \subset \Ex{7}$ so that the globally defined vectors arrange into 
\begin{equation}
\begin{aligned}
\rep{56}  & = \rep{28} \oplus \rep{28^\prime} = \rep{21} \oplus\rep{7} \oplus\rep{21^\prime} \oplus\rep{7^\prime} \\ 
K_I  & = (K_{ij} , K^{ij} ) = (K_{ab} , K_{a8}, K^{ab}, K^{a8} ) \, , 
\end{aligned}
\end{equation}
with  $i,j = ,1,8$ and  $a,b = 1, \dots, 7$. The non-zero components of $X_{IJ}{}^K$ in \eqref{N8alg} have a simple expression in terms of $\SL{8, \mathbb{R}}$ indices \cite{Cassani:2016ncu}
\begin{equation}
\begin{aligned}
    X_{[i i^\prime] [j j^\prime]}{}^{[k k^\prime]} & = -   X_{[i i^\prime] \, \quad [j j^\prime]}^{\quad  [k k^\prime]} = 8 \delta_{[i}^{[k} \theta_{i^\prime ] [ j} \delta_{j^\prime]}^{k^\prime]} \, , \\
 X^{[i i^\prime] \, \quad [j j^\prime]}_{\quad  [k k^\prime]} & =    -  X^{[i i^\prime] [k k^\prime]}{}_{[j j^\prime]} = 8  \delta_{[j}^{[i} \xi^{i^\prime ] [ k} \delta_{j^\prime]}^{k^\prime]}   \, , 
\end{aligned}
\end{equation}
where the tensors
\begin{equation}
    \theta_{ij} = \frac{1}{2 R}  \begin{pmatrix}
        \id_7 & \\ & 0
    \end{pmatrix}
    \qquad  \mbox{and} \qquad  \xi^{ij} = \frac{m}{2}  \begin{pmatrix}
        0_7 & \\ & 1
    \end{pmatrix}
\end{equation}
give the components of the embedding tensor corresponding to an element of the $\rep{28}$ and the singlet in the decomposition of the $\rep{36}$ of $\SL{8, \mathbb{R}}$ under $\SL{7, \mathbb{R}}$ \cite{Guarino:2015vca, Cassani:2016ncu}.  They correspond to the gauging of the ${\rm ISO}(7)$ group as it can be seen from 
the covariant derivative in 4 dimensions
\begin{equation}
     D_\mu  = 
    \nabla_\mu - g A^{ab}_\mu t_{[a}{}^c \theta_{b]c}  + g (\theta_{ab} A_\mu^{a 8} - \frac{m}{2} A_{\mu a 8}) t_8^b
\label{iso7g}
\end{equation}
where $t_a^c$ and $t_8^b$ are  $\SL{8, \mathbb{R}}$ generators,  and $t_{ab} =t_{[a}{}^c \theta_{b]c}$ gives the embedding of the $\SO{7}$ generators. 
From \eqref{iso7g} one easily sees that, while the $\SO{7}$ gauging is purely electric, the $\mathbb{R}^+$ is dyonic for non-zero Roman mass $m$.

\subsection{\texorpdfstring{$\mathcal{N} = 6$}{}  supergravity}
\label{sec:n6}

We find one truncation with $\mathcal{N} = 6$ supersymmetry corresponding to the generalised structure  $\Gst = \SU{2}$.\footnote{We also find $\mathcal{N} = 6$ truncations corresponding to $\Gst=\U{1}$ and $\Gst=\mathbb{Z}_2$. They have the same field content as the $\Gst = \SU{2}$ one.} 
From the commutant of $\Gst$ in $\SU{8}$ we recover the R-symmetry
of  $\mathcal{N} = 6$ supersymmetry  
 \begin{equation}
\label{N6-R}
   \Com{\SU{2}_S}{\SU{8}} =   \SU{6}_R \times \U{1}_R \, . 
\end{equation}

The structure group embeds 
in $\Ex{7}$ as 
\begin{equation}
\label{N6-E}
    \Ex{7} \to \SU{2}_S \times \SOs{12} \, , 
\end{equation}
where ${\rm G}_{\rm iso} = \SOs{12} $ is the global isometry group.
The $\Gst$-structure is determined by invariant generalised vectors  only. From the decomposition of the generalised tangent bundle
under \eqref{N6-E}
\begin{equation}
\begin{aligned}
    \rep{56} & \to (\rep{2} , \rep{12}) \oplus (\rep{1} , \rep{32^\prime})\\  
\end{aligned} 
\end{equation}
we see that there are 32 generalised vectors $K_I$ transforming in the $\rep{32^\prime}$ of the isometry group $\SOs{12}$. The vectors $K_I$ are normalised to $G(K_I, K_J) = \delta_{IJ}$, where again $G$ is the generalised metric.

By further decomposing under the R-symmetry $\SOs{12} \supset \SU{6} \times \U{1}$, the invariant vectors split into the 
 singlet and $\rep{15}$ representations  (and their conjugates)  $\SU{6}$
\begin{equation}
  K^0 \in \rep{1}_{-6} \, \quad  K^{[ij]} \in \rep{15}_{2} \qquad 
 K'^0 \in \rep{1}_{6} \, \quad  K'^{[ij]} \in \rep{\overline{15}}_{-2} \, . 
\end{equation}
They give the 16  vectors  of the truncated theory and their magnetic duals
\begin{equation}
    A_\mu^I = (A_\mu^0, A_\mu^{[ij]}, A_{\mu 0}, A_{\mu[ij]}) \, . 
\end{equation}
The scalar manifold is obtained from  \eqref{scalarman} with  $G= \SOs{12}$ and $H = \SU{6}\times\U{1}$, the R-symmetry
\begin{equation}
\label{N6-sm} 
\mathcal{M} = \frac{\SOs{12}}{\SU{6}\times\U{1}} \, . 
\end{equation}

From representation theory one can check that this truncation can be obtained as a consistent truncation of the $\mathcal{N}=8$ theory  where only the $\SU{2}_S$-singlets are kept. 

\medskip

The data above reproduce the field content of 4$d$  $\mathcal{N} = 6$ supergravity, namely 1 graviton, 6 gravitini, 16 vectors, 26 Majorana spin 1/2 fields and 30 scalars, all organised in a single multiplet of the $\SU{6} \times \U{1}$ R-symmetry group. The scalars parametrise the manifold  \eqref{N6-sm}. 

The gauging of the scalar isometries are given by the embedding tensor  
$\Theta_I{}^\alpha$ 
\begin{equation}
    D_\mu = \nabla_\mu - g A^I_\mu \Theta_I{}^\alpha t_\alpha \, ,
\end{equation}
where $t_\alpha$ ($\alpha=1, \dots, 66$) are the $\SOs{12}$ generators and $A_\mu^I$ are the 16 electric vectors and their magnetic duals ($I= 1, \dots ,32$).
The linear and quadratic constraints imply that $\Theta_I{}^\alpha$
transforms in the $\rep{352}$ of $\SOs{12}$.
This is exactly the representation of singlet intrinsic torsion in  the decomposition of the $\rep{912}$  under \eqref{N6-E}.
From the structure of the vector fields, we see that the largest subgroup of the scalar isometries that can be gauged is an electric (magnetic) $\SO{6} \times \SO{2}$. 

\medskip

 $\mathcal{N} =6$ gauged supergravity with $\SO{6}$ gauge group was obtained 
 as the consistent truncation of type II supergravity on $AdS_4 \times \mathbb{CP}^3$ in  \cite{Nilsson:1984bj}, where the relation to the maximally supersymmetric truncation of 11-dimensional supergravity  was also discussed.

\subsection{\texorpdfstring{$\mathcal{N} = 5$}{}  supergravity}

Also for $\mathcal{N} = 5$ supersymmetry  we find only one truncation, which reproduces the field content and embedding tensor of  $\mathcal{N} = 5$ supergravity.  The R-symmetry is $\SU{5} \times \U{1}$ and the  fields are arranged into a single gravity multiplet containing  1 graviton, 5 gravitini, 10 vectors, 11 Majorana spin 1/2 fields and 10 scalars parameterising the   manifold 
\begin{equation}
\label{N5-sm}
 \mathcal{M}= \frac{\SU{1,5}}{\SU{5}\times\U{1}}  \, .
\end{equation}

The truncation corresponds to the  generalised structure $\Gst = \SU{3}$. From the embedding  in $\SU{8}$ and $ \Ex{7}$
\begin{equation}
\label{N5-Rs}
\begin{aligned} 
    \SU{8} &  \supset  \SU{3}_S  \times \SU{5}_R \times \U{1}_R  \\
    \Ex{7} & \supset \SU{3}_S \times \SU{1,5}   \, ,    
\end{aligned} 
\end{equation}
we find that the R-symmetry is $\SU{5} \times \U{1}$, as expected,  and  the scalar isometries are  ${\rm G}_{\rm iso} =  \SU{1,5}$. 
Combining \eqref{N5-Rs} with \eqref{scalarman}  we recover the scalar manifold \eqref{N5-sm}.

From the decomposition of the fundamental of $\Ex{7}$ under \eqref{N5-Rs} 
\begin{equation}
\begin{aligned} 
    \rep{56}  & =  (\rep{3} , \rep{6})  \oplus  (\rep{\bar 3} , \rep{\bar 6}) \oplus  (\rep{1} , \rep{20}) \\
    &   =  [ (\rep{3} , \rep{5})_{1}  \oplus  (\rep{3} , \rep{1})_{-5} 
    \oplus  (\rep{\bar 3}  , \rep{\bar 5})_{-1}  \oplus  (\rep{\bar 3} , \rep{1})_{5} ]   \oplus 
[    (\rep{1}, \rep{10})_{-3} \oplus  (\rep{1} , \rep{\overline{10}})_{3}]
    \end{aligned} \, , 
\end{equation}
it follows that the $\Gst$-structure is defined by 20 invariant generalised vectors
\begin{equation}
  \{  K_I \} = \{ K_{[mnp]} \}   \qquad m,n,p = 0, \dots, 5
    \end{equation}
transforming in the $\rep{20}$ of the global isometry group and satisfying 
the normalisation condition $G(K_I, K_J) = \delta_{IJ}$, with $G$ the generalised metric. Splitting the $\SU{1,5}$ indices into $0$ and $i,j=1, \dots, 5$, the vectors splits into $K_{[ij]0}$ and
$K_{[ijk]}$, giving  the 10 vectors of the truncated theory in the $\rep{10}_{-3}$ of the R-symmetry group and their magnetic duals
\begin{equation}
A_\mu^I = ( A_\mu^{[ij]} \,  A_{\mu [ij]}) \qquad i,j =1, \ldots, 5 \, .
\end{equation}

The  intrinsic torsion contains two $\SU{3}$ singlets in the $\rep{70}$ and  $\rep{\overline{70}}$  of $\SU{5,1}$ 
\begin{equation}
\begin{aligned} 
    W_{\rm int} &  =   \rep{70} \oplus   \rep{\overline{70}} \, , 
\end{aligned} 
\end{equation}
corresponding to the  components of the embedding tensor
\begin{equation}
    \Theta_I{}^\alpha = ( \theta_{[mm],p}, \theta^{[mn],p} ) \qquad m,n,p= 0, \dots, 5 \, ,
\end{equation}
with $\theta_{[mn,p]}= \theta^{[mn,p]} =0$.  From the representations of the generalised vectors it follows that the largest compact gauging is an electric (magnetic) subgroup $\SO{5}$ of ${\rm G}_{\rm iso} =  \SU{5,1}$.

\medskip

$\mathcal{N}=5$ supergravity can also be  obtained as a truncation of  $\mathcal{N}=8$ gauged supergravity where only the $\Gst$-singlets are kept. 
An  example with gauge group $\SO{5}$, the largest possible compact gauging,  was constructed directly in four dimensions in \cite{deWit:1981yv}. No explicit  truncation with just $\mathcal{N}=5$ is known to us.

\subsection{\texorpdfstring{$\mathcal{N} = 4$}{} supergravity}
\label{sec:N4}

For $\mathcal{N} = 4$ supergravity  two kinds of multiplet are possible:  the graviton and  vector multiplets.
The former consists of 
 1 graviton, 4 gravitini, 6 vectors, 4 Majorana spin 1/2 fields and 2 scalars, while the latter are formed by 1 vector, 4 Majorana spin 1/2 fields and 6 scalars, all transforming in representation of the  $\SU{4}_R \sim \SO{6}_R$ R-symmetry.  In a theory with $n_{\rm V}$  vector multiplets the
scalar manifold is given by 
 \begin{equation}
 \label{N4-sm}
\mathcal{M} = \frac{\SO{6,n_{\rm V}}}{\SO{6}_R \times\SO{n_{\rm V}}}\times \frac{\SL{2, \mathbb{R}}}{\SO{2}} \, , 
\end{equation}
where the first factor is parametrised by scalars in the  vector multiplets and the second by those in the gravity multiplet. The global symmetry group is 
${\rm G}_{\rm iso} = \SO{6,n_{\rm V}} \times \SL{2,\mathbb{R}}$.

The gaugings of the theory 
\begin{equation}
D_\mu = \nabla_\mu - g A_\mu{}^{I a} \, f_{a I}{}^{JK} t_{JK} + g A_\mu{}^{I (a} \epsilon^{b ) c} \xi_{c I} t_{ab}  \, , 
\end{equation}
with $\nabla_\mu$  the spin-connection and $t_{IJ}$  and $t_{a b}$   the generators of $\SO{6,n_{\rm V}}$ and $\SL{2,\mathbb{R}}$, respectively, 
are determined by the embedding tensor \cite{Schon:2006kz} 
\begin{equation}
\label{N4-et}
\Theta_I{}^{\alpha} =  (  \xi_{a I}  \, ,   f_{a IJK}  ) \qquad 
a =1,2 \quad I=1,\dots, 6+n_{\rm V} \, , 
\end{equation}
whose components  transform  as doublets of  $\SL{2, \mathbb{R}}$ and as the fundamental and the three-index anti-symmetric representations of $\SO{6,n_{\rm V}}$, respectively.

 \medskip

We find six inequivalent truncations with $\mathcal{N} = 4$ supersymmetry associated to the structure groups 
\begin{equation}
\Gst  =  \Spin{6-n_{\rm V}} \, ,  \qquad n_{\rm V} = 0, \dots, 6 \, , 
\end{equation}
with $ \Spin{1}=\Spin{0}= \mathbb{Z}_2$ (see Appendix \ref{explicit_ex_subsec} for more details on the continuous structures). 
In what follows we will define the structure in terms of the corresponding orthogonal groups, since these are the ones acting on the invariant generalised vectors. The $\Gst$-structures correspond to the  embeddings\footnote{For $n_{\rm V}= 0,2,4$ the truncations correspond to regular branching while for $n_{\rm V} = 1,3$ they arise from non-regular ones. }
\begin{equation}
\label{N4-em}
\begin{aligned} 
    \Ex{7} & \supset \SO{6-n_{\rm V}}_S \times \SO{6,n_{\rm V}} \times \SL{2, \mathbb{R}} \,   \\
    \SU{8} & \supset \SO{6-n_{\rm V}}_S \times \SO{6} \times \SO{n_{\rm V}} \times \U{1} \, . 
    \end{aligned}  
\end{equation}

The generalised  $\SO{6-n_{\rm V}}$-structure is defined by an $\SL{2, \bbR}$ doublet of  $6+ n_{\rm V}$ generalised vectors 
in the decomposition of the generalised tangent bundle
\begin{equation}
\begin{aligned} 
\label{N4-vec}
    \rep{56} & =  [ (\rep{6} - n_{\rm V} , \rep{1}, \rep{2}) \oplus  (\rep{1} , \rep{6} +n_{\rm V} , \rep{2} ) ]  
     \oplus [ (\rep{S}_{\SO{6-n_{\rm V}}} , \rep{S}_{\SO{6,n_{\rm V}}} ,\rep{1} ) \oplus c. c. ]  \, ,
      \end{aligned} 
\end{equation}
where  $\rep{S}_{\SO{6-n_{\rm V}}}$  and  $\rep{S}_{\SO{6,n_{\rm V}}}$ denote the spinorial representiations of $\SO{6-n_{\rm V}}$ and $\SO{6,n_{\rm V}}$ respectively. The $K_{I a}$ singlets  satisfy the compatibility conditions
\begin{equation}
    s(K_{I a}, K_{J b}) = \kappa^2 \delta_{IJ} \epsilon_{ab} \qquad 
    \forall \,  I,J = 1, \dots, 6+  n_{\rm V} \, \quad  \forall \, a,b=\pm       \, ,
\end{equation}
where $s(V,V^\prime)$ is the $\Ex{7}$
simplectic invariant,  $\epsilon_{ab}$ is the $\SL{2, \mathbb{R}}$ invariant antisymmetric tensor and $\kappa = (\det T^*M)^{1/2}$ (see also \cite{Malek:2017njj}).
The generalised vectors give
$6+n_{\rm V}$ electric vectors and $6+n_{\rm V}$ magnetic vectors in the truncated theory
\begin{equation}
    A_\mu^{I \alpha}= (A_\mu^{I +} , A_\mu^{I -} ) \qquad I= 1, \ldots, 6 + n_{\rm V} \, , 
\end{equation}
where $\pm$ denote the charges under $\SO{2}\subset \SL{2, \mathbb{R}}$. 
Moreover, under $\SO{6,n_{\rm V}} \supset \SO{6} \times \SO{n_{\rm V}}$, 
the singlet vectors decompose as $\rep{6} +  n_{\rm V} =(\rep{6}, 1) \oplus (\rep{1}, n_{\rm V})$
\begin{equation}
\begin{aligned} 
A_\mu^{I +} & = ( A_\mu^{i +} , A^+_{\mu a} ) \\
A_\mu^{I -}& = ( A_\mu^{i -} , A^-_{\mu a} )
\end{aligned}   \qquad 
\begin{aligned}  &i= 1, \ldots, 6 \\
 &a =1, \dots, n_{\rm V} 
 \end{aligned}
\end{equation}
where the vectors in the $\rep{6}$ belong to the gravity multiplet and singlets to the vector multiplets.

The scalar isometries are 
${\rm G}_{\rm iso} = \SO{6,n_{\rm V}} \times \SL{2, \mathbb{R}}$, 
and,  from \eqref{N4-em} and \eqref{scalarman} we reproduce the scalar manifold   \eqref{N4-sm}. 

The $\Gst$-singlets in  the intrinsic torsion reproduce the embedding tensor  \eqref{N4-et} 
\begin{equation}
\begin{aligned}
    W_{\rm int}  =   (\rep{6 + n_{\rm V}}, \rep{2}) \oplus  ( \rep{X}_{[IJK]},\rep{2})  \, , 
    \end{aligned} 
\end{equation}
 where the first term transforms in the fundamental of $\SO{6+ n_{\rm V}}$ and   the second in  three-index anti-symmetric representation. Both are doublets of $\SL{2, \mathbb{R}}$. From the number of vectors we see that the maximal compact gauging is the electric (magnetic) subgroup $\SO{4} \times \U{1}$ of the R-symmetry. 

\medskip

Four-dimensional $\mathcal{N}=4$ gauged supergravities without and with vector multiplets were constructed since the seventies \cite{Das:1977uy,Cremmer:1977tc,Cremmer:1977tt,Freedman:1978ra,deRoo:1984zyh,Bergshoeff:1985ms,deRoo:1985jh,Bergshoeff:1985ms} and the general reformulation in terms of the embedding tensor can be found in \cite{Schon:2006kz,DallAgata:2023ahj}.  
Some examples of the truncations discussed above can be  found in the literature. Pure supergravity ($n_{\rm V}=0$) with 
$\SO{4}$ gauging is obtained truncating 11-dimensional supegravity on $S^7$  \cite{Cvetic:1999au} or as a reduction of IIB supergravity on $S^1 \times S^5$ \cite{Guarino:2024gke}.
$\mathcal{N}=4$ supergravity  with $n_{\rm V} = 3$ vector multiplets (and gauge group $\SO{3} \ltimes_{\rep{3} \oplus \rep{3}} {\rm nil}_{(6,3)}$) was derived in \cite{Cassani:2011fu} as the  universal truncation of 11-dimensional supergravity on  tri-Sasakian manifolds. 
The derivation in the context of EGG can be found in \cite{Duboeuf:2023dmq}. 
In \cite{Cassani:2011fu} an $\mathcal{N}=4$ supergravity with $n_{\rm V} = 4$ was found  when the tri-Sasakian manifold is taken to be $N^{010}$. The extra vector multiplet corresponds to a Betti-multiplet arising from the non-trivial cohomology of $N^{010}$.

\subsection{\texorpdfstring{$\mathcal{N} = 3$}{} supergravity}
\label{sec:N3}

The field content of $\mathcal{N}=3$ supergravity \cite{Castellani:1985ka}  is again given by one gravity multiplet, consisting  of  1 graviton, 3 gravitini, 3 vectors and 1 Majorana spin 1/2 field,
and $n_{\rm V}$ vector multiplets containing  1 vector, 4 Majorana spin 1/2 fields and 6 scalars.
The R-symmetry is $\SU{3}\times \U{1}$ and the scalars parameterise the coset
 \begin{equation}
\mathcal{M} = \frac{\SU{3,n_{\rm V}}}{\Suni{\U{3}\times\U{n_{\rm V}}}} \, ,
\end{equation}
where the denominator is locally isomorphic to $S\U{3}\times\U{n_{\rm V}}$.

The gaugings of the scalar isometry group $\SU{3,n_{\rm V}}$ are given by the embedding tensor
\begin{equation}
\label{N3-embt}
    \Theta_I{}^\alpha = (\theta_{IJ}{}^K, \theta^{IJ}{}_K ) \, ,
\end{equation}
where $I,J,K = 1, \dots, 3 + n_{\rm V}$ and  $\theta_{IJ}{}^K = \theta_{[IJ]}{}^K$. 
\medskip

$\mathcal{N}=3$ truncations correspond to generalised structures $\Gst \subseteq \SU{5}$, where $\SU{5}$ is the commutant of the R-symmetry in $\SU{8}$ 
\begin{equation}
\label{N3-comp}
\SU{8} \supset   \SU{5}_S \times \SU{3}   \times \U{1}  \, . 
\end{equation}

The largest structure, $\Gst = \SU{5}$,
corresponds to a truncation to  pure supergravity. It is defined by six invariant generalised vectors  
\begin{equation}
    \{K_I\} = \{K_{i}, K^i \} \, , \qquad i=1,2,3 \, ,
\end{equation} 
satisfying 
\begin{equation}
    s(K_I, K_J) = \kappa^2 \delta_{IJ} \, .
\end{equation}
Decomposing  the fundamental  of $\Ex{7}$  under 
\begin{equation}
\label{N3-E7}
 \Ex{7} \supset  \SU{5}_S \times \U{1} \times \SU{3}   \,,
 \end{equation}
the six singlet vectors arrange into two triplets of the R-symmetry group
\begin{equation}
\label{N3-dec}
\begin{aligned} 
 \rep{56} & = (\rep{5},\rep{3})_1 \oplus (\rep{1},\rep{3})_{-5} \oplus (\rep{10}, \rep{1})_{-3} \oplus c.c  \\
\end{aligned} 
\end{equation}
corresponding to the three vectors in the gravity multiplet and their duals.
Combining \eqref{scalarman} with 
\eqref{N3-E7} and \eqref{N3-comp}, it is easy to verify that there are no scalars. Thus we recover the field content of pure supergravity.

\medskip

For $\Gst \subset \SU{5}$ extra vector multiplets are possible. Since the vectors in the vector multiplets are singlets of the R-symmetry, 
they correspond to $\Gst$-singlets in $(\rep{10}, \rep{1})_{-3}$ and $(\rep{\overline{10}}, \rep{1})_3$ of \eqref{N3-dec}. Together with those in the gravity multiplets they  transform in  the fundamental and antifundamental of the global $\SU{3,n_{\rm V}}$ symmetry group 
\begin{equation}
( \rep{3} + n_{\rm V} ) \oplus (\overline{\rep{3}+n_{\rm V}})  \, . 
\end{equation}
Under $\SU{3} \times \SU{n_{\rm V}} \times \U{1}$ the $3+n_{\rm V}$ vectors split as $(\rep{3}, \rep{1})_{-1}$ and $(\rep{1}, n_{\rm V})_{3/n_{\rm V}}$ (plus their conjugates) and  give the electric and 
magnetic vectors of the truncated theory
\begin{equation}
\{A^I_\mu \} = \{A_\mu^{ij} , A^a_\mu , A_{\mu ij} , A_{\mu a} \}  \qquad i=1,2,3, \,\, a= 1, \dots, n_{\rm V} \, .
\end{equation}

Decomposing the intrinsic torsion under $\Gst \times \SU{3, n_{\rm V}}$
we find that the $\Gst $-singlets transform in the 
\begin{equation}
     \rep{N}  \oplus  \frac{\rep{N} (\rep{N}-2) (\rep{N}+ 1)!}{2 \rep{N}!}  \qquad \quad  (\rep{N} = \rep{3} + n_{\rm V})
\end{equation}
of the global symmetry group ${G}_{\rm iso} = \SU{3,n_{\rm v}}$. These representations correspond to the $\SU{3,n_{\rm v}}$ irreducible representations of \eqref{N3-embt}.

\medskip

The list of allowed truncation is given in the table below 

\begingroup
\renewcommand{\arraystretch}{1.5}
\begin{table}[H]
\begin{center}
\begin{tabular}{|c|c|c|}
\hline
$n_{\rm V}$ & $G_S$  & $\mathcal{M}$\\
\hline
 0  &  $\SU{5}$ &  $\mathbb{{1}}$ \\  
 1  & $ \SU{3} \times \SU{2}$ &  $\frac{\SU{3,1}}{\SU{3}\times\U{1}}$   \\ 
 2 & $\U{1}^2$ &  $\frac{\SU{3,2}}{\SU{3}\times\SU{2} \times \U{1}}$   \\ 
 3  & $ \U{1}$ & $\frac{\SU{3,3}}{\SU{3}\times\SU{3} \times \U{1}}$   \\ 
 4  & $ \mathbb{Z}_6$ & $\frac{\SU{3,4}}{\SU{3}\times\SU{4} \times \U{1}}$   \\
\hline
\end{tabular}
\end{center}
\caption{$\mathcal{N}=3$ truncations and associated $\Gst$-structures}
\label{N=3tr}
\end{table}
\endgroup

Four-dimensional gauged $\mathcal{N}=3$ supergravity coupled to vector mutiplets was constructed in \cite{Castellani:1985ka}, and explicit gaugings inside $\SO{3,n_{\rm V}}$ were studied  in \cite{Karndumri:2016miq}. We are not aware of exciplit truncations to  $\mathcal{N}=3$ supergravity that are not 
subtruncations of a more supersymmetric theory.

\subsection{\texorpdfstring{$\mathcal{N} = 2$}{} supergravity}
\label{sec:N2}

In $\mathcal{N} = 2$ supergravity  the fields are arranged into a graviton multiplet, $n_{\rm V}$ vector multiplets and $n_{\rm H}$ hyper-multiplets all carrying a representation of the $\SU{2}_R \times \U{1}_R$ R-symmetry. The graviton multiplet contains the metric, the graviphoton, and an $\SU{2}_R$ doublet of gravitini. The vector multiplets consist of a vector, 2 spin 1/2 and 1 complex scalar, while the hypermultiplets contain 2 spin 1/2 fermions and 4  real scalar fields.  

The scalars in the vector multiplets parametrise a special K\"ahler manifold $\mathcal{M}_{\rm V}$ of complex dimension $n_{\rm V}$ and those in the hypermultiplets 
parametrise a quaternionic K\"ahler manifold $\mathcal{M}_{\rm H}$ of real dimension $4 n_{\rm H}$. Together, the scalar manifold is given by the product
\be
\calM = \calM_{\rm V}\times \calM_{\rm H} \, . 
\ee

 The isometry group of the scalar manifolds also splits into two factors
 ${\rm G_{\rm iso}} = G_{\rm V}\times G_{\rm H}$ acting on the scalars in the vector and in the hypermultiplets, respectively. 

The gauging of the scalar isometries can be expressed in terms of the embedding tensor, which consists of two parts
\begin{equation}
\label{N2-et}
(\Theta_{\tilde I}{}^a , \Theta_{\tilde I}{}^A)    
\end{equation}
corresponding to symmetries of the vector and hyper-multiplet scalars.
In the above equation ${\tilde I}= 0,  \hat{0}, \ldots, 2 n_{\rm V}$ runs over the number of electric and magnetic vectors in the theory,  while $a=1, \dots \dim G_{\rm V}$ and $A= 1, \dots \dim G_{\rm H}$ span the generators of $G_{\rm V}$ and $G_{\rm H}$. 
The embedding tensor determines the combination of Killing vectors on $\calM_{\rm V}$ and $\calM_{\rm H}$ that are gauged  
\begin{equation}
 k^i_{\tilde I} = \Theta_{\tilde I}{}^a k_a^i(\phi) \qquad k^x_{\tilde I} = \Theta_{\tilde I}{}^A k^x_A(q) \, , 
\end{equation}
where $\phi^i$, $i = 1, \dots, n_{\rm V}$, 
are the scalars in the vector multiplets and $q^x$, $x= 1, \dots, 4 n_{\rm H}$ those in the hypermultiplets. 
On the scalars the gaugings are defined via the covariant derivatives
\begin{equation}
\begin{aligned}
    &\mathcal{D}_\mu \phi^i = \partial_\mu \phi^i + i g k^i_{\tilde I} A^{\tilde I}_\mu \, , \\
    &  \mathcal{D}_\mu q^x = \partial_\mu q^x + i g k^x_{\tilde I} A^{\tilde I}_\mu  \, . 
    \end{aligned}
\end{equation}

\medskip

In generalised geometry the largest structure group that is compatible with $\mathcal{N}=2$ supersymmetry in four dimensions is $\Gst= \SU{6}$, the commutant in  $\SU{8}$ of the R-symmetry. The decomposition of the
spinorial representation of $\SU{8}$ under the embedding $\SU{8}  \supset  \SU{6}  \times \SU{2} \times \U{1}$ 
 \begin{equation}
\label{N2-R}
\rep{8}   \to (\rep{6}, \rep{1})_{1} \oplus  (\rep{1} , \rep{2})_{-3} \, ,
\end{equation}
contains the two $\SU{6}$-singlets corresponding to  the supersymmetry parameters of the truncated theory.

The $\SU{6}$-structure is defined by two generalised vectors $K, \hat{K} \in \Gamma(E)$ and a triplet of weighted adjoint elements $J_\alpha \in \Gamma ((\det T^*M)^{1/2} {\rm ad}F)$, with $\alpha =1,2,3$, defining a highest root $\su(2)$ subalgebra of $\mathfrak{e}_{7(7)}$ \cite{Ashmore:2015joa}. Together they satisfy 
\begin{equation}
\begin{aligned}
\label{eq:gencomp}
    J_\alpha \cdot K = J_\alpha \cdot \hat{K} =0  \qquad \qquad 
    Tr(J_\alpha, J_\beta) = -  2 \sqrt{q(K)} \delta_{\alpha \beta} \, ,
\end{aligned}
\end{equation}
with $q(K)$  the quartic invariant of $\Ex{7}$. 
The vectors $K$ and $\hat{K}$ 
are the only singlets in the decomposition under  $\Ex{7} \supset \SU{6}  \times \SU{2} \times \U{1}$  of the fundamental of  $\Ex{7}$ 
\begin{equation}
\label{SU6-N2-a}
    \rep{56}  = (\rep{6}, \rep{2})_{-2} \oplus  (\rep{15}, \rep{1})_{2} \oplus  (\rep{1}, \rep{1})_{-6} \oplus c.c \, ,
\end{equation}
and  give the graviphoton $A^0_\mu$  and its magnetic dual.
The singlets in the adjoint
\begin{equation}
\begin{aligned}
\label{SU6-N2-b}
     \rep{133} & = (\rep{35}, \rep{1})_{0}  \oplus  (\rep{1}, \rep{3})_{0} \oplus (\rep{1}, \rep{1})_{0}  \\
     & \quad  \oplus [    (\rep{6}, \rep{2})_{4} \oplus  
(\rep{15}, \rep{1})_{-4} \oplus c.c]  \oplus (\rep{20}, \rep{2})_{0}    
\end{aligned}
\end{equation}
correspond to the generators of $\SU{6} \times \SU{2} \times \U{1}$, where the $\SU{2}$ R-symmetry is generated by the invariant tensors $J_\alpha$. 

Moreover, since $\Com{\SU{6}}{\Ex{7}} = \Com{\SU{6}}{\SU{8}} = \SU{2}\times \U{1}$, 
from \eqref{scalarman} it follows that the scalar manifold is trivial.
As expected, this corresponds to the truncation to pure supergravity.

\medskip

In order to obtain truncations with vector and/or hypermultiplets, the
generalised structure must be a subgroup of $\SU{6}$.
Generically the $\Gst$ structure is defined by a set of generalised $\Gst$-invariant vectors and adjoint elements 
\begin{equation}
    \{ K_{\tilde I}, J_A \} \qquad \tilde{I}=  0, \hat{0}, \ldots, 2 n_{\rm V}  \qquad A = 1,  \ldots, \dim G_{\rm H}
\end{equation}
satisfying
\begin{equation}
\label{eq:n2comp}
 J_A  \cdot  K_{\tilde I}  = 0 
\end{equation}
for any  $\tilde{I}  =  0, \hat{0}, \ldots, 2 n_{\rm V} $ and $A = 1 \ldots \dim G_{\rm H}$.
The generalised vectors $K_{\tilde I}$ transform as a vector of the  $\Sp{2+ 2 n_{\rm V},\mathbb{R}}$ symplectic group and satisfy
\begin{equation}
    s( K_{\tilde I} , K_{\tilde J} ) =  \kappa^2 \Omega_{\tilde{I} \tilde{J}} 
\end{equation}
where $s(\cdot, \cdot)$ is the $\Ex{7}$ symplectic invariant (see for instance \cite{Ashmore:2015joa})  and 
$\Omega_{\tilde{I} \tilde{J}} $
is  the $\Sp{2 + 2 n_{\rm V}, \mathbb{R}}$ invariant matrix
\begin{equation}
 \Omega =    \begin{pmatrix}
    0 & \id_{n_{\rm V}+1} \\
   -  \id_{n_{\rm V}+1} &  0
\end{pmatrix} \, . 
\end{equation}

Condition \eqref{eq:n2comp} implies  that the extra singlet generalised vectors 
must be invariant under the $\SU{2}$ R-symmetry and therefore must come 
from  the $\Gst$-singlets in the $(\rep{15}, \rep{1})_{2}$ and its conjugate in \eqref{SU6-N2-a} 
\begin{equation}
 \{ K_{\tilde I} \}  =  \{ K, \hat{K}, K_i, \hat{K}_i \}  \qquad i = 1, \dots, n_{\rm V} \, .
\end{equation}
They give the vectors in the vector multiplets and their magnetic duals
\begin{equation}
 A^{\tilde I} =   \{ A^0_\mu, A_{0 \mu}, A^i_\mu, A_{i \mu} \}   \qquad i = 1, \dots, n_{\rm V} \, .
\end{equation}

The invariant adjoint elements generate the group $G_{\rm H}$ 
\begin{equation}
[J_A,J_B] = \kappa f_{A B}{}^C J_C \, , 
\end{equation}
of the isometries of  the hypermultiplet scalar manifold and can be normalised as
\begin{equation}
    \tr(J_A J_B) = \kappa^2 \eta_{AB} \, , 
\end{equation}
where $\eta_{AB} $ is a diagonal matrix with $-1$ and $+1$ entries in correspondence with compact and non-compact generators of $G_{\rm H}$, respectively. The group $G_{\rm H}$ always contains the $\SU{2}$ R-symmetry, as it can be seen from \eqref{SU6-N2-b}, with the other invariant adjoint elements coming from the $\Gst$-singlets in  the  $(\rep{20}, \rep{2})_{0}$.

The scalar manifold is again given by \eqref{scalarman}.
For $\Gst$-structures that give strict $\mathcal{N}=2$ supersymmetry, 
one can show that the coset manifold \eqref{scalarman}
factorises \cite{Josse:2021put}. Indeed,  the commutant $\Com{\Gst}{\Ex{7}}$ cannot contain elements that change the structure while leaving the generalised metric invariant.  This means that the elements of $\Com{\Gst}{\Ex{7}}$ must split into two groups
\begin{equation}
\Com{\Gst}{\Ex{7}} = \Com{\Gst}{G_{J_A}} \times   \Com{\Gst}{G_{K_I}} 
\end{equation}
where $\Com{\Gst}{G_{K_I}}$ is the subgroup  of $\Ex{7}$ that leaves invariant all generalised vectors $K_I$
while  $\Com{\Gst}{G_{J_A}}$ is the one leaving fixed the adjoint elements $J_A$.\footnote{In the same way, we denote by $\Com{\Gst}{H_{K_I}}$ and $\Com{\Gst}{H_{J_A}}$ denote subgroups of $\SU{8}$ that leave invariant all generalised vectors $K_I$ and all  adjoint elements $J_A$, respectively. } 
Then the scalar manifold factorises as expected 
\begin{equation}
\label{N2-scalarm}
\begin{aligned}
    \calM  &  = 
\frac{\Com{\Gst}{G_{J_A}}}{\Com{\Gst}{H_{J_A}}} \times  \frac{\Com{\Gst}{G_{K_I}}}{\Com{\Gst}{H_{K_I}}}  \\
&  =  \frac{G_{\rm V}}{H_{\rm V}} \times \frac{G_{\rm H}}{H_{\rm H}} =  \calM_{\rm V}\times \calM_{\rm H} \, .
\end{aligned}
\end{equation}
In the above expressions  $G_{\rm V}$ and $G_{\rm H}$ ($H_{\rm V}$ and $H_{\rm H}$) are the groups that  remain after cancellation of possible 
common factors between the numerators and denominators, and correspond to the isometries of the vectors and hypermultiplet scalars. 
From $\eqref{N2-scalarm}$  it follows that the number of non-compact invariant adjoint singlets determine the number of hypermultiplets of the truncated theory. 

\medskip

The two components of the embedding tensor in \eqref{N2-et} are reflected in the torsion of the $\Gst$-structure, which also has two components 
\begin{equation}
\label{eq;torsionLDN24d}
    \begin{aligned}
    L_{K_{\tilde I}} K_{\tilde J} &  = - T_{\rm int}(K_{\tilde I}) \cdot K_{\tilde J}  = t_{{\tilde I} {\tilde J}}{}^{\tilde K} K_{\tilde K} \,  ,  \\
 L_{K_{\tilde I}} J_{A}  & = - T_{\rm int}(K_{\tilde I}) \cdot J_{A}  = p_{{\tilde I} A}{}^{B} J_{B}  \, , 
    \end{aligned}
\end{equation}
where the matrices 
$(t_{\tilde I})_{\tilde J}{}^{\tilde K}$ and 
$(p_{\tilde I})_{A}{}^{B}$  are  constant and give the elements of Lie algebrae of $G_{\rm V}$  and $G_{\rm H}$ respectively. 

For pure $\mathcal{N}=2$ supergravity, the intrinsic torsion of the $\SU{6}$ structure contains two singlet representations 
\begin{equation}
    W_{\rm int} \supset  (\rep{1}, \rep{3})_{-3}  \oplus (\rep{1},\rep{3})_{3}          
\end{equation}
transforming in the adjoint of $\SU{2}_R$. The only non-zero component in \eqref{eq;torsionLDN24d} is $p_{\tilde{I} \alpha}{}^\beta$  with  $\tilde{I} = (0, \hat{0})$ labelling the graviphoton and its dual. It  corresponds to the FI-terms for the gauging of the R-symmetry action on the fermions.  
For $\Gst \subset \SU{6}$ extra singlets appear in the intrinsic torsion giving a large variety of possible gaugings. We comment further on this in the rest of the section.

\medskip
We find truncations containing only vectors multiplets or hypermultiplets, and some truncations with both.

\subsubsection{Truncations with only vector multiplets}
\label{sec:onlyvec}

Truncations with no hypermultiplets correspond to 
$\Gst$-structures with only a triplet of singlet invariant adjoint elements $J_\alpha$, $\alpha= 1,2,3$, which generate the $\SU{2}$ R-symmetry of the truncated theory. 
This means that the relevant $\Gst$-structures embed as
\begin{equation}
\begin{aligned}
 &   \Gst \subset \SOs{12} \sim {\rm Spin}^*(12) \subset \Ex{7} \, ,  \\
 &   \Gst \subset \SU{6} \subset \SU{8} \, ,
 \end{aligned}
\end{equation}
where $\SOs{12}$ is the stabiliser of the triplet of $J_\alpha$.

The group $\SOs{12}$ embeds in $\Ex{7}$ as 
\begin{equation}
\label{eq:veconlysp}
    \SOs{12} \times \SU{2}_R \subset \Ex{7} \, .
\end{equation} 
The compatibility condition \eqref{eq:gencomp} implies that the  generalised vectors $K_{\tilde I}$  are invariant under the $\SU{2}_R$ generated by $J_\alpha$. It follows that in the decomposition of the   fundamental of $\Ex{7}$ under  \eqref{eq:veconlysp}
\begin{equation}
\label{eq:fundStoSO12}
    \rep{56} = (\rep{12} ,\rep{2}) \oplus (\rep{32^\prime} ,\rep{1})
\end{equation}
the extra $\Gst$-invariant vectors can only come from the $(\rep{32^\prime} , \rep{1})$ component. Thus there is an upper 
 bound of $n_{\rm V} \leq15$ vector multiplets.  Decomposing the $(\rep{32^\prime} , \rep{1})$ under $\SU{6} \times \U{1}$ we recover the two components 
  $(\rep{15}, \rep{1})_{2}  \oplus c.c$ of 
\eqref{SU6-N2-a}. 

The torsion in the $\rep{912}$ decomposes under \eqref{eq:veconlysp} as
\begin{equation}
    \rep{912} = (\rep{351}^\prime, \rep{1} ) \oplus (\rep{32^\prime},\rep{3} ) \oplus (\rep{220},\rep{2} ) \oplus (\rep{12},\rep{2} )  \, . 
\end{equation}
For truncations with only vector multiplets, one can check that  the  intrinsic torsion is contained in the components $(\rep{351}^\prime, \rep{1} ) \oplus (\rep{32^\prime},\rep{3})$, where the former provides the gauging of the vector multiplet isometries and the latter of the $\SU{2}_R$ symmetry. 
We will not discuss the intrinsic torsion and the possible gaugings in more details.

\medskip

We find truncations corresponding to all scalar manifolds listed in Table \ref{hmn2}  but the last one. This is expected since the groups at the  numerator and denominator cannot be contained in $\Ex{7}$ and $\SU{8}$, respectively. 

\begin{enumerate}
\item We find a class of truncations with scalar manifold 
\begin{equation}
 \mathcal{M}_{\rm V} =  \frac{\SU{1,n_{\rm V}}}{\SU{n_{\rm V}} \times \U{1}} 
\end{equation}
with $n_{\rm V} = 1, \dots, 4$, vector multiplets. Together with the graviphoton and their magnetic duals, the vectors   transform in the $(\rep{1} +n_{\rm V}) \oplus (\rep{1} +n_{\rm V})^\prime$ representations of the global isometry group. \\
The structure group for $n_{\rm V} =1$ is 
\begin{equation}
\Gst = \SU{4} \times \SU{2} \, , 
\end{equation}
while for $n_{\rm V} =2,3,4$ it reduces to  
\begin{equation}
\Gst = \SU{5-n_{\rm V}} \times \U{1} \, , 
\end{equation}
with  $\SU{5-n_{\rm V}}  \subset \SU{4}$.

\item 
We also find the family of truncations with scalar manifolds 
\begin{equation}
 \mathcal{M}_{\rm V} =  \frac{\SL{2,\mathbb{R}}}{\SO{2}} \times \frac{\SO{2,n_{\rm V}-1}}{\SO{2} \times\SO{n_{\rm V}-1}}  
\end{equation}
and $n_{\rm V} = 1, \dots, 5$ vector multiplets, 
corresponding to the  generalised structure groups
\begin{equation}
    \Gst = \Spin{7-n_{\rm V}} \times \SU{2} \, . 
\end{equation}
The $n_{\rm V}$ vectors coming from the $(\rep{15}, \rep{1})_2$ and  $(\rep{\overline{15}}, \rep{1})_2$ in \eqref{SU6-N2-a} combine with the two singlets giving the graviphoton and its magnetic dual into the  $(\rep{2}, \rep{1} + n_{\rm V})$ representation of the 
$ \SL{2, \mathbb{R}} \times \SO{2,n_{\rm V}-1}$ global isometry.

\item 
There is a  truncation with $n_{\rm V}=1$ vector multiplet and the  scalar manifold 
\begin{equation}
\mathcal{M}_{\rm V} = \frac{\SU{1,1}}{\U{1}}  \, . 
\end{equation}
 It  is associated  to the structure group
\begin{equation}
    \Gst=\USp{6} \subset \SU{6} \, . 
\end{equation} 
Under this embedding the $(\rep{15}, \rep{1})_2$ splits into $(\rep{14}, \rep{1})_2 \oplus  (\rep{1}, \rep{1})_2$, giving, together with its conjugate, an extra vector and its magnetic dual. Together with the graviphoton and its magnetic dual they transform in the $\rep{4}$
of the global symmetry group $\SU{1,1}$.
It corresponds to the third line of Table \ref{hmn2}.

\item  We also find a  truncation with $n_{\rm V}=6$ vector multiplets and scalar manifold 
\begin{equation}
\label{Usp6coset}
\mathcal{M}_{\rm V} = \frac{\USp{6}}{\U{3}} \, .
\end{equation}
It  is associated to  $\Gst=\SU{2}$, which embeds in $\SU{6}$ as 
$\SU{2}_S \times \SU{3} \subset \SU{6}$. 
The decomposition $(\rep{15}, \rep{1})_2 = (\rep{6}, \rep{1}) \oplus (\rep{3}, \rep{3})$ gives the six extra vectors. Togheter with the graviphoton and their magnetic duals, they transform in the $\rep{14^\prime}$ of the global isometry group $\USp{6}$.

\item 
The truncation with scalar manifold 
\begin{equation}
\mathcal{M}_{\rm V} = \frac{\SU{3,3}}{S[\SU{3}\times \SU{3}]}
\end{equation}
is obtained from a $\Gst= \U{1}$ structure, embedded in 
$\SU{6} \supset \SU{3}\times \SU{3}\times  \U{1}$. The theory has $9$ vector multiplets, whose vectors come from the decomposition of the  $(\rep{15}, \rep{1})_{2} \ni  (\rep{3} ,\rep{3})_{0,2}$ under this breaking. The 9 vectors, the graviphoton and their magnetic dual arrange in the $\rep{20}$ of the $\SU{3,3}$ global isometry group.

\item 
Finally, we also find a truncation with $n_{\rm V} = 15$ vector multiplets  with scalar manifold 
\begin{equation}
\label{maxvectr}
\mathcal{M}_{\rm V} =\frac{\SOs{12}}{\SU{6}\times\U{1}} \, . 
\end{equation}
It corresponds to a $\mathbb{Z}_2$ structure, where the  $\mathbb{Z}_2$ is in the centre of $\SOs{12}$, since it must commute with $\SU{2}_R$
\begin{equation}
 \Com{\SU{2}_R \times \SOs{12}}{\Ex{7}} = C(\SOs{12}) \, . 
\end{equation}
When embedded in $\SU{6}$, the 
 $\mathbb{Z}_2$ structure acts as a reflection on the fundamental of $\SU{6}$. The vectors in $(\rep{15}, \rep{2})_2$ of \eqref{SU6-N2-a} can be seen as six-dimensional 2-forms and therefore are left invariant  by the 
$\mathbb{Z}_2$ action.
Together with the graviphoton and their magnetic duals, they transform in the $\rep{32}^\prime$ of the global isometry group  (see also  Section \ref{sec:discretest}).
\end{enumerate}

In Table \ref{table:TableN=2NH=0} below we summarise the truncations we find according to the number of vector multiplets. 
The symbol $\star$ denote the $\Gst$-structures coming from special branchings (see Section \ref{Section_Algorithm}). For any truncation, we list the largest $\Gst$-structure that allows for it.

\begingroup
\renewcommand{\arraystretch}{1.5}
\begin{table}[H]
\begin{center}
\begin{tabular}{|c|c|c|}
\hline
$n_{\rm V}$ & $G_S$ & $\mathcal{M}_v$ \\
\hline
\multirow{2}{*}{$1$}& $\SU{4}\times \SU{2}$ & $\frac{\SU{1,1}}{\U{1}}$     \\ \cline{2-3}
& $\USp{6}^\sast$ & $\frac{\SU{1,1}}{\U{1}}$    \\
\cline{1-3}
\multirow{3}{*}{$2$}& $\Spin{5}\times \SU{2}^\sast$ & $\left(\frac{\SU{1,1}}{\U{1}}\right)^2$     \\\cline{2-3}
& $\SU{3} \times \U{1}$ & $\frac{\SU{2,1}}{\SU{2}\times\U{1}}$      \\
\cline{1-3}
\multirow{2}{*}{$3$}
& $\SU{2}\times \U{1}$ & $\frac{\SU{3,1}}{\SU{3}\times\U{1}}$     \\\cline{2-3}
& $\Spin{4}\times\SU{2}$ & $\frac{\SO{2,2}}{\SO{2}\times\SO{2}}\times \frac{\SU{1,1}}{\U{1}}$    \\
\cline{1-3}
\multirow{2}{*}{$4$}& $\U{1}$ & $\frac{\SU{4,1}}{\SU{4}\times\U{1}}$    \\\cline{2-3}
& $\Spin{3}\times\SU{2}^\sast$ & $\frac{\SO{3,2}}{\SO{3}\times\SO{2}}\times \frac{\SU{1,1}}{\U{1}}$      \\
\cline{1-3}
$5$ & $\Spin{2}\times\SU{2}$ &$\frac{\SO{4,2}}{\SO{4}\times\SO{2}}\times \frac{\SU{1,1}}{\U{1}}$    \\
\cline{1-3}
$6$ &  $\SU{2}^\sast$& $\frac{\Sp{6}}{\U{3}}$    \\
\cline{1-3}
$9$ &  $\U{1}$ & $\frac{\SU{3,3}}{\SU{3}\times\SU{3}\times \U{1}}$    \\
\cline{1-3}
$15$ &  $\mathbb{Z}_2$ & $\frac{\SOs{12}}{\SU{6}\times\U{1}} $    \\
\hline
\end{tabular}
\end{center}
\caption{$\mathcal{N}=2$ truncations with $n_{\rm H}=0$.}
\label{table:TableN=2NH=0}
\end{table}
\endgroup

\medskip

The only examples of truncation with only vector multiplets are provided by the STU models. These can be obtained truncating  4-dimensional $\mathcal{N}=8$ supergravity by imposing the invariance under the $\U{1}^4$ Cartan generators of the $\SO{8}$ gauge group \cite{Duff:1999gh}. The 4-dimensional theory is gravity coupled to  three vector multiplets, where the scalar parameterise $\mathcal{M}_{\rm V} = \left(\frac{\SU{1,1}}{\U{1}} \right)^3$.
The STU model has various realisations both in  11-dimensional  and  type II supergravity (see for instance \cite{Andrianopoli:1997wi,Bertolini:1999uz, Azizi:2016noi}).
In our classification it corresponds to the truncation in item 2)
with $n_{\rm V}=3$ and can also be obtained as a subtruncation of the theory in item 5) with $\Gst=\U{1}$.

\subsubsection{Truncations with only hypermultiplets}

In order to have truncations with only  hypermultiplets, the $\Gst$-structures must only admit  the  invariant generalised vectors $K$ and $\hat{K}$. This corresponds  to the embeddings
\begin{equation}
\begin{aligned}
\label{decE6}
 &   \Gst \subset {\rm E}_{6(2)} \times\U{1} \subset \Ex{7}\, ,  \\
 &   \Gst \subset \SU{6} \times  \SU{2}  \times \U{1} \subset \SU{8} \, ,
 \end{aligned}
\end{equation}
where ${\rm E}_{6(2)}$ is the stabiliser of the vector $X = K + i \hat{K}$ in the $\rep{56}$ 
\begin{equation}
\label{eq:fundStoE6}
    \rep{56} = \rep{1}_3 \oplus \rep{27}_{-1} \oplus c.c \, . 
\end{equation}
The hypersmultiplet scalars are given 
by the non-compact generators of $\Ex{7}$ that are singlet of both 
 $\Gst$ and  the  $\U{1}$ R-symmetry and transform non-trivially under the $\SU{2}$ R-symmetry. This means that, in the decomposition of the adjoint of $\Ex{7}$ under $\SU{6} \times \SU{2} \times \U{1} \subset {\rm E}_{6(2)} \times\U{1}  \subset \Ex{7}$,  the extra singlets must be in the  
\begin{equation}
\begin{aligned}
 (\rep{20},\rep{2})_0 \in \rep{78}_0 \, , 
 \end{aligned}
\end{equation}
giving a maximum number of $n_{\rm H} =10$ hypermultiplets. 

For truncations with only hypermultiplets the
intrinsic torsion takes a very simple form. Since the only vectors are the graviphoton and its magnetic dual, only abelian gauging of the hypermultiplet isometries are possible. It is straightforward to check that for all the truncations listed below the only components of the intrinsic torsion in  \eqref{eq;torsionLDN24d} are $p_{0A}{}^B$ and 
$p_{\hat{0}A}{}^B$ transforming in the adjoint representation of the hyperscalar isometry group.

\medskip

As in the previous section, we present the truncations we find according to the geometry of the scalar manifolds as given in  Table \ref{hmn2}.

\begin{enumerate}
    \item A first family of truncations has scalar manifold 
\begin{equation}
\label{novecf1}
  \mathcal{M}_{\rm H} = \frac{\SU{2,n_{\rm H}}}{S[\U{2} \times \U{n_{\rm H}}]} \qquad \quad n_{\rm H}=1,2,3 \, . 
\end{equation} 
For $n_{\rm H}=2,3$ the corresponding 
$\Gst$-structure is 
\begin{equation}
\Gst= \SU{2}^{4 -n_{\rm H}} \times  \U{1} \, , 
\end{equation}
and, for $n_{\rm H}=1$, it enhances to 
$\Gst= \SU{3} \times  \SU{3}$.

\item A second family of truncations has scalar manifold
\begin{equation}
   \mathcal{M}_{\rm H} =\frac{\SO{4, n_{\rm H}}}{\SO{4} \times\SO{n_{\rm H}}} 
\end{equation} 
for $n_{\rm H}=2, \dots, 6$. The corresponding $\Gst$-structures are listed below. For $n_{\rm H}=5$ the truncation automatically enhances to  $n_{\rm H}=6$  (see Section \ref{sec:discretest}). 
\begingroup
\renewcommand{\arraystretch}{1.5}
\begin{table}[H]
\begin{center}
\begin{tabular}{|c|c|c|c|c|}
\hline
$n_{\rm H}$ & $2$ & $3$  & $4$  & $6$  \\
\hline 
$G_S$ & $\SU{2} \times \SU{2} \times \U{1}$ & $\SU{2}\times   \U{1} $ &  $\U{1}\times \U{1}$ & $\U{1}$ \\
\hline 
\end{tabular}
\end{center}
\end{table}
\endgroup

\item The scalar manifold 
\begin{equation}
\label{G2coset}
\mathcal{M}_{\rm H} = \frac{{\rm G}_{2(2)}}{\SU{2} \times \SU{2}}
\end{equation} 
corresponds to a truncation with $n_{\rm H} =2$ hypermultiplets. The truncation comes from a $\SU{3}$-structure, which embeds as $\SU{3}_S \times  {\rm G}_{2(2)} \subset  {\rm E}_{6(2)}$ and
$\SU{3}_S \times \SU{2}  \subset   \SU{6}$.

\item The truncation with scalar manifold 
\begin{equation}
\label{E6coset}
\mathcal{M}_{\rm H} =\frac{{\rm E}_{6(2)}}{\SU{2} \times \SU{6}}
\end{equation} 
is obtained with a $\Gst = \mathbb{Z}_3$ structure (see Section \ref{sec:discretest} for an explicit expression). The only $\mathbb{Z}_3$-invariant terms in the fundamental of $\Ex{7}$ are the 
$\rm{E}_{6(2)}$-singlets, so that there are no vector multiplets. On the contrary,  $\mathbb{Z}_3$ acts trivially on the 
whole $(\rep{20}, \rep{2})$ in the adjoint, thus giving 
$n_{\rm H} =10$ hypermultiplets. 
It is easy to verify that the  $\mathbb{Z}_3$ singlets in the $(\rep{20}, \rep{2})$ span the coset \eqref{E6coset}. 
\end{enumerate}

We do not find some of  the quaternionic scalar manifolds listed in Table \ref{hmn2}.
Clearly the scalar manifolds 
\begin{equation}
    \begin{aligned}
  \calM_{\rm H} & = \frac{{\rm E}_{7(-5)}}{\SU{2} \times \SO{12}} 
  \qquad {\rm and} \qquad \calM_{\rm H} & = \frac{{\rm E}_{8(-24)}}{\SU{2} \times \Ex{7}}
    \end{aligned}
\end{equation}
are too big to be realised in truncations of $\Ex{7}$ generalised geometry. On the contrary, the manifolds 
\begin{equation}
\label{F4man}
  \calM_{\rm H}  = \frac{{\rm F}_{4(4)}}{\SU{2} \times \USp{6}}  \quad  n_{\rm H} = 7 
\end{equation}
and 
\begin{equation}
\label{Uspman}
   \calM_{\rm H}  = \frac{\USp{2, 2 n_{\rm H}}}{\USp{2} \times \USp{2 n_{\rm H}}}  \quad  n_{\rm H} \leq 3 
\end{equation}
are in principle allowed. However, they cannot be obtained in a consistent truncation since the
$\Gst$-structures that are  compatible with them always contain extra singlets and hence give larger truncations. 

Consider first the manifold \eqref{F4man}. To obtain it, $\Gst$ must be discrete since  \eqref{scalarman} implies
\begin{equation}
\Gst = \Com{\USp{6}\times \SU{2} \times \U{1}}{\SU{8}} \, .
\end{equation}
Moreover, from the embedding of $\USp{6}$ in $\SU{6}$ and  Schur's lemma  it follows that  $\Gst$
can only be of the form $(a \id_6 , b \id_2)$, and so it also commutes with $\SU{6}$,  with a number of adjoint singlets allowing to reconstruct the full coset \eqref{E6coset} and not simply \eqref{F4man}.

The same reasoning holds for the scalar manifolds \eqref{Uspman}. Shur's lemma guarantees that $\USp{2n}$ always enhances to $\SU{2n}$. So the scalar manifolds corresponding to $n_{\rm H}=3$ and $n_{\rm H} =2$, are both enhanced to the truncation \eqref{E6coset}, while that with $n_{\rm H}=1$  enhances to the truncation in \eqref{novecf1} with $n_{\rm H} = 2$.

\begingroup
\renewcommand{\arraystretch}{1.5}
\begin{table}[H]
\begin{center}
\begin{tabular}{|c|c|c|}
\hline
$n_{\rm H}$ & $G_S$& $\mathcal{M}_h$  \\
\hline
$1$ & 
$\SU{3}\times \SU{3}$ & $\frac{\SU{2,1}}{\SU{2}\times\U{1}}$  \\
\cline{1-3}
\multirow{2}{*}{$2$} & $\SU{3}^\sast$ & $\frac{{\rm G}_2}{\SO{4}}$   \\\cline{2-3}
& $\SU{2}\times \SU{2} \times \U{1}$  &  $\frac{\SU{2,2}}{\SU{2} \times \SU{2} \times \U{1}} $    \\
\cline{1-3}
\multirow{2}{*}{$3$} &  $\SU{2} \times \U{1}^\sast$  & $\frac{\SO{4,3}}{\SO{4} \times \SO{3}}$ \\
\cline{2-3}
& $ \SU{2}\times \U{1}$   & $\frac{\SU{2,3}}{\SU{2} \times \SU{3} \times\U{1}} $     \\
\cline{1-3}
$4$ &  $\U{1}^2$  &  $\frac{\SO{4,4}}{\SO{4} \times \SO{4}}$   \\
\cline{1-3}
6 & $\U{1}$  &  $\frac{\SO{4,6}}{\SO{4} \times \SO{6}}$  \\
\cline{1-3}
10 & $\mathbb{Z}_3$  &  $\frac{\mathrm{E}_{6(2)}}{\SU{6} \times \SU{2}}$  \\
\hline
\end{tabular}
\end{center}
\caption{$\mathcal{N}=2$ truncations with $n_{\rm V}=0 $. For any truncation, we list the largest $\Gst$-structure that allows for it.}
\label{MainProcedure_TableN=2Nv=0}
\end{table}
\endgroup

\subsubsection{Truncations with vector and hypermultiplets}

We also find  a number of truncations with both vector and hypermultiplets.

\medskip

The truncations associated to 
continuous $\Gst$-structures are quite limited and are listed in Table \ref{MainProcedure_Table-nv-nh}. 
Several such truncations already appeared in the literature. For instance, we recover all $\mathcal{N}=2$ consistent truncations of 11-dimensional supergravity on coset manifolds derived in \cite{Cassani:2012pj}. 
The truncation with $n_{\rm V} = n_{\rm H} = 1$ and
$\Gst= \SU{3}$ corresponds to the universal truncation  on a SE$_7$ manifold 
originally derived in \cite{Gauntlett:2009zw}.
A particular example is the truncation on $S^7 = \frac{\SU{4}}{\SU{3}}$.\footnote{This truncations was extended in \cite{Gauntlett:2009bh} to include a skew-wiffling mode. In the context 
EGG  both truncations have been reproduced in \cite{Blair:2024ofc}.} 
The truncation on $V_{5,2}$ with $n_{\rm V}=1$ and $n_{\rm H}=2$
is obtained from a  $G_S = \SO{3}$ structure, while that on $M^{110}$ with  $n_{\rm V} =2 $ and $n_{\rm H}=1$ corresponds to $G_S = \SU{2}\times \U{1}$. 
The truncations with $n_{\rm V}=3$ and $n_{\rm H}=1$ on 
$Q^{111}$ and $N(k,l)$ are associated to a $G_S = \U{1}^2$  and the truncation on $N(1,-1)$ with $n_{\rm V}=5$ and $n_{\rm H}=1$  is given by a $\U{1}$-structure.

The $\SO{3}$-structure 
with $n_{\rm V}=1$ vector and $n_{\rm H}=2$ hypermultiplets is also 
associated to the consistent truncations  derived in \cite{Donos:2010ax} of 11-dimensional supergravity on $\Sigma_3 \times S^4$, where $\Sigma_3 =H_3/\Gamma,S^3/\Gamma$ or $\mathbb{R}^3/\Gamma$, with $\Gamma$ a discrete group of isometries.

We also recover the consistent type IIA truncations on coset manifolds discussed in \cite{Cassani:2009ck}.
The coset manifolds are $\frac{G_2}{\SU{3}}$, $\frac{\Sp{2}}{\SU{2}\times \U{1}}$ and $\frac{\SU{3}}{\U{1} \times \U{1}}$, and the truncated theories contain $n_{\rm H}=1$ hypermultiplets and $n_{\rm V}=1$, $n_{\rm V}=2$
and $n_{\rm V}=3$ vector multiplets, respectively. 
They correspond to the $\Gst$-structure 
$\SU{3}$, $\SU{2}\times \U{1}$ and $\U{1}^2$ 
in Table \ref{MainProcedure_Table-nv-nh}.

\medskip

The set of possible truncations enlarges if we consider discrete structure groups or direct products of a continuous and a discrete factor. 
For truncations of this kind we just give some examples so that, differently from the case of continuous structure groups, our list is not exhaustive. 

Discrete structures appear when, starting from a truncation with only vector (or hyper) multiplets, we add extra hyper (vectors) to have 
 a maximum matter content (see Section \ref{sec:discretest}). In this way, 
we find the two truncations in Table \ref{N=2TruncationsFromPureDiscreteStructures}, with purely discrete structure groups. 

\begingroup
\renewcommand{\arraystretch}{1.5}
\begin{table}[H]
\begin{center}    \hspace*{-0.3cm}
\small
\begin{tabular}{|c|c|c|c|}
\hline
 $G_S$ & $n_V$ & $n_{\rm H}$ & $\calM_V \times \calM_{\rm H}$\\
 \hline
 $\mathbb{Z}_4$ & $1$ & $6$ & $\frac{\SU{1,1}}{\U{1}} \times \frac{\SO{4,6}}{\SO{4} \times \SO{6}}$ \\
  \hline
 $\mathbb{Z}_3$ & $9$ & $1$ & $\frac{\SU{3,3}}{\SU{3} \times \SU{3} \times \U{1}} \times \frac{\SU{2,1}}{\SU{2} \times \U{1}}$ \\
\hline
\end{tabular}
\end{center}
\caption{Discrete groups with $\calN=2$.}
\label{N=2TruncationsFromPureDiscreteStructures}
\end{table}
\endgroup

Another family of truncations associated to a discrete structure is  obtained starting from $\mathcal{N}=4$ truncations and imposing a $\Gst$-structure that is the (direct) product of the orginal  $\Gst$-structure times a discrete factor. A 
 non-exhaustive list of truncations of this type is provided in Table \ref{MainProcedure_TableN=2FromN=4}. 

\begingroup
\renewcommand{\arraystretch}{1.5}
\begin{table}[H]
\begin{center}    \hspace*{-0.3cm}
\small
\begin{tabular}{|c|c|c|c|}
\hline
 $G_S$ & $n_V$ & $n_{\rm H}$ & $\calM_V \times \calM_{\rm H}$\\
\hline
 $\SU{2} \ltimes \mathbb{Z}_2$ & $1$ & $3$ & $\frac{\SU{1,1}}{\U{1}} \times \frac{\SO{4,3}}{\SO{4} \times \SO{3}}$ \\
\hline
 $\U{1} \times \mathbb{Z}_2$ & $1$ & $4$ & $\frac{\SU{1,1}}{\U{1}} \times \frac{\SO{4,4}}{\SO{4} \times \SO{4}}$ \\
 \hline
 $\U{1} \times \mathbb{Z}_2$ & $2$ & $3$ & $ \left( \frac{\SU{1,1}}{\U{1}} \right)^2 \times \frac{\SO{4,3}}{\SO{4} \times \SO{3}}$ \\
\hline
 $\SU{2} \ltimes \mathbb{Z}_2$ & $2$ & $2$ & $ \left( \frac{\SU{1,1}}{\U{1}} \right)^2 \times \frac{\SO{4,2}}{\SO{4} \times \SO{2}}$ \\
\hline
 $\U{1} \times \mathbb{Z}_2$ & $3$ & $2$ & $ \left( \frac{\SU{1,1}}{\U{1}} \right)^3 \times \frac{\SO{4,3}}{\SO{4} \times \SO{3}}$ \\
 \hline
\end{tabular}
\end{center}
\caption{Discrete groups with $\calN=2$ from $\calN=4$.}
\label{MainProcedure_TableN=2FromN=4}
\end{table}
\endgroup

The truncations with $n_{\rm V}=2$ and $n_{\rm H}=2$ on $\frac{\Sp{2}}{\Sp{1}}$ and  $n_{\rm V}=3$ and $n_{\rm H}=2$ on $N(1,1)$ mentioned in \cite{Cassani:2012pj} are in this class. They correspond to the   last two  lines of  Table  \ref{MainProcedure_TableN=2FromN=4}. 
The $\frac{\Sp{2}}{\Sp{1}}$ theory 
is obtained imposing a  $\mathbb{Z}_2$ invariance on  the $\mathcal{N}=4$ theory with $n_{\rm V}=3$ vector multiplets of Section  \ref{sec:N4}. Similarly, the  $N(1,1)$ is given by a  $\mathbb{Z}_2 \times \U{1}$ structure and comes from the  $\mathcal{N}=4$ theory with $n_{\rm V}=4$ vector multiplets.

In all these cases the embedding tensor is easily derived decomposing the $\rep{912}$ under $\Gst \times G_{\rm iso}$. We will not discuss it in this paper.

\begin{landscape}
\begingroup
\renewcommand{\arraystretch}{3}
\begin{table}[H]
\small
\begin{center} 
\hspace*{-1.2cm}
\begin{tabular}{|c|c|c|c|c|c|c|}
\hline
{\diagbox{$n_{\rm V}$}{$n_{\rm H}$}}  & \multicolumn{2}{|c|}{$1$} & \multicolumn{2}{|c|}{$2$}  & \multicolumn{2}{|c|}{$3$} \\
\cline{1 - 7}
 \multirow{3}{*}{$1$}  & $G_S = \SU{3}^\sast$ & \multirow{3}{*}{\begin{minipage}[c]{3cm} $\mathcal{M}_{\rm V}=\frac{\SU{1,1}}{\U{1}}$ 
 
 \vspace{0.1cm}
 
 $\mathcal{M}_{\rm H}=\frac{\SU{2,1}}{\SU{2}\times\U{1}}$ 
 \end{minipage}} & $G_S = \SO{3}^\sast$  &  {\begin{minipage}[c]{3cm} 
 $\mathcal{M}_{\rm V}= \frac{\SU{1,1}}{\U{1}}$ 
 
 \vspace{0.1cm}
 
 $\mathcal{M}_{\rm H}=\frac{{\rm G}_2}{\SO{4}}$ 

 \vspace{0.1cm}
 
 \end{minipage}} & \multirow{3}{*}{$G_S = \U{1}$} & 
 \multirow{3}{*} {\begin{minipage}[c]{4.5cm} $\mathcal{M}_{\rm V}= \frac{\SU{1,1}}{\U{1}}$ 

\vspace{0.2cm}
 
 $\mathcal{M}_{\rm H}=\frac{\SU{2,3}}{\SU{2}  \times \SU{3} \times\U{1}}$  
 \end{minipage}} \\
 \cline{2-2} \cline{4-5} 
&  $G_S = \SU{2}\times \SU{2}\times\U{1}$ &  & $G_S = \U{1}^2$ & \multirow{2}{*}{\begin{minipage}[c]{4.5cm} $\mathcal{M}_{\rm V}= \frac{\SU{1,1}}{\U{1}}$  

\vspace{0.1cm}

 $\mathcal{M}_{\rm H}=\frac{\SU{2,2}}{\SU{2} \times \SU{2} \times \U{1}}$ 
 \end{minipage}} &  &  \\ 
\cline{2-2} \cline{4-4} 
&  $G_S = \U{1}$ & & $G_S = \U{1}$ &  &  &  \\ 
 \cline{1-7} 
 \multirow{2}{*}{$2$}  & $G_S = \SU{2}\times \U{1}^\sast$ &  {\begin{minipage}[c]{3cm} $\mathcal{M}_{\rm V}= \left(\frac{\SU{1,1}}{\U{1}}\right)^2$ 

\vspace{0.1cm}
 
 $\mathcal{M}_{\rm H}=\frac{\SU{2,1}}{\SU{2}\times\U{1}}$ 

 \vspace{0.2cm}
 
 \end{minipage}} &  \multirow{2}{*}{$G_S =  \U{1}$} & 
 \multirow{2}{*} {\begin{minipage}[c]{4.5cm} $\mathcal{M}_{\rm V}= \frac{\SU{2,1}}{\SU{2} \times \U{1}}$ 

\vspace{0.1cm}
 
 $\mathcal{M}_{\rm H}=\frac{\SU{2,2}}{\SU{2} \times \SU{2} \times\U{1}}$ 
 \end{minipage}} &  \multirow{2}{*}{--} &   
 \multirow{2}{*}{--} \\
 \cline{2-3} 
&  $G_S = \SU{2}\times\U{1}$ &  {\begin{minipage}[c]{3cm} $\mathcal{M}_{\rm V}= \frac{\SU{2,1}}{\SU{2} \times \U{1}}$ 

\vspace{0.1cm}

 $\mathcal{M}_{\rm H}=\frac{\SU{2,1}}{\SU{2}\times\U{1}}$ 

 \vspace{0.2cm}
 
 \end{minipage}} & & &  &  \\ 
 \cline{1-7} 
3 & $G_S = \U{1}^2$ &  {\begin{minipage}[c]{3cm} $\mathcal{M}_{\rm V}= \left(\frac{\SU{1,1}}{\U{1}}\right)^3$ 

\vspace{0.1cm}

 $\mathcal{M}_{\rm H}=\frac{\SU{2,1}}{\SU{2}\times\U{1}}$ 

 \vspace{0.2cm}
 
 \end{minipage}} & -- &  -- & -- & -- \\
 \cline{1-7} 
5 & $G_S = \U{1}$ &  {\begin{minipage}[c]{4.5cm} $\mathcal{M}_{\rm V}= \frac{\SO{4,2}}{\SO{4} \times \SO{2}} \times \frac{\SU{1,1}}{\U{1}} $ 

\vspace{0.1cm}

 $\mathcal{M}_{\rm H}=\frac{\SU{2,1}}{\SU{2}\times\U{1}}$

 \vspace{0.2cm}
 
 \end{minipage}} & -- & -- & -- & -- \\
  \hline
\end{tabular}
\end{center}
\caption{$\mathcal{N}=2$ truncations with  $n_{\rm V}$ and $n_{\rm H}$}
\label{MainProcedure_Table-nv-nh}
\end{table}
\endgroup
\end{landscape}

\section{Scanning through supersymmetry and field content}
\label{Section_Algorithm}

In this section we provide some details about the derivation of the results discussed in Section \ref{sec:scan}. 
As already mentioned in the previous sections, the classification consists in  determining  all generalised  $\Gst$-structures with singlet intrinsic torsion that  give rise to inequivalent truncated theories with different amount of supersymmetry.

Since we are interested in supersymmetric truncations, the allowed $\Gst$-structures must be subgroups of $\SU{8}$, the double cover of the maximally compact subgroup of $\Ex{7}$, under which the spinors of the theory transform. 
We first consider continuous $\Gst$-structures
and then, at the end of the section, we comment on discrete ones. 

\medskip 

 We first focus on truncations to pure supergravity with $ 2 \leq  \mathcal{N} < 8$. For these cases, for any amount of supersymmetry $\mathcal{N}$, the corresponding structure 
$\Gst^{max}$ is the largest generalised structure compatible with the given amount of supersymmetry, and it is determined  by 
\begin{equation}
\Gst^{max}  =  \Com{G_R}{\SU{8}} \, , 
\end{equation}
 where $G_R$ is the R-symmetry group. Both $G_R$ and the corresponding $\Gst^{max}$ are listed in Table \ref{tab:my_label}. 
The idea is to find  the explicit embedding of $\Gst^{max}$ in the $\Ex{7}$ and $\SU{8}$ generators given in Appendix \ref{PreliminariesE77_Mth}, for any fixed $\mathcal{N}$.

An economic way to do so is to use the fact that all the branchings in Table \ref{tab:my_label}  
\begin{equation}
    \SU{8} \supset \Gst^{max} \times G_R \, , 
\end{equation}
correspond to 
maximal regular subalgebrae  $\mathfrak{h}$ of $\su(8)$.\footnote{In this paper we need to distinguish between regular and special maximal subalgebrae of a Lie group $G$  \cite{Slansky:1981yr, Yamatsu:2015npn}.
Let  $G$  be a Lie group and $\mathfrak{g}$  its Lie algebra.  The subalgebra $\mathfrak{h} \subset \mathfrak{g}$ is a maximal regular subalgebra of $\mathfrak{g}$ if it has the same  rank as  $\mathfrak{g}$,  if not,  $\mathfrak{h}$  is called special.  
We are interested in the embedding of the algebra $\mathfrak{g}_S$ of the $\Gst$-structure group into $\mathfrak{h}$.
We call the embedding  $\mathfrak{g}_S \subset \mathfrak{h}$ 
a regular branching or a special branching depending on whether  $\mathfrak{h}$ is a  regular or special (maximal) subalgebra. } 
Then we find it useful to derive the allowed $\Gst^{max}$-structures just by looking at their Cartan subalgebrae. 

The idea is to construct a generic $\U{1}$-structure as a linear combination of the Cartan subalgebra  of $\SU{8}$
\begin{equation}
\label{genU1}
   \mu_{\overrightarrow{\lambda}}   = \sum_{i=1}^7 \lambda_i  H_{i}  \qquad  \lambda_i \in \mathbb{N} \, , 
\end{equation}
where $H_i$ are the Cartan generators  of $\SU{8}$.\footnote{All Cartan algebrae are isomorphic, so it is enough to study just one of them. We take $(H_i)^\alpha{}_\beta=i\delta^\alpha_i \delta_{i\beta} - \tfrac{i}{8} \delta^\alpha_\beta$ (see Appendix \ref{PreliminariesE77_Mth} for our conventions).}
We let the number $\lambda_i$ run over $n_i = 0, \ldots, N$ with 
$N \in \mathbb{N}$. A priori  the coefficient $\lambda_i$  could be real  numbers. However, since the generators $H_{i}$ only have rational entries and multiplication by a global factor will not change the 
$\U{1}$-structure, we can restrict our study to $\lambda_1, \dots , \lambda_7 \in \mathbb{Z}$. Then, using the freedom in the tracelessnes condition, it is possible to consider only positive $\lambda_i$.  
Any set  $\overrightarrow{\lambda}=\{n_1, \ldots, n_7\}$ defines a different $\U{1}$-structure, $ \mu_{\overrightarrow{\lambda}}$.

Now we look for singlets under $ \mu_{\overrightarrow{\lambda}}$
in the spinorial and adjoint of $\SU{8}$, and in 
the fundamental and adjoint representations of $\Ex{7}$ 
\begin{equation}
\label{singcond-U1}
\begin{aligned}
\mu_{\overrightarrow{\lambda}} \cdot \epsilon = 0 \, , \\
  \mu_{\overrightarrow{\lambda}}  \cdot V = 0\,, \\
 \mu_{\overrightarrow{\lambda}}  \cdot R = 0 \, , \\
\mu_{\overrightarrow{\lambda}} \cdot R_{\SU{8}} = 0 
\,, 
\end{aligned}
\end{equation}
where $\epsilon \in \rep{8}$ and $R_{\SU{8}} \in \rep{63}$ of $\SU{8}$, while $V \in \rep{56}$ 
and $R \in \rep{133}$ of $\Ex{7}$. The explicit expressions for the various representations are given in Appendix \ref{PreliminariesE77_Mth}.

The first equation \eqref{singcond-U1} determines the number $\mathcal{N}$ of supersymmetries preseved by the theory, whereas the remaining three determine the bosonic field content. More precisely the second equation gives the number $2 n_{\rm V}$ of invariant generalised vectors, namely the vectors of the truncated theory, together with their magnetic duals. The third equation determines the number $n_{\rm ad F}$ of $\Ex{7}$ generators that commute with the structure, while the fourth gives the number $n_c$ of compact ones. 
The difference $n_{\rm ad F} - n_c$  between the $\Ex{7}$ and the $\SU{8}$-singlets gives the number of scalars in the truncated theory. 

With the help of Mathematica, we then look for solutions of \eqref{singcond-U1} and classify them  according the number of singlets in the various representations
\begin{equation}
\label{singlet-cont}
  \mu_{\overrightarrow{\lambda}} \quad \leftrightarrow \quad    \{\mathcal{N}, 2 n_{\rm V}, n_{\rm ad F}, n_c \} \, . 
\end{equation}
 If two $\U{1}$-structures $ \mu_{\overrightarrow{\lambda}}$ give the same singlet content, they are considered as equivalent. Notice that from the knowledge of  the $\U{1}$ generator we
can  reconstruct the full structure group by
 looking at the commutation relations among the $\U{1}$  singlets in the adjoint of $\SU{8}$. 

\medskip 

For $\mathcal{N} \geq 5$ we find only one solution for each amount of supersymmetry, characterised by the singlets

\begin{table}[H]
    \centering
    \begin{tabular}{c|c|c|c}
 $\mathcal{N}$  & 8 & 6 & 5  \\ 
 \hline 
 $n_{\rm V}$ & 28 & 16 & 10 \\ 
 \hline
 $n_{\rm ad F} - n_c$ & 70 & 30  & 10   \\
    \end{tabular}
\end{table}

\noindent The  $\mathcal{N}=8$ truncation is trivially realised as it corresponds to the  identity structure  \eqref{genU1}, while for  $\mathcal{N}=6$ and 
$\mathcal{N}=5$ we recover $\Gst^{max} = \SU{2}$ and  $\Gst^{max} = \SU{3}$. 
As expected,  the solutions are unique
 since the theories only contain the gravity multiplet. The singlets in the table above 
 reproduce the field content of the gravity multiplet and the dimension of the associated scalar manifolds. 

\medskip 

For $2 \leq \mathcal{N} \leq 4$  we find several independent solutions $ \mu_{\overrightarrow{\lambda}}$ for each  $\mathcal{N}$. The solutions
 corresponding to $\Gst^{max}$ are singled out by the values 

\begin{table}[H]
    \centering
    \begin{tabular}{c|c|c|c}
 $\mathcal{N}$  &  4 & 3 & 2 \\ 
 \hline 
 $n_{\rm V}$ &  6 & 3  & 1  \\ 
 \hline
 $n_{\rm ad F} - n_c$  & 2  &  0 &  0  \\
    \end{tabular}
\end{table}

\noindent  As in the previous cases we can reconstruct the full structures $\Gst^{max} = \SU{8 - \mathcal{N}}$. 

For any given $\mathcal{N}$, the other solutions $ \mu_{\overrightarrow{\lambda}}$ correspond to truncations with extra matter fields. The values $n_{\rm V}$, $n_{\rm adF}$
and $n_c$ allow to derive their field content, while 
the  $\Gst \subset \Gst^{max}$ can be reconstructed from the corresponding $\U{1}$ generator.

However, these solutions do not exhaust all possible truncations with  $2 \leq \mathcal{N} \leq 4$  supersymmetry and continuous $\Gst$-structures. This is because the trick of looking only at the Cartan subalgebrae holds only for subgroups coming from regular subalgebrae of $\SU{8}$.\footnote{For any regular structure $\Gst \subset \Gst^{max}$, the roots of $\Gst$ are a subset of those of $\Gst^{max}$. Thus, as it happens with the roots, the weights of  $\Gst$ are a subset of those of $\Gst^{max}$ as well. The $\U{1}$ representing $\Gst$ is regular in both $\Gst$ and $\Gst^{max}$ and, therefore, it preserves information about the weights. This allows to reproduce the $\Gst$-structure field content with a $\U{1}$-structure. For special cases roots and weights can not be found as subsets of the ones of $\Gst^{max}$, so one is forced to study them independently.}
For $\Gst$-structures coming from special subalgrebrae, we  have not been able to find a similar algorithmic procedure. 

For this reason, we proceeded to a systematic scan of the different subgroup 
$\Gst \subset \Gst^{max}$ for fixed $\mathcal{N}$ and looked for solutions of equations \eqref{singcond-U1} where now $ \mu_{\overrightarrow{\lambda}}$ is replaced by a generic element of the  algebra $\Gst$, embedded in $\Ex{7}$ and $\SU{8}$ via $\Gst^{max}$. 
This analysis gives the results discussed in Section \ref{sec:scan}. As an example, the explicit derivation for  $\mathcal{N}=4$ is given in Appendix \ref{explicit_ex_subsec}.

\subsection{Discrete structures}
\label{sec:discretest}

The knowledge of the explicit embedding of $\Gst^{max}$ allow us to study also some discrete structures.

Discrete structures are subtler and classifying them all is out of the scope of this paper.
However, some cases can be easily studied.
The analysis is the same as for continuous groups, with the difference that  now we have to solve 
\begin{equation}
\begin{aligned}
& g \cdot \epsilon = \epsilon\, ,  \\
& g \cdot R_{\SU{8}} = R_{\SU{8}} \,  , \\
& g \cdot V = V \, ,  \\
& g \cdot R = R \, , 
\end{aligned}
\label{Preliminaries_Singlets_SingletsEquationGroup}
\end{equation}
where $g$ is any element of the discrete group. 

\medskip

A first instance where discrete $\Gst$-structures  appear is in truncations with $ 2 \leq \mathcal{N} \leq 4$ supersymmetry and the largest number of matter multiplets.

 Consider first the truncation to  $\mathcal{N}=2$ supergravity with $n_{\rm V} =15$ vector multiplets and no hypermultiplets  (see Section \ref{sec:onlyvec}). 
Recall that truncations with only vector multiplets are given by $\Gst$-structures 
\begin{equation}
    \Gst \subset \SU{6} \subset \SOs{12} \, , 
\end{equation}
where $\SOs{12}$ is  the stabiliser of the triplet of adjoint singlets generating the $\SU{2}$ R-symmetry. 
Morevoer, the compatibility condition \eqref{eq:gencomp} implies 
that the singlets generalised vectors can only belong the representation $(\rep{32^\prime} , \rep{1})$  in the decomposition \eqref{eq:fundStoSO12}
of the fundamental of $\Ex{7}$ under 
$\SOs{12} \times \SU{2}_R$, which in turn splits as 
\begin{equation}
  (\rep{32^\prime} ,\rep{1}) =  (\rep{1}, \rep{1})_{-6} \oplus(\rep{1}, \rep{1})_{6} \oplus (\rep{15}, \rep{1})_{2} \oplus (\rep{\overline{15}}, \rep{1})_{-2} 
\end{equation}
under $ \Ex{7} \supset \SOs{12} \times \SU{2}_R \supset \SU{6} \times \U{1}_R \times \SU{2}_R $.

In order to preserve all 15 vectors, the 
group $\Gst$ must commute with the whole $\SU{6}$. Since  we want $\Gst$ to be  also a subgroup of $\SU{6}$, we expect it to be an element of the center of $\SU{6}$.
It is easy to check that $\Gst = \mathbb{Z}_2$
\begin{equation}
  g_{\mathbb{Z}_2} =    \begin{pmatrix}
       - \id_6 &  \\ 
         &  \id_2
  \end{pmatrix}  \in \SU{8} \,.
    \label{gZ2}
\end{equation} 

Consider now a generalised vector in the  notation of  
 \eqref{vecSU6spl}
\begin{equation}
 V^{\alpha\beta}=(V^{mn},V^{mi},V^{ij}) \, , 
 \label{GenVectSU6xSU2Branched}
\end{equation}  
where $\alpha=(m,i)$,  $i=1,2\in\SU{2}$ and $m=1,\dots, 6\in \SU{6}$.
The  action of  $g_{\mathbb{Z}_2}$ on $V$
leaves 
invariant the components $V^{ij}$ and $V^{mn}$, corresponding to the generalised vectors $K$ and $\hat{K}$ and the $15$ vector multiplets, respectively.

Using  the expressions in Appendix \ref{PreliminariesE77_Mth}  for the generators of $\SOs{12}$ in terms of those of $\SU{6}$, one can check that  $\Gst=\mathbb{Z}_2$ belongs to the center of $\SOs{12}$.\footnote{It is interesting to note that the same $\mathbb{Z}_2$ is contained in all the $\Gst$ of Table \ref{table:TableN=2NH=0}  and, in all cases it belongs to the center. }
Thus from \eqref{scalarman}, we recover, as expected, the scalar manifold 
\begin{equation}
\calM_{\rm V}= \frac{ \SOs{12}}{\SU{6}\times \U{1}} \, . 
\end{equation}

\medskip

The truncation to  $\mathcal{N}=2$ supergravity with  no vector multiplets and $n_{\rm H} = 10$ hypermultiplets is obtained along the same lines.
In this case we look for a subgroup
\begin{equation}
    \Gst \subset \SU{6} \subset {\rm E}_{6(2)} \, , 
\end{equation}
with $E_{6(2)}$ the stabiliser of the generalised vectors $K$ and $\hat{K}$, and 
\begin{equation}
    \Ex{7}\supset \mathrm{E}_{6(2)} \times \U{1}_R \,.
    \label{E62EmbeddingInE77}
\end{equation}
The extra hypermultiplets come from the component $(\rep{20}, \rep{2})_0$ in the decomposition of the $\rep{78}_0$ under $\SU{6} \times \SU{2}_R$ 
\begin{equation}
    \rep{78}_0 = (\rep{35}, \rep{1})_0 \oplus (\rep{1}, \rep{3})_0 \oplus (\rep{20}, \rep{2})_0 \, . 
\end{equation}
The structure that leaves invariant the whole $(\rep{20}, \rep{2})_0$ is now $\Gst = \mathbb{Z}_3$, which is again in the centre of $\SU{6}$ and ${\rm E}_{6(2)}$,  and 
embeds in $\SU{8}$ as 
\begin{equation}
  g_{\mathbb{Z}_3} = e^{\tfrac23 \pi \;\mu_{\U{1}_R}}=   \begin{pmatrix}
        e^{\tfrac{2\pi}{3} i} \id_6 &  \\ 
         &  \id_2
  \end{pmatrix} \,,
    \label{gZ3}
\end{equation} 
where $\mu_{\U{1}_R}$ is given in \eqref{app:U1R}.\footnote{As it happens for truncations with only vector multiplets, the same $\mathbb{Z}_3$ is contained in all the $\Gst$ of Table \ref{MainProcedure_TableN=2Nv=0}  and, in all cases, it belongs to the centre. } The action of  $g_{\mathbb{Z}_3}$ on a generalised vector \eqref{GenVectSU6xSU2Branched} leaves invariant only the component  $V^{ij}$, namely the generalised vectors $K$ and $\hat{K}$. Using again \eqref{scalarman}, the scalar coset coset manifold is 
\begin{equation}
   \calM_{\rm H}= \frac{ \mathrm{E}_{6(2)}}{\SU{6}\times \SU{2}}\,.
\end{equation}

\medskip

Discrete structures can also
be used to construct $\mathcal{N}=2$ truncations with a maximal amount of  vector and hypermultiplets. The idea is to start with a truncation with only vectors (hypers) and to enhance it  to a truncation preserving the same number of vectors (hypers) and a maximum number of hypers (vectors) with a discrete $\Gst^\prime$.   

To see how it works consider the truncation of Section \ref{sec:onlyvec} with only one vector multiplet and scalar manifold
\begin{equation}
\label{discMv1}
 \calM_{\rm V}=\frac{\SU{1,1}}{\U{1}} \, . 
\end{equation}
The corresponding structure is $\Gst=\SU{4}\times \SU{2}$ and it embeds in $\SU{8}$ as  
\begin{equation}
    \SU{8} \supset \SU{4}\times \SU{2} \times \SU{2}_R \times \U{1}_R \, .
\end{equation}
In order to leave invariant  the same amount of vectors, the new structure must 
be a subgroup of $\SU{6}$ that also commutes 
with $\SU{4}\times \SU{2}$. These 
requirements lead to a discrete structure of the form
\begin{equation}
\label{gdiscMv1}
  g_{\Gst^\prime} =  \begin{pmatrix}
       e^{\theta i} \id_4 & & \\
        &  e^{-2\theta i} \id_2 &  \\
        & &  \id_2
  \end{pmatrix} \,.
\end{equation}
By construction
\begin{equation}
    \Com{\Gst^\prime}{\SU{8}} = \SU{4}\times \SU{2} \times \SU{2}_R \times \U{1}_R \, , 
\end{equation}
which now becomes the denominator of the full scalar manifold. The $\U{1}_R$ is the 
denominator of  \eqref{discMv1},  while 
\begin{equation}
    \SU{4}\times \SU{2} \times \SU{2}_R \sim \SO{6} \times \SO{4}
\end{equation}
is the denominator of the quaternionic K\"ahler manifold
\begin{equation}
   \mathcal{M}_{\rm H} =  \frac{\SO{4,6}}{\SO{4}\times \SO{6}} \, .
\end{equation}
By embedding \eqref{gdiscMv1} in $\Ex{7}$ it is straightforward to verify that, for  $\theta=\frac\pi2$,  we have 
\begin{equation}
    \Com{\Gst^\prime}{\Ex{7}} = \SU{1,1}\times \SO{4,6} \, . 
\end{equation}
This is the truncation in the first line of Table \ref{N=2TruncationsFromPureDiscreteStructures}. The one in the second line is 
 obtained similarly.  

\medskip 

The same approach could in principle be applied to any truncation with only vector or hypermultiplets by taking higher-order $\mathbb{Z}_p$-structures. However it is beyond the scope of this paper to perform a complete analysis of such truncations. 

We simply want to mention that  there are cases where we know from the start that truncations of this kind cannot be constructed. The obstruction comes again from Schur's lemma. 

Consider, as an example, the truncation
with $n_{\rm V}=2$ and $n_{\rm H}=0$, associated to  $\Gst=\USp{4}\times \SU{2}$.
We could try to extend the truncation by adding $n_{\rm H}=5$ hyper-multiplets to obtain the scalar manifold
\begin{equation}
\label{uspdiscr}
   \calM_{\rm V}\times \calM_{\rm H}=\left(\frac{\SU{1,1}}{\U{1}}\right)^2\times \frac{\SO{4,5}}{\SO{4}\times \SO{5}} \, , 
\end{equation}
with $\SO{5} \sim \USp{4}$ and $\SO{4} \sim \SU{2} \times \SU{2}_R$.
However, since the embedding of  $\USp{4}$ in $\SU{4}$ is such that 
the $\rep{4}$ of $\SU{4}$ goes into the $\rep{4}$ of $\USp{4}$, Schur's lemma implies that any group element commuting with $\Gst=\USp{4}\times \SU{2}$ commutes with $\SU{4}\times \SU{2}$ as well. Thus, instead of the truncation in \eqref{uspdiscr} we end up again with the truncation of  the previous example.

We checked that this behaviour is true whenever  the original $\Gst$-structure is a $\USp{n}$ group. In addition, this behaviour holds in all cases where the original $\Gst$-structure corresponds to a special branching of $\SU{6}$ where the fundamental of $\SU{6}$ decomposes into just one irreducible representation of $\Gst$.
As an example, consider the branching 
\begin{equation}
    \SU{2}\times \SU{3} \subset \SU{6}\,,
\end{equation} 
in which the fundamental of $\SU{6}$ breaks as 
$ \rep{6} = (\rep{2},\rep{3})$.
This branching leads to either the truncation with  structure $\Gst = \SU{2}$, $n_{\rm V}=6$  vector multiplets and coset manifold $\frac{\Sp{6}}{\U{3}}$ in \eqref{Usp6coset} or the
truncation with $\Gst =\SU{3}$, $n_{\rm H}=2$ hypermutliplets and coset manifold $\frac{{\rm G}_2}{\SO{4}}$ in \eqref{G2coset}. 
However, it is not possible to find a discrete structure leading to a truncation with vector and hypermultiplets with scalar manifold
$\calM_{\rm V}\times \calM_{\rm H} = \frac{\Sp{6}}{\U{3}}\times \frac{{\rm G}_2}{\SO{4}}$. Indeed 
Schur's lemma
implies that any discrete structure commuting with $\SU{2}\times \SU{3}$ will commute with $ \SU{6}$ as well, leaving the two possibilities in 
\eqref{maxvectr} or \eqref{E6coset}.

Let us stress that this is just an observation derived from examples, which could provide a hint at how some discrete structures can be found. 
It is not meant to be a formal proof.

\section{Review of consistent truncations to 5, 6 and 7 dimensions}
\label{sec:higherD}

The classification of the supergravity theories that can be obtained as consistent truncations of 11/10-dimensional supergravity can be performed along the same lines at least for truncations to  $D \geq 4$ dimensions. Of particular interest are the truncations to 5,6 and 7 dimensions since they provide valuable tools in many instances of the gauge/gravity duality. 

The difference with respect to the analysis of Section \ref{sec:scan}  is in the exceptional group $\Ex{d}$ determining the theory, which changes depending on the dimension of the compactification manifold.
The generalised structure groups, 
the (double cover of) the maximal compact subgroups
and the relevant tensor bundles for truncations to 5, 6 and 7 dimensions are listed in Table \ref{table:higherED} below

\begingroup
\renewcommand{\arraystretch}{1.5}
\begin{table}[h!]
\begin{center}
\begin{tabular}{c|c|c|c|c|c||c|c}
$D$ & $\Ex{d}$ & $E$ & ${\rm ad}F$  & $N$  & $W$ & $\tilde{H}_d$ & $\mathcal{S}$  \\
\hline
5 &  $\Ex{6}$  & $ \rep{27}_1$ &  $\rep{78}_0 $ & $\rep{27}^\prime_2$  & $ \rep{351}^\prime_{-1}$  & $\USp{8}$ & $ \rep{8}$\\
\hline
6 &  $\Spin{5,5}$  &
$ \rep{16}^c_{-1}$ &  $\rep{45}_0 $ & $\rep{10}_2$  & $ \rep{144}^c_{-1}$  &   $\USp{4} \times \USp{4}$ &
$ (\rep{4}, \rep{1}) \oplus   (\rep{1}, \rep{4}) $  \\
\hline 
7 &  $\SL{5,\mathbb{R}}$  & $ \rep{10}_1$ &
$\rep{24}_0$ & $\rep{5}_2$  & $ \rep{40}_{-1} \oplus \rep{15}^\prime_{-1}$   & $\USp{4} $ & $ \rep{4}$
\end{tabular}
\end{center}
\caption{Exceptional geometries for truncations to 5,6 and 7 dimensions}
\label{table:higherED}
\end{table}
\endgroup

Truncations to $D=5,6,7$ within 
generalised geometry have already been  studied in the literature  (see for instance \cite{Cassani:2019vcl, Lee:2014mla,Hohm:2014qga,Baguet:2015sma,Ciceri:2016dmd,Cassani:2016ncu,Malek:2016bpu,Malek:2017njj,Malek:2018zcz,Malek:2019ucd,Cassani:2020cod,Josse:2021put}). 
In this section, for completeness, we summarise the main results, present them all in the language of Exceptional Generalised Geometry  and complete them when needed.

\subsection{Truncations to 5 dimensions}

 Consistent truncations to 5 dimensions are associated to generalised structures in $\Ex{6}$ generalised geometry and have been studied in \cite{Lee:2014mla, Hohm:2014qga, Malek:2017njj,Cassani:2020cod, Josse:2021put}.

In 5 dimensions supergravity theories exist with $2 \leq \mathcal{N} \leq 8$ supercharges. The supersymmetry parameters $\epsilon_i$, with $i=1, \dots, \mathcal{N}$, transform in the fundamental of the $\USp{\mathcal{N}}$  R-symmetry group and are symplectic (pseudo) Majorana spinors: $\epsilon^i = \Omega^{ij} \epsilon^c_j$, where $\Omega^{ij}$ is the $\USp{\mathcal{N}}$  symplectic invariant. Thus only even numbers of supercharges 
are allowed.

Truncations to 
$\mathcal{N} \geq 4$ supergravity are associated to generalised $\Gst$-structures defined only by globally invariant vectors, while for  $\mathcal{N} = 2$ supersymmetry invariant adjoint elements are also needed.

\subsubsection{\texorpdfstring{$\mathcal{N} = 8$}{} supergravity }

Five-dimensional ungauged supergravity with maximal symmetry was constructed in \cite{Cremmer:1980gs} and its gauging have been studied in several papers (see for instance \cite{Gunaydin:1984qu,Gunaydin:1985cu, Pernici:1985ju}). The field content consists of the graviton, 8 gravitini, 27 vectors, 28 gravitini and 42 scalars. The fermions transform in the $\rep{8}$ and
$\rep{48}$ of $\USp{8}$, while the vector fields transform in the fundamental of $\Ex{6}$. The scalar parameterise the manifold
\begin{equation}
\label{n85dsc}
    \mathcal{M} = \frac{\Ex{6}}{\USp{8}} \, . 
\end{equation}

\medskip 

The truncation corresponds to a generalised Scherk-Schwarz reduction, where the 
$\rep{27}$ globally generalised vectors $K_I$ , $I=1, \dots, 27$, define a Leibniz parallelisation \cite{Lee:2014mla, Hohm:2014qga} 
\begin{equation}
    L_{K_I} K_J = X_{IJ}{}^K K_K \,  , 
\end{equation}
with  $X_{IJ}{}^K$ constant and again $[X_I, X_J] =  - X_{I J}{}^K X_K$. 
The vectors are normalised to $G(K_I, K_J) = \delta_{IJ}$, with $G$ the generalised metric. 
They give the 27 vectors of the truncated theory, while from \eqref{scalarman} one recovers the scalar manifold \eqref{n85dsc}.

The intrinsic torsion $X_{I J}{}^{K}$ transforms in the $\rep{351} $ and gives the embedding tensor of the truncated theory.

\medskip 

$\mathcal{N}=8$ supergravity with gauge group $\SO{6}$ \cite{Gunaydin:1984qu,Pernici:1985ju, Gunaydin:1985cu} is obtained 
as  the truncation of type IIB on $AdS_5 \times S^5$, whose  consistency   was proven only thanks to the results of \cite{Lee:2014mla, Hohm:2014qga}. In this case, decomposing the singlet intrinsic torsion under $\Ex{6} \supset   \SL{6,\mathbb{R}} \times\SL{2, \mathbb{R}}$ 
\begin{equation}
    \rep{351} = (\rep{21},\rep{1})+(\rep{15},\rep{3})+(\rep{84},\rep{2})+(\rep{6},\rep{2})+(\rep{105},\rep{1}) \, , 
\end{equation}
one can show \cite{Lee:2014mla} that $X_{IJ}{}^K$ belongs to the component $(\rep{21},\rep{1})$ and takes the form
\begin{equation}
\begin{aligned}
    X_{[i i^\prime] [j j^\prime] }{}^{[k k^\prime]} & = R^{-1} (\delta_{ij} \delta_{i^\prime j^\prime}^{k k^\prime} - \delta_{i^\prime j} \delta_{i j^\prime}^{k k^\prime} - 
\delta_{ij^\prime } \delta_{i^\prime j}^{k k^\prime} + \delta_{i^\prime j^\prime} \delta_{ij}^{k k^\prime} ) \, , \\
X_{[i i^\prime] \beta  k }{}^{\gamma j } & = R^{-1} (\delta_{ik} \delta_{i^\prime}^{j} - \delta_{i^\prime k} \delta_{i}^{j}  ) \delta_{\beta}^{\gamma}  \, ,
\end{aligned}
\end{equation}
where $i,j \dots = 1, \dots, 6$  are $\SL{6, \bbR}$ indices and $R$ is the radius of $S^5$. The tensors above   give  the  embedding tensor for the $\SO{6}$ electric gauging. 
 The consistency of other compact and non-compact gaugings can also be obtained in this way \cite{Baguet:2015sma}.

\subsubsection{\texorpdfstring{$\mathcal{N} = 6$}{} supergravity} 

$\mathcal{N}=6$ pure supergravity with gauge group  $\SU{3} \times \U{1}$ was constructed in \cite{Gunaydin:1985cu}
as a consistent truncation of  the 
$\mathcal{N}=8$ theory.  The fields are arranged in the graviton multiplet, which consists of the graviton, 6 gravitini, 15 vectors, 20 spin 1/2 and 14 scalars. 

The $\mathcal{N}=6$ theory corresponds to a truncation with a generalised $\Gst=\SU{2}$ structure, defined by $15$ invariant vectors $K_I$, as it can be seen 
from the embedding 
\begin{equation}
\begin{aligned}
    \Ex{6} &  \supset \SUs{6} \times 
    \SU{2} \\
    \rep{27} &  = (\rep{\overline{15}} , \rep{1} )\oplus (\rep{1}, \rep{2}) \, . 
\end{aligned}
\end{equation}
The generalised vectors are normalised to $G(K_I, K_J) = \delta_{IJ}$, with $G$ the generalised metric, and  give the 15 vectors in the  graviton multiplet.
Using  again \eqref{scalarman} with $\Com{\SU{2}}{\USp{8}} =\USp{6}$, we recover the scalar manifold
\begin{equation}
    \mathcal{M}=\frac{\SUs{6}}{\USp{6}} \, . 
\end{equation}
The possible gaugings are encoded in the singlet intrinsic torsion transforming in 
\begin{equation}
    W_{\rm int} = \rep{\overline{105}} \oplus \rep{\overline{21}}
\end{equation}
of the $\SUs{6}$ global symmetry group.

The $\mathcal{N}=6$ theory of \cite{Gunaydin:1985cu} is believed to describe a subsector of the chiral primary operators of $\mathcal{N}=4$ Super Yang-Mills \cite{Ferrara:1998zt}, which is identified using the $\SL{2, \mathbb{R}}$ symmetry. It is still unclear whether a supergravity truncation can be constructed.

\subsubsection{\texorpdfstring{$\mathcal{N} = 4$}{} supergravity }

For $\mathcal{N}=4$ supersymmetry there are two kinds of multiplet: the graviton multiplet, containing the metric,
4 gravitini, 6 vectors, 4  spin 1/2 fermions and 1 real scalar, and 
vector multiplets, consisting of 1 vector, 4 spin 1/2 fermions and 5 real scalars, each.
The scalar manifold is the coset
\begin{equation}
\label{eq:d5scalm}
    \mathcal{M} = \SO{1,1} \times \frac{\SO{5,n_{\rm V}}}{\SO{5} \times \SO{n_{\rm V}}} \, , 
\end{equation}
where $\SO{1,1}$  is parameterised by  the scalar in the graviton multiplet and the other factor by those in the vector multiplets. 

The gauging of the global isometry group ${\rm G}_{\rm iso} = \SO{1,1} \times\SO{5,n_{\rm V}}$
\begin{equation}
    D_\mu = \nabla_\mu - g A_\mu{}^I 
    (f_I{}^{JK} t_{JK} + \xi^J t_{JK} + \xi_I t_0)  - g A_\mu{}^0 \xi^{IJ} t_{IJ} \, , 
\end{equation}
where  $t_{IJ}$ and $t_0$ are the generators of  $\SO{5,n}$ and $\SO{1,1}$, 
are determined by the embedding tensor \cite{Schon:2006kz}, with components 
\begin{equation}
 \Theta_I{}^\alpha  =  ( \xi_{I},  \xi_{[IJ]}, f_{[IJK]} )
    \qquad \quad I,J, K = 1, \ldots, 5+n_{\rm V} \, ,
\end{equation}
where the bracket denotes full antisymmetrisation. 

\medskip

In the context of Exceptional Geometry and Exceptional Field Theory, 
consistent truncations to 
$\mathcal{N}=4$ supergravity theories have been studied in \cite{Malek:2017njj,Cassani:2019vcl}. 

There exists only one family\footnote{As discussed in  \cite{Cassani:2019vcl} the  $\SO{5}$ subgroups $\Gst= \SU{2} \times \U{1}$ and $\Gst= \U{1}^2$ also lead 
to an $\mathcal{N}=4$ truncation.
They are both subgroups of $\Gst=\SO{4}$ and one can show that they give the same truncated theory as the $\SO{4}$ case.}  of truncations corresponding to 
\begin{equation}
    \Gst = \Spin{5-n_{\rm V}} \subseteq \USp{4} \subset \USp{8} \, , 
\end{equation}
where $n_{\rm V}= 0, \ldots 5$ is the number of vector multiplets, and  $\Spin{1}=\Spin{0} = \mathbb{Z}_2$. 

The $\Gst$-structure is  defined by 
$6 + n_{\rm V}$ invariant generalised vectors 
according to 
\begin{equation}
\begin{aligned}
    \Ex{6} &  \supset  \SO{1,1} \times \SO{5,n} \times \SO{5-n} \\ 
    \rep{27} &  \ni (\rep{1}, \rep{1})_{-4}
 \oplus (\rep{5} + n, \rep{1})_2  = \{K_0, K_I \} \qquad I=1, \dots, 5+n_{\rm V} \, , 
\end{aligned}
\end{equation}
 satisfying 
the compatibility conditions (see \cite{Cassani:2019vcl} for more details)
\begin{equation}
\begin{aligned}
& c(K_0,K_0,V) = 0 \, , \\ 
& c(K_0,K_I,K_J) = \eta_{IJ}  \kappa^2   \\ 
& c(K_I,K_J,K_K) = 0  
\end{aligned}
\qquad 
\begin{aligned} & \forall \,V\in\Gamma(E)  \\
 & \forall I,J,K = 1, \dots, 5+n_{\rm V}
 \end{aligned}
\end{equation}
where $c(V,V',V'')$ is the $\Ex{6} $
cubic invariant, 
$\eta_{IJ}={\rm diag}(-\id_5, \id_{n_{\rm V}})$ is the  flat $\SO{5,n_{\rm V}}$ metric.

The generalised vectors $\{K_{\tilde I}\}$ with $\tilde{I}=0,\dots 5$ 
and $\tilde{I}=6 \dots 5+ n_{\rm V}$ 
determine the vectors in the gravity multiplets and  those in the vector multiplets, respectively. 
From \eqref{scalarman} one recovers the scalar manifold \eqref{eq:d5scalm}.

The singlet intrinsic torsion has components
\begin{equation}
    W_{\rm int} = ( \rep{5} \oplus n_{\rm V} )_{-4} \oplus( \rep{X}_{IJ})_2 \oplus (\rep{X}_{IJK})_2
\end{equation}
where $\rep{X}_{IJ}$ and $\rep{X}_{IJK}$ denote the antisymmetric two and three-tensors representation of $\SO{5+ n_{\rm V}}$ and the subscripts give the $\SO{1,1}$ charge.
The generalised Lie derivative among the singlet generalised vectors
\begin{equation}
    L_{K_{\tilde I}} K_{\tilde J} =  \Theta_{\tilde I} \cdot K_{\tilde  J}   = \Theta_{\tilde I}{}^{\hat{\alpha}}(t_{\hat{\alpha}})_{\tilde J}  {}^{\tilde K} K_{\tilde K} 
=  X_{ \tilde{I} \tilde {J}}{}^{\tilde K}  K_{\tilde K} \, , 
\end{equation}
with $(X_{\tilde I})_{\tilde J}{}^{\tilde K} = X_{\tilde{I} \tilde{J}}{}^{\tilde K}$ and 
$[X_{\tilde I}, X_{\tilde J}]
       =  - X_{\tilde{I} \tilde{J}}{}^{\tilde K} X_{\tilde K} $,
reproduces the embedding tensor 
of the truncated theory 
\begin{equation}
X_{IJ}{}^K  = - f_{IJ}{}^K - 
 \frac{1}{2} \eta_{IJ} \xi^K + \delta_{[I}{}^J \xi_{K]} \,  \qquad 
 X_{I0}{}^0 = \xi_I  \, \qquad 
 X_{0 I}{}^J = - \xi_{I}{}^J \,. 
\end{equation}

\medskip 

There are several examples of 
half-maximal truncations to 5 dimensions.
For instance, the truncation 
\cite{Cassani:2010uw, Gauntlett:2010vu}  of type IIB supergravity on squashed Sasaki-Einstein manifolds to gravity coupled to  two vector  multiplets and gauge group  $\mathrm{Heis}_3 \times U(1)$
is reproduced with a generalised $\Gst= \SO{3}$ structure \cite{Cassani:2019vcl}, which corresponds to the ordinary $\SU{2}$ structure of the original truncation. 

A generalised $\SU{2}$ structure can also be used to derive a consistent truncation on  $\beta$-deformed  Sasaki-Einstein manifolds \cite{Cassani:2019vcl} to give a continuous family of $\mathcal{N}=4$ theories   with two vector multiplets and  $U (1) \times  \mathrm{Heis}_3$  gauging. This family  contains  the  truncation  to pure gauged supergravity  of \cite{Liu:2019cea}.

A generalised $\U{1}$ structure leads to the most general truncation of 11-dimensional supergravity around  the Maldacena-Nun\~ez  solution with $\mathcal{N}=4$ supersymmetry \cite{Maldacena:2000mw}. The truncated theory \cite{Cassani:2019vcl} consists of  half-maximal supergravity coupled to three vector multiplets and with  $\U{1} \times {\rm ISO}(3)$ gauge group, which 
 reproduces the reduction of 7-dimensional gauged supergravity of 
\cite{Cheung:2019pge}. 
On the other hand, if one includes the trombone symmetry a larger truncation around the Maldacena-Nun\~ez solution to maximal supergravity can be obtained \cite{Bhattacharya:2024tjw, Varela:2025xeb}. In this case there is no Lagrangian for the truncated theory.

There are also examples of truncations to  ungauged supergravity. For instance, a 
$\Gst=\SO{3}$ on $K3 \times T^2$  gives a consistent truncation of 11-dimensional supergravity  to  ungauged supergravity with two vector multiplets \cite{Malek:2017njj}.

\subsubsection{\texorpdfstring{$\mathcal{N} = 2$}{} supergravity} 

In 5 dimensions  $\mathcal{N}=2$ supergravity coupled to matter multiplets contains  the gravity multiplet, vector, tensor and hypermultiplets \cite{Gunaydin:1999zx, Ceresole:2000jd, Bergshoeff:2002qk}. 
The gravity multiplet consists of the graviton, 2 gravitini transforming as a doublet of the R-symmetry group $\SU{2}_R$ and the graviphoton. A vector multiplet consists of  a vector, 2 spin-1/2 fermions in the fundamental of $\SU{2}_R$ and a complex scalar, while in a tensor multiplet the vector is replaced by a two-form, which is dual to a vector.
 Thus  vector and tensor multiplets have the same  number of degrees of freedom.  The scalars of the vector and tensor multiplets are grouped together and  parametrise a very special real manifold $\mathcal{M}_{\rm VT}$. Finally, a hypermultiplet consists of  4 real scalars and an R-symmetry doublet of spin-1/2 fermions. The scalars of the hypermultiplets parameterise a quaternionic K\"ahler manifold $\mathcal{M}_{\rm H}$. 
The isometry group splits into the isometries of the vector/tensor and hypermultiplet scalar manifolds
\begin{equation}
    {\rm G}_{\rm iso} = {\rm G}_{\rm VT} \times  {\rm G}_{\rm H} \, . 
\end{equation}
The Killing vectors determining the gaugings are the combinations 
\begin{equation}
\label{5dKilling}
    k_{\tilde I}^i(\phi) = \Theta_{\tilde I}^a k_a^i(\phi) 
    \qquad k_{\tilde I}^x(q) = \Theta_{\tilde I}^A k_A^x(q)
\end{equation}
of the vectors  generating  $G_{\rm iso}$ via the embedding tensor 
\begin{equation}
    \Theta_{\tilde I}{}^{\alpha} = ( \Theta_{\tilde I}^a ,  \Theta_{\tilde I}^A ) \, , 
\end{equation}
 where the indices $a$ and $A$ run over the dimensions of 
  ${\rm G}_{\rm VT}$ and ${\rm G}_{\rm H}$, respectively. In \eqref{5dKilling} $\phi^i$ with  $i=1,\dots, n_{\rm VT}$ and $q^x$ with $x =1, \dots, ,4 n_{\rm H}$ denote the scalars in the vector and hypermultiplets,  respectively.

\medskip 

The classification of   $\mathcal{N}=2$ supergravity  theories in 5 dimensions that have an 11/10-dimensional origin can be found in \cite{Josse:2021put}. Here we summarise the main results. 

In order to have $\mathcal{N}=2$ supersymmetry the generalised structures must be 
\begin{equation}
    \Gst \subseteq \USp{6} =  \Com{\SU{2}_R}{\USp{8}} \, , 
\end{equation}
where $\USp{8}$ is the double cover of the maximal compact subgroup of $\Ex{6}$.

\medskip 

$ \Gst =\USp{6}$ gives the truncation to pure supergravity and it is defined by a singlet generalised vector $K \in \rep{27}$  of positive norm with respect to 
the $\Ex{6}$ cubic invariant
\begin{equation}
    c(K,K,K) = 6 \kappa^2 \, ,  \qquad \kappa^2 \in \Gamma(\det T^* M) \, ,
\end{equation}
 and
 a triplet of singlet weighted adjoint elements $J_\alpha$ defining a highest weight $\su(2)$ subalgebra of $\ex{6}$. 
The pair $(K, J_\alpha)$ satisfies
\begin{equation}
\begin{aligned}
    &  J_\alpha \cdot K =0  \, ,  \\
    &  \tr(J_\alpha J_\beta ) = - c(K,K,K) \delta_{\alpha \beta} \, , 
\end{aligned}
\end{equation}
with $\alpha = 1,2,3$. The singlet generalised vector gives the graviphoton of the truncated theory, while the singlets $J_\alpha$ are the generators of the $\SU{2}$ R-symmetry. From \eqref{scalarman}
it follows that the scalar manifold is trivial, as expected. 

\medskip  

Also in this case, there are truncations with only vector/tensor multiplets and only hypermultiplets, and truncations with both. The generalised structures are defined by a set of singlet generalised vectors $K_{\tilde I}$ with ${\tilde I}=0, \ldots n_{VT}$ and a set of singlet adjoint tensors $J_A$, with $A = 1, \ldots  {\dim}{\rm G}_{\rm H}$, satisfying
\begin{equation}
\label{compN25d}
\begin{aligned}
    &  J_A \cdot K_{\tilde I} = 0 \,  ,\\
&  c(K_{\tilde I},K_{\tilde J},K_{\tilde K}) = 6 \kappa^2 C_{{\tilde I}, {\tilde J}, {\tilde K}} \, , \\
& [ J_A, J_B] = \kappa  f_{AB}{}^C J_C \, , \\
& \tr(J_A , J_B) =  \kappa^2  \eta_{AB} \, , 
\end{aligned} \qquad
\begin{aligned} 
& \forall\,  \tilde{I}, \tilde{J}, \tilde{K} = 0, \ldots n_{VT} \\
& \forall \,  A,B = 1, \ldots  {\dim}{\rm G}_{\rm H}
\end{aligned}
\end{equation}
where $C_{{\tilde I} {\tilde J} {\tilde K}}$ is a symmetric, constant tensor (it gives the tensor of the same name in the truncated theory), $f_{AB}{}^C$ are the structure constants of $\mathfrak{g}_{\rm H}$, and $\eta_{AB}$ is a diagonal matrix with entries -1 and +1 for the  compact and non-compact generators of $G_{\rm H}$, respectively.

The scalar manifold is given by \eqref{scalarman} and it factorises  in 
\begin{equation}
   \mathcal{M} = \mathcal{M}_{\rm VT} \times  \mathcal{M}_{\rm H} = \frac{G_{\rm VT}}{H_{\rm VT}} \times \frac{G_{\rm H}}{H_{\rm H}}  \, . 
\end{equation}
By construction, all scalar manifolds are homogeneous and  symmetric spaces.

The intrinsic torsion transforms in $\rep{351}$ of $\Ex{6}$ and the $\Gst$- singlet components are determined  by 
\begin{equation}
\label{eq;torsionLD5}
    \begin{aligned}
    L_{K_{\tilde I}} K_{\tilde J} &  = - T_{\rm int}(K_{\tilde I}) \cdot K_{\tilde J}  = t_{{\tilde I} {\tilde J}}{}^{\tilde K} K_{\tilde K} \,   \\
 L_{K_{\tilde I}} J_{A}  & = - T_{\rm int}(K_{\tilde I}) \cdot J_{a}  = p_{{\tilde I} A}{}^{B} J_{B}  \, , 
    \end{aligned}
\end{equation}
with  $(t_{\tilde I})_{\tilde J}{}^{\tilde K}$ and 
$(p_{\tilde I})_{A}{}^{B}$  constant matrices  giving 
the two components of the embedding tensor. They also determine 
the elements of Lie algebrae of $G_{\rm VT}$  and $G_{\rm H}$ respectively. A detailed analysis of the intrinsic torsion and the possible gaugings can be found in  \cite{Josse:2021put}.

Truncations with only vector/tensor multiplets are associated to generalised structures that only admit as adjoint singlets the triplet $J_\alpha$ defining the $\SU{2}$ R-symmetry. This means that 
\begin{equation}
    \Gst  \subset \SUs{6} \, ,
\end{equation}
where $\SUs{6}$ is the stabiliser of the  $J_\alpha$'s. The $\Gst$-structures are defined by the set of singlets
\begin{equation}
(K_{\tilde I}, J_\alpha) \qquad \alpha=1,2,3 \quad {\tilde I} = 0, \ldots ,n_{\rm VT} \, ,
\end{equation} 
where,  by the compatibility condition \eqref{compN25d}, the generalised vectors $K_{\tilde I}$  belong to the component $(\rep{15}, \rep{1})$ in the breaking of the $\rep{27}$ under $\Ex{6} \supset \SUs{6} \times \SU{2}_R$. 
In the truncated theory, the vector $K_0 = K$ gives the graviphoton while the  other $K_I$, with $I=1 \ldots n_{\rm VT}$, give the vectors in the vector/tensor multiplets. 

The truncations  obtained in \cite{Josse:2021put}
are listed in Table \ref{table:vtN25d}. 
\begingroup
\renewcommand{\arraystretch}{1.5}
\begin{table}[h!]
\begin{center}
\begin{tabular}{|c|c|c|}
\hline
$n_{\rm VT}$ & $\Gst$ & $\mathcal{M}_{\rm VT}$   \\
\hline
$1 \dots 6$ &  $\Spin{6 - n_{\rm VT}}$  & $ \mathbb{R}^+ \times \frac{\SO{n_{\rm VT}-1,1}}{\SO{n_{\rm VT}-1}}$ \\
\hline
5 &  $\SU{2}$  & $ \frac{\SL{3, \mathbb{R}}}{\SO{3}}$   \\
\hline 
8 &  $\U{1}$  &  $\frac{\SL{3, \mathbb{C}}}{\SU{3}}$ \\
\hline 
14 & $\mathbb{Z}_2$ & $\frac{\SUs{6}}{\USp{6}}$ \\
\hline
\end{tabular}
\end{center}
\caption{$\mathcal{N}=2$ truncations with $n_{\rm H}=0$}
\label{table:vtN25d}
\end{table}
\endgroup

For truncations with only hypermultiplets the only singlet in $E$ must be the vector $K$. Thus the associated to generalised structures must be 
\begin{equation}
    \Gst \subset {\rm F}_{4(4)} \, , 
\end{equation}
where now ${\rm F}_{4(4)}$ is the stabiliser of 
the generalised vector $K$. The $\Gst$-structures are defined by the singlet $K$ and a set of adjoint tensors $J_A$, with $A = 1, \ldots  {\dim}G_{\rm H}$.  

There are only two possible truncations 
\begingroup
\renewcommand{\arraystretch}{1.5}
\begin{table}[h!]
\begin{center}
\begin{tabular}{|c|c|c|}
\hline
$n_{\rm H}$ & $\Gst$ & $\mathcal{M}_{\rm H}$   \\
\hline
1  &  $\SU{3}$  & $ \frac{\SU{2,1}}{\SU{2} \times \U{1}}$ \\
\hline
2 &  $\SO{3}$  & $ \frac{{\rm G}_{2(2)}}{\SO{4}}$   \\
\hline 
\end{tabular}
\end{center}
\caption{$\mathcal{N}=2$ truncations with $n_{\rm V}=0$.  }
\label{table:vh25d}
\end{table}
\endgroup

As mentioned in \cite{Josse:2021put}, there are two other symmetric spaces\footnote{${\rm SO}_0(4, n_{\rm V})$ denotes
the  connected component of $\SO{4, n_{\rm V}}$.}
\begin{equation}
    \begin{aligned}
 \mathcal{M} & =  \frac{{\rm SO}_0(4, n_{\rm V})}{\SO{4} \times \SO{n_{\rm V}}}    \\
\mathcal{M} & =  \frac{{\rm F}_{4(4)}}{\USp{6} \times \SU{2}} 
    \end{aligned}
\end{equation}
which could, in principle, correspond to a consistent truncation with hypermultiplets. They should correspond to substruncations  of half-maximal and maximal supergravity with a discrete $\Gst$-structure.
However, the requirement that $\Gst \subset \USp{8}$ and Schur lemmma are enough to show that such truncations are not allowed. 

\medskip

Finally, the truncations with both 
 vector/tensor and hypermultiplets,
 are simply\footnote{The truncation with $n_{\rm V} = 3$ automatically enhances to $n_{\rm V} = 4$.} 
\begingroup
\renewcommand{\arraystretch}{1.5}
\begin{table}[H]
\begin{center}
\begin{tabular}{|c|c|c|c|c|}
\hline
$n_{\rm VT}$ & $n_{\rm H}$ & $\Gst$ & $\mathcal{M}_{\rm VT}$ & $\mathcal{M}_{\rm H}$  \\
\hline
1  &  1 &  $\SU{2} \times \U{1}$  & $\mathbb{R}^+$ &
$ \frac{\SU{2,1}}{\SU{2} \times \U{1}}$ \\
\hline
2  &  1 &  $\U{1}$  & $\mathbb{R}^+ \times \frac{\SO{1,1}}{\SO{3}} $ &
$ \frac{\SU{2,1}}{\SU{2} \times \U{1}}$ \\
\hline
4  &  1 & $\U{1}$ & $\mathbb{R}^+ \times \frac{\SO{3,1}}{\SO{3}}$  & $ \frac{\SU{2,1}}{\SU{2} \times \U{1}}$  \\
\hline 
\end{tabular}
\end{center}
\caption{$\mathcal{N}=2$ truncations with vector/tensor and hypermultiplets}
\label{table:vthN25d}
\end{table}
\endgroup

\medskip

To our knowledge there are no examples of truncations with only vector multiplets nor only hypermultiplets.

On the other hand, all cases in Table \eqref{table:vthN25d} correspond 
known truncations of 11-dimensional supergravity around $AdS_5$ solutions. 
The  $\Gst=\U{1}$ structure in the third line gives the truncation 
of 11-dimensional supergravity around the Maldacena-Nu\~nez solution \cite{Maldacena:2000mw} with $\mathcal{N} = 2$ supersymmetry \cite{Cassani:2020cod}.  The resulting 5-dimensional theory contains 4 vector and 1 hypermultiplet and has gauge group $\SO{3} \times \U{1}$.
The first line should correspond to the subtruncation of this theory to one vector and one hypermultiplet and gauging $\U{1} \times \mathbb{R}^+$ derived in \cite{Faedo:2019cvr}. 
Finally, the other  $\Gst=\U{1}$ structure gives the truncation 
of 11-dimensional supergravity around the BBBW solutions \cite{Bah:2012dg}. This is an infinite family of the $\mathcal{N}=2$ solutions generalising the Maldacena-Nu\~nez one.
The truncated theory contains two vectors and one hypermultiplet, with gauge group $\U{1} \times \mathbb{R}^+$ \cite{Cassani:2020cod}. This truncation  extends the one of \cite{Szepietowski:2012tb}.

\subsection{Truncations to 6 dimensions}

In 6 dimensions the spinorial representation is reducible and the 
supersymmetry parameters are symplectic Majorana-Weyl spinors. 
The amount of allowed supersymmetry is $(\mathcal{N}_-, \mathcal{N}_+)$  with the supersymmetry parameters transforming in the fundamental of the  $\USp{\mathcal{N}_+} \times \USp{\mathcal{N}_-}$ R-symmetry group.
Depending on the values of $\mathcal{N}_-$ and $\mathcal{N}_+$ one can construct chiral and non chiral theories.

For truncations to  6 dimensions the relevant exceptional group  is $\Ex{5} = \Spin{5,5} \sim \SO{5,5}$ \cite{Malek:2016bpu,Cassani:2016ncu, Malek:2018zcz, Malek:2019ucd}, and the $\Gst$-structures leading to minimal supergravities with $(\mathcal{N}_-, \mathcal{N}_+)$ supersymmetry are  \cite{Cassani:2019vcl} 
\begingroup
\renewcommand{\arraystretch}{1.5}
\begin{table}[H]
\begin{center}
\begin{tabular}{c|c|c}
$(\mathcal{N}_-, \mathcal{N}_+)$ &  $\Gst$ & ${\rm G}_R = \Com{\Gst}{\USp{4} \times \USp{4}}$ \\
\hline
$(\rep{2}, \rep{2})$  &  $\id$ &  $\USp{4} \times \USp{4}$  \\
\hline
$(\rep{2}, \rep{1})$  &  $\SU{2}$ &  $\USp{4} \times \SU{2}$  \\
\hline
$(\rep{2}, \rep{0})$  &  $\USp{4}$ &  $\USp{4}$  \\
\hline
$(\rep{1}, \rep{1})$ &  $\SU{2} \times \SU{2}$ &  $\SU{2} \times \SU{2}$  \\
\hline
$(\rep{1}, \rep{0})$  &  $\SU{2} \times \USp{4}$ &  $\SU{2}$  \\
\end{tabular}
\end{center}
\caption{Structure and R-symmetry groups for six-dimensional truncations}
\label{table:6dtr}
\end{table}
\endgroup

\subsubsection{Maximal supergravity}

There are three possible maximal supergravity algebrae in six dimensions: the non-chiral $\mathcal{N}= (2,2)$ algebra and two chiral ones, $\mathcal{N}= (4,0)$ and $\mathcal{N}= (3,1)$. 

The full non-linear $\mathcal{N}= (2,2)$ theory has been constructed in \cite{Tanii:1998px}. It contains  the graviton, 5 two-forms, 8 gravitini, 16 vectors, 40 spin 1/2 fields and 25 scalars
which parameterise the coset
\begin{equation}
\label{6dmaxsc}
    \mathcal{M} = \frac{\SO{5,5}}{\SO{5} \times \SO{5}} \, , 
\end{equation}
where $\SO{5,5}$ is the global isometry group and $\SO{5} \times \SO{5} \sim \USp{4}_l \times \USp{4}_r$\footnote{The subscript $l$ and $r$ denote left and right chirality.} its maximal compact subgroup.   The fermions are left- and right-handed symplectic Majorana-Weyl. In particular the supersymmetry parameters transform in the $(\rep{4},\rep{1}) \oplus (\rep{1},\rep{4})$ of $\USp{4} \times \USp{4}$.

The gauging of the theory are given in terms of the embedding tensor 
\begin{equation}
 D_\mu = \nabla_\mu - g A^I_\mu \Theta_I{}^{AB} t_{AB} 
\end{equation}
where $t_{AB} = t_{[AB]}$  with $A,B = 1, \dots, 10$ are the $\SO{5, 5}$ generators. The embedding tensor  $\Theta$ can be written in terms
of a tensor $\theta_{JA}$ transforming in the  $\rep{144^c}$ of $\SO{5,5}$ 
\begin{equation}
\label{max6det}
   \Theta_I{}^{AB} = - \theta^{L[A } \gamma^{B]}{}_{LI } 
\end{equation}
where $\gamma_{AIJ}$ are the $\SO{5,5}$ gamma matrices and the tensor and  $\theta^{IA}$ satisfies 
$\gamma_{AIJ} \theta^{JA}  = 0$ \cite{Bergshoeff:2007ef}. The theory with $\SO{5}$ gauge group was obtained in \cite{Cowdall:1998rs} as a circle reduction of $\SO{5}$ maximal supergravity in 7 dimensions. The classification of all possible gaugings can be found in \cite{Bergshoeff:2007ef}.

The chiral theories $\mathcal{N}= (4,0)$ and $\mathcal{N}= (3,1)$ are more exotic since  the graviton is replaced by tensor fields with mixed symmetries and self-duality conditions. For both theories only the linearised actions are  known (see  \cite{Hull:2000zn,Hull:2000rr,Bertrand:2022pyi, Bertrand:2020nob}  for more details).

\medskip
Only the non-chiral theory $\mathcal{N}= (2,2)$ can be obtained as a consistent truncation.
Recall that the supersymmetry parameters of the truncated theory are given by $\Gst$-singlets in the generalised spinor bundle $\mathcal{S}$ and that the R-symmetry is $\Com{\Gst}{\tilde{H}_d}$. 
Since maximally supersymmetric truncations are  associated to a
$\Gst= \id$, the supersymmetry parameters are given by the full generalised spinor bundle and transform as 
\begin{equation}
    (\rep{4},\rep{1}) \oplus  (\rep{1}, \rep{4})
\end{equation}
of the  $\USp{4} \times \USp{4}$ R-symmetry group (see Table \ref{table:higherED}). This structure is only compatible 
with  $\mathcal{N}= (2,2)$ supersymmetry. 

The $\Gst= \id$ is associated to  a genaralised Leibiniz parallelisation defined by 16 generalised vectors $K_I$, $I=1, \dots, 16$, transforming in the spinorial representation of $\SO{5,5}$. The $K_I$ are  normalised with the generalised metric $G(K_I, K_J) = \eta_{IJ}$, with $ \eta_{IJ}$ the 
$\SO{5,5}$ invariant metric. They  give the 16 vectors of the truncated theory. 

The projection  on the bundle $N$ of the symmetric product of the singlet generalised vectors defines 10 singlet tensors $Q_M$, which,  in the truncated theory, give  the 5 two-forms and their duals. 
The scalar manifold is given by \eqref{scalarman} and reproduces 
\eqref{6dmaxsc}.

The vectors realise a Leibniz algebra
\begin{equation}
    L_{K_I} K_J = X_{IJ}{}^K K_K \,, 
\end{equation}
where the constant intrinsic torsion  
\begin{equation}
\begin{aligned}
     X_{IJ}{}^K & = - \tfrac14 \Theta_I{}^{AB} \gamma_{AB}{}^K{}_J  
     \end{aligned}
\end{equation}
  transforms in the $\rep{144}^c$ of $\SO{5,5}$ and gives the embedding tensor \eqref{max6det}.

\medskip

The $\SO{5}$ gauged supergravity of \cite{Cowdall:1998rs} is obtained as 
a truncation of massless type IIA on $S^4$ \cite{Cowdall:1998rs,Cvetic:2000ah} and has been rederived using  generalised geometry in \cite{Cassani:2016ncu}. 
It is convenient to consider the embedding of $S^4$ in $\bbR^5$ and then decompose $\SO{5,5}$ under $\SL{5, \bbR}$. Then  
 non-zero component of the intrinsic torsion are 
\begin{equation}
\begin{aligned}
 X_{[i i^\prime] [j j^\prime]}{}^{[k k^\prime]} & \sim R^{-1} ( \delta_{i j}  \delta_{j^\prime i^\prime}^{k k^\prime} - \delta_{i j^\prime}  \delta_{j^\prime i}^{k k^\prime} -
 \delta_{i^\prime j}  \delta_{j^\prime i}^{k k^\prime} 
 + \delta_{i^\prime j^\prime}  \delta_{j i}^{k k ^\prime} )
 \, , \\
 X_{[i i^\prime]  j }{}^{k } & \sim  - R^{-1}  ( \delta_{j i^\prime } \delta_i^k - \delta_{j i} \delta_{i^\prime}^k ) \, , 
 \end{aligned}
\end{equation}
with $R$ the radius of $S^4$ and $\delta_{ij} \sim \theta_{ij}$ are the non trivial components of the tensor $\theta^{IA}$ in \eqref{max6det}. The tensors $X_{[i i^\prime] [j j^\prime]}{}^{[k k^\prime]}$ and $X_{[i i^\prime]  j }{}^{k }$
reproduce the $\SO{5}$ gauge algebra.

\subsubsection{\texorpdfstring{$\mathcal{N} = (2,1)$}{} supergravity}

There also exists an $\mathcal{N}=(2,1)$ supergravity \cite{Julia81,DAuria:1997caz,Bergshoeff:1986hv}. The R-symmetry is $\USp{4} \times \USp{2}$ and the fields are arranged in the graviton multiplet consisting of  the graviton, 1 self-dual and 5 anti-self dual two forms in $\rep{5}$ of $\USp{4}$, a $\USp{2}$ doublet of positive chirality gravitini and 
4 doublets of negative chirality gravitini, 
8 vectors in $(\rep{4,2})$ of the R-symmetry, 10 
positive chirality and 4 negative  chirality  spin1/2 spinors  in the  $(\rep{5,2})$ and $(\rep{4,1})$ of $\USp{4} \times \USp{2}$ respectively, and 5 scalars, which are neutral under  $\USp{2}$.  The theory is anomalous. 

\medskip

The field content of the  $\mathcal{N}=(2,1)$ theory is easily deduced from Exceptional Generalised Geometry.  The theory corresponds to a $\Gst= \USp{2}$ structure embedded in $\SO{5,5}$ as 
\begin{equation}
\begin{aligned}
    \SO{5,5} & \supset \SO{5,1} \times \USp{2}_R \times \USp{2}_S \, .
\end{aligned}
\end{equation}
It gives 8 singlet generalised vectors $K_I$ and 6 singlet tensors $Q_M$ in the bundle $N$
\begin{equation}
    \begin{aligned}
    \rep{16} & \to (\rep{4}, \rep{1} , \rep{2}) \oplus  (\rep{4}, \rep{2} , \rep{1}) \, , \\
        \rep{10} & \to (\rep{6}, \rep{1} , \rep{1})  \oplus  (\rep{1}, \rep{2} , \rep{2}) \, .
    \end{aligned}
\end{equation}
corresponding to the vectors and tensors in the graviton multiplet of the truncated theory. The scalar manifold is given by \eqref{scalarman}
\begin{equation}
    \mathcal{M} = \frac{ \SO{5,1}}{\SO{5}} \, . 
\end{equation}

\subsubsection{Half-maximal supergravity}

In six dimensions half-maximal supergravity can be chiral and non chiral.

The non chiral theory 
has $\mathcal{N}=(1,1)$ supersymmetry with the supersymmetry parameteres transforming as left- and right-handed doublets of the $\SO{4}_R \sim \SU{2} \times \SU{2}$ R-symmetry. 
There are two types of multiplets, the graviton and  vector multiplets.   The graviton multiplet contains the graviton,  4 gravitini,  4 vectors, 1 two-form, 4 spin 1/2 fermions and 1 scalar,  while a vector multiplet consists of  1 vector,  4  1/2 fermions and 4 scalars.
All fermions are symplectic Majorana-Weyl and transform in $(\rep{2},\rep{1}) \oplus (\rep{1},\rep{2})$
of the R-symmetry group.\footnote{
For  Anti de Sitter backgrounds, which are relevant for gauged supergravity, the description in terms of chiral fermions does not hold:  each pair of chiral spinors is arranged into an  8-dimensional pseudo-Majorana spinor. 
At the same time the R-symmetry is broken to the diagonal $\SU{2}$ in $\SU{2} \times \SU{2}$. }

For $n_{\rm V}$ vector multiplets, the dilaton and the scalars in the vector multiplets  parameterise the coset 
\begin{equation}
    \mathcal{M} = \frac{\SO{4, n_{\rm V}}}{\SO{4} \times \SO{n_{\rm V}}} \times \mathbb{R}^+  \, . 
\end{equation}

The most general form of the gauged theory was constructed in \cite{DAuria:2000afl}. It has gauge group ${\rm F}(4)$ and generalises  \cite{ Giani:1984dw,Romans:1985tw} (see also \cite{Karndumri:2016ruc} for more recent results). 

Even if a fully $\SO{4, n_{\rm V}} \times \mathbb{R}^+$ covariant formulation of six-dimensional gauged supergravity has not been constructed,  the components of the embedding tensors have been identified \cite{Bergshoeff:2007vb} in the Kac-Moody approach to supergravity
\begin{equation}
\label{embt6n11KM}
 \Theta_I^{JK}  = f_{I}{}^{JK} + \delta_I^{[J} \xi^{K]}  \, , \qquad   \Theta_I^{0}   =  \xi_I \, ,
\end{equation} 
where $I,J,K = 1, \dots, 4 + n_{\rm V}$ and $\Theta_I^{JK}$ and  $\Theta_I^{0}$  give the gaugings of ${\rm G} \subset\SO{4, n_{\rm V}} $ and $\mathbb{R}^+$, respectively.

The chiral theory \cite{Townsend:1983xt, Romans:1986er} has supersymmetry 
$\mathcal{N}= (2,0)$. The supersymmetry parameters are left-handed symplectic Majorana-Weyl  transforming in the $\rep{4}$ of the $\USp{4}_R$ R-symmetry group. The field consists of the gravity multiplet (the graviton, 4 left-handed gravitini and 5 two-forms) coupled to $n_{\rm T}$ tensor multiplets (1 anti self-dual two-form, 4 right-handed symplectic Majorana spinors and 5 scalars).  The scalars parameterise the manifold
\begin{equation}
    \mathcal{M} = \frac{\SO{5,n_{\rm T}}}{\SO{5} \times \SO{n_{\rm T}}}
\end{equation}
The supersymmetry transformations and field equations for the graviton multiplet coupled to $n_{\rm T}$ tensor multiplets are given in \cite{Romans:1986er,Riccioni:1997np}. Being chiral, the theory is anomalous.
The only anomaly free theory has  $n_{\rm T} =21$ tensor multiplets and it can be obtained as a reduction of type IIB on $K3$ \cite{Townsend:1983xt}.

\medskip 

In the context of generalised geometry, 
truncations to half-maximal supergravity have been studied in \cite{Malek:2017njj,Malek:2018zcz,Malek:2019ucd}. We review them in the conventions of Exceptional Generalised Geometry. 

The truncation to minimal $\mathcal{N}=(1,1)$ supergravity is associated to the structure 
\begin{equation}
    \Gst  =  \SO{4}  \sim \SU{2}_l  \times \SU{2}_r
\end{equation}
where each of the two $\SU{2}$ factors embeds in one 
$\USp{4}$ as 
$\SU{2}_S \times \SU{2}_R \subset \USp{4}$  (see Table \ref{table:6dtr}).
From the embedding
\begin{equation}
    \begin{aligned}
        \SO{5,5}  & \supset  ( \SU{2}_S \times \SU{2}_R ) \times  ( \SU{2}_S \times \SU{2}_R ) \times \mathbb{R}^+  \\
    \rep{16}&  \to  (\rep{1} , \rep{2}, \rep{1} , \rep{2} )_1 \oplus   (\rep{2} , \rep{1}, \rep{2} , \rep{1} )_1    \oplus (\rep{1} , \rep{1}, \rep{2} , \rep{2} )_{-1} 
    \oplus (\rep{2} , \rep{2}, \rep{1} , \rep{1} )_{-1}  \, ,  \\
    \rep{10} & \to (\rep{1} , \rep{1}, \rep{1} , \rep{1} )_2 \oplus    (\rep{1} , \rep{1}, \rep{1} , \rep{1} )_{-2}   \oplus (\rep{1} , \rep{2}, \rep{2} , \rep{1} )_{0} 
    \oplus (\rep{2} , \rep{1}, \rep{1} , \rep{2} )_{0}  \, ,
    \end{aligned}
\end{equation}
it follows that the
structure is defined by 2 singlets $Q$ and $\hat{Q}$ in the bundle $N$  and 
4 singlet generalised vectors $K_I$, $I=1, \dots, 4$, transforming   in the $\rep{1}$ and $\rep{4}$  of the $\SO{4}_R \sim \SU{2}_R \times \SU{2}_R$ R-symmetry group.

The singlets $K_I$ give the 4 vectors in the gravity multiplets, while $Q$ and $\hat{Q}$ give the two-form in the same multiplet and its dual. 
From  
$\Com{\Gst}{\Spin{5,5}} = \mathbb{R}^+$ and 
\eqref{scalarman}, one recovers the scalar manifold of minimal supergravity
 \begin{equation}
    \mathcal{M} = \mathbb{R}^+ \, . 
\end{equation}

To have extra vector multiplets the structure is further reduced to the diagonal
$\SU{2}$ in  $\Gst = \SO{4}$ and to subgroups thereof. The allowed truncation are listed in Table \ref{table:vecN116d}. Notice that the $n_{\rm V} =3$ case is missing as it enhances to $n_{\rm V} =4$. 
\begingroup
\renewcommand{\arraystretch}{1.5}
\begin{table}[H]
\begin{center}
\begin{tabular}{|c|c|c|c|}
\hline
$n_{\rm V}$ & 1  & 2  & 4   \\
\hline
 $\Gst$   &  $\SU{2}$  &  $\U{1}$   &
$\mathbb{Z}_2$   \\
\hline
\end{tabular}
\end{center}
\caption{Generalised structure groups for truncation with $n_{\rm V} \neq 0$}
\label{table:vecN116d}
\end{table}
\endgroup

The structures are defined by the pair of singlet $Q$ and $\hat{Q}$ in the bundle $N$ and by $4+ n_{\rm V}$ generalised vectors $K_I$ transforming in $(\rep{4+n_{\rm V}})_{-1}$ of the global isometry group ${\rm G}_{\rm iso} = \SO{4, n_{\rm V}} \times \mathbb{R}^+$.

The tensor $Q$, $\hat{Q}$ and $K_I$ defining the generalised structure   satisfy  the algebraic conditions \cite{Malek:2017njj}
\begin{equation}
    \begin{aligned}
      &  \eta (Q,Q) =  \eta (\hat Q, \hat Q)  = 0 \, ,  \\
      &   \eta( Q , \hat{Q})  = \kappa^2  \, , \\
      & K_I  \times_N  K_J  =  \eta_{IJ }  \hat{Q} \, , 
    \end{aligned}
    \qquad I=1, \dots, 4 + n_{\rm V} \, . 
    \label{compcondN116d}
\end{equation} 
where $\eta(\cdot, \cdot)$ is the $\SO{5,5}$ invariant (see \cite{Ashmore:2015joa}) while  $\eta_{IJ }$ is the $\SO{4,n_{\rm V}}$ invariant metric, as well as the differential ones \cite{Malek:2017njj}
\begin{equation}
    \begin{aligned}
        & L_{K_I} K_J = - X_{IJ}{}^K K_K \, ,  \\
& L_{K_I} Q = X_I Q  \, ,  \\
& \partial \times_E  Q = \tilde{X}^I K_I \, ,
 \end{aligned}
\end{equation}
with $X_{IJK} =  X_{IJ}{}^L \eta_{LK} =  X_{[IJK]}$. The singlet components of the intrinsic 
torsion,  $(X_{IJK}, X_I, \tilde{X}_I)$,  transform in the
\begin{equation}
    W_{\rm int} = \rep{X}_{IJK , -1} \oplus  \rep{N}_{-1} \oplus \rep{N}_{3}
\end{equation}
of the global isometry group ${\rm G}_{\rm iso} = \SO{4, n_{\rm V}} \times \mathbb{R}^+$ and give the components of the embedding tensor. With respect to \eqref{embt6n11KM}, there are two indipendent vector components.\footnote{We thank
G. Bossard for confirming this results. See also Section \ref{7dhalfmax}.}

\medskip

Minimal supergravity theory with ${\rm}F(4)$ gauge group was obtained as a consistent truncation of massive type IIA supergravity on (the upper hemisphere of)  $S^4$ in \cite{Cvetic:1999un}, and also 
as a consistent truncation of type IIB on a $S^2$ warped over a Riemann surface in \cite{Hong:2018amk}. The latter truncation extends those in \cite{Jeong:2013jfc}. 

In the context of Exceptional Field Theory the truncation of type IIB on $AdS_6$ times the warped product of $S^2$ and a Riemann surface to minimal ${\rm}F(4)$ supergravity 
 was recovered in \cite{Malek:2018zcz}. In 
\cite{Malek:2019ucd} the truncation was extended to extra vector multiplets. The truncations with $n_{\rm V}=1$ and $n_{\rm V}=2$ vector multiplets correspond to the $\SU{2}$ and $\U{1}$ structures of Table \ref{table:vecN116d}. On the other hand,  the truncations in \cite{Malek:2019ucd} with 3 vector multiplets transforming as triplets of the $\SU{2}_R$ symmetry and with 3 vector multiplets in the $\rep{3}$ and one in the $\rep{1}$ of $\SU{2}_R$ cannot be obtained in our analysis since all the extra vectors are singlets of the $\SU{2}$ R-symmetry.  
It would be interesting to understand what is the origin of this mismatch. 

Truncations to minimal ungauged supergravities are obtained from 11-dimensional supergravity on $K3 \times S^2$ or type IIA on $K3$ \cite{Malek:2017njj}.

\medskip

The chiral $\mathcal{N}=(2,0)$ theories  
correspond to structures that embed uniquely in one of the two $\USp{4}$ factors. The allowed $\Gst$-structures are listed in the table below  
\begingroup
\renewcommand{\arraystretch}{1.5}
\begin{table}[H]
\begin{center}
\begin{tabular}{|c|c|c|c|c|c|}
\hline
$n_{\rm T}$ & 0  & 1  & 2 & 3 & 5  \\
\hline
 $\Gst$  &  $\USp{4}$   &  $\SU{2} \times \SU{2}$  & 
 $\SU{2}_{\rm diag}$ &  $\U{1}$ &  $\mathbb{Z}_2$   \\
\hline
\end{tabular}
\end{center}
\caption{Generalised structure groups for truncation to chiral $\mathcal{N}=(2,0)$ supergravity}
\label{table:ten026d}
\end{table}
\endgroup
Decomposing the bundles $E$ and $N$ under 
\begin{equation}
 \SO{5,5} \supset \Gst \times  \SO{5,n_{\rm T}}
\end{equation}
with ${\rm G}_{\rm iso}= \SO{5,n_{\rm T}}$ the global isometry group, one sees that there are no singlets generalised vectors and $5 + n_{\rm T}$ singlets $Q_i$, $i = 1, \dots 5 + n_{\rm T}$, in $N$.
They transform in the fundamental of ${\rm G}_{\rm iso}$ and satisfy 
\begin{equation}
    \eta(Q_i , Q_j) = \delta_{ij}
\end{equation}
with $\eta(\cdot , \cdot)$ the $\SO{5,5}$ invariant. 

The tensors $Q_i$ with $i = 1, \dots 5$ give the 5 tensors in the gravity multiplet, while each of those with  $i = 6, \dots 5 + n_{\rm T}$ give the tensor in the tensor multiplets. From \eqref{scalarman}, the scalar manifold is 
\begin{equation}
    \mathcal{M} = \frac{\SO{5, n_{\rm T}}}{\SO{5} \times \SO{n_{\rm T}}} \, . 
\end{equation}

It is easy to see that for any $n_T=0, \dots 5$, there are no $\Gst$-singlets in the intrinsic torsion \cite{Malek:2017njj}. 
This implies that only truncations to ungauged supergravity are possible, thus giving only Minkowski vacua. 
We are not aware of explicit examples of such truncations.

\subsubsection{Miminal supergravity: \texorpdfstring{$\mathcal{N} = (1,0)$}{} }

The  generic $\mathcal{N}=(1,0)$ chiral supergravity contains, beside the gravity multiplet, $n_{\rm T} $ tensor, $n_{\rm V}$ vector  and $n_{\rm H}$ hypermultiplets. 
The gravity multiplet consists of the  graviton, a $\USp{2}_R$ doublet of left-handed gravitini and an anti-self-dual two-form.  Each  tensor multiplet is formed by a self-dual two-form,  an 
$\USp{2}_R$ doublet of right-handed spin 1/2 fermions and a scalar. The scalar in the $n_{\rm T}$ tensor multiplets parameterise the coset space 
\begin{equation}
\label{6dN1tm}
    \mathcal{M}_{\rm T} = \frac{\SO{1,n_{\rm T}}}{\SO{n_{\rm T}}}
    \, . 
\end{equation}

Each vector multiplet contains a vector and a $\USp{2}_R$ doublet of left-handed spin 1/2 fermions.
A hypermultiplet consists of a right-handed  spin 1/2 fermion and 4 real scalars. 
The scalars in the $n_{\rm H}$ hypermultiplets parameterise a quaternionic  K\"ahler manifold of negative curvature. 
The full list of allowed  manifolds can be found in \cite{Ale75,deWit:1991nm}. 
In consistent truncation we are only interested homogeneous symmetric spaces 
\footnote{The other homogeneous spaces in in \cite{Nishino:1984gk,Nishino:1986dc,Ale75,deWit:1991nm} are clearly too big to be obtained from $\Ex{5}$ geometry. }
\begin{equation}
\begin{aligned}
    \mathcal{M} & = \frac{\USp{2n_{\rm H},2}}{\USp{2n_{\rm H}} \times \USp{2}} \, , 
\\ 
 \mathcal{M} & = \frac{\SU{n_{\rm H},2}}{\SU{n_{\rm H}} \times \SU{2} \times \U{1}} \, , 
\\ 
  \mathcal{M} & = \frac{\SO{n_{\rm H},4}}{\SO{n_{\rm H}} \times \SO{4} } \, . 
  \end{aligned}
\end{equation}

For theories with $n_{\rm T} >1$
tensor multiplets, the (anti-)self-duality condition of the tensors does not allow for a Lagrangian formulation, but the equations of motion, supersymmetry variations and pseudo-Lagrangian were derived in \cite{Romans:1986er,Nishino:1997ff,Ferrara:1997gh,Riccioni:2001bg}.
A standard 
Lagrangian formulation is possible for $n_{\rm T} = 1$ since the self-dual two-form in the tensor multiplet  combines with the anti-self-dual two-form in the gravity multiplet to give  a two-form with no self-duality property.

While the gauging of the isometries of the hypermultiplet sector takes the ordinary form 
\begin{equation}
   \mathcal{D}_\mu \phi^x = \partial_\mu \phi^x  + i g k^x_{I} A^{I}_\mu  \, . 
\end{equation}
where $\phi^x$ are the scalars in the hypermultiplets and  $k^x_{I}$ are Killing vector fields, the self-duality of the two-forms in the tensor multiplets  makes the gauging of the isometries of the tensor scalar manifold more involved. 

The full gaugings were studied in \cite{Gunaydin:2010fi} in the case of magical supergravities\footnote{Magical supergravities are a class of theories characterised by a fixed number of tensor and vector multiplets and arbitrary number of hypermultiplets. In six dimensions magical supergravities exist for 
$(n_{\rm T} ,  n_{\rm V} ) = (2,2), (3,4), (5,8), (9,16)$.}  in terms of an embedding tensor that takes the form\footnote{The embedding tensor could a priori contain an extra term gauging the symmmetries of the theory that only act on the vectors. However consistency of the gauging sets it to zero. See \cite{Gunaydin:2010fi} for more details} 
\begin{equation}
    \Theta_I{}^\alpha = (\Theta_I^{ij} , \Theta_I{}^{AB} ) 
    \label{etN1d6}
\end{equation}
where $\Theta_I^{ij} = - \gamma_{I J}^{[i} \theta^{j]J}$ and $\gamma_{IJ}^{i} \theta_i^J = 0$, with $\gamma_{IJ}^i$ the $\SO{5,5}$ gamma metrices. 
The first term in \eqref{etN1d6} determines the gaugings of the tensor multiplet isometry group and the second one those of the hypermultiplet scalars. 

The construction via the embedding tensor shows that the non-abelian gauge algebra only closes for the values $n_{\rm T}= 2, 3, 5, 9$ \cite{Gunaydin:2010fi}. In all other cases the non-abelian gauging of the tensors isometry groups are not known.

For a generic number of tensor, vector and hypermultiplet the $\mathcal{N}=(1,0)$ suffers of  gravitational, gauge and mixed anomalies \cite{Townsend:1983xt} (see \cite{Randjbar-Daemi:1985tdc,Avramis:2005qt, Avramis:2005hc,Suzuki:2005vu} for anomaly cancellation in gauged supergravity and \cite{Kumar:2009ae,Taylor:2010wm,Kumar:2010am, Bonetti:2011mw, Kim:2024eoa} for more recent references). 
The Green-Schwarz mechanism for anomaly cancellation constrains the number of multiplet to satisfy \cite{Randjbar-Daemi:1985tdc} 
\begin{equation}
    n_{\rm H} - n_{\rm V} + 29 n_{\rm T} = 273 \, .
\end{equation}

\medskip

It is easy to see that  none of the theories that can be obtained as consistent truncations  satisfy the anomaly cancellation condition. 
However, for completeness, we still present the results of our classification.

In Exceptional Geometry, minimal $\mathcal{N}=(1,0)$ supergravity 
corresponds to the generalised structures 
\begin{equation}
    \Gst \subseteq \USp{2} \times \USp{4} \, . 
\end{equation}

The structures are defined by the a set of singlets in the bundles $N$, $E$, $ (\det T^* M)^{1/2} {\rm ad}F$, respectively 
\begin{equation}
    \{Q_i,  K_I, J_A \} \,  .
\end{equation}

In the truncated theory the singlets $Q_i$,  
with $i=0, \ldots, n_{\rm T}$, determine the two-forms in the gravity and the tensor multiplets, $K_I$  the vectors ($I=1, \ldots, n_{\rm V}$), while the scalars in the hypermultiplets are associated to $J_A$, with  $A= 1, \ldots, {\rm dim} G_{\rm H}$, where $G_{\rm H}$ is the isometry group of the hyperscalars.
As we will show below, the embedding of the $\Gst$-structures in $\SO{5,5}$ puts bounds on the allowed number of multiplets.

Truncations to the gravity multiplet only are associated to 
$\Gst=  \USp{2} \times \USp{4}$.  As discussed in \cite{Ashmore:2015joa}, the structure is defined by 
a single tensor $Q$ and 
a triplet  $J_\alpha$, $\alpha =1,2,3$, defining a highest weight  $\mathfrak{su}(2)$ subalgebra of  $\SO{5,5}$. The tensors $Q$ and $J_\alpha$ satisfy
\begin{equation}
    \begin{aligned}
       &  \eta(Q,Q) = \kappa^2 \, ,  \\ 
       & J_\alpha \cdot Q = 0  \, , \\
       & \tr(J_\alpha, J_\beta) = - \eta(Q,Q) \delta_{\alpha \beta} \,, 
    \end{aligned}
    \label{acN1d6}
\end{equation}
where $\eta(\cdot, \cdot)$ is the $\SO{5,5}$ invariant and $\cdot$ denotes the adjoint action. 
The triplet $J_\alpha$ are the generators of the 
$\USp{2}$ R-symmetry. 
Decomposing the intrinsic torsion under $\SU{2}_R \times \USp{2} \times   \USp{4}$ \cite{Ashmore:2015joa}
\begin{equation}
W_{\rm int} = (\rep{1}, \rep{2}, \rep{4})\oplus  (\rep{2}, \rep{1}, \rep{4})\oplus  (\rep{2}, \rep{1}, \rep{4}) \oplus  (\rep{3}, \rep{2}, \rep{4})
\end{equation}
one sees that there are no $\Gst$-singlets. This implies 
\begin{equation}
    \partial \times_E Q = 0 \, , 
\end{equation}
so that  only truncations to
ungauged supergravity are possible.

For truncations with tensor, vector and hypermultiplets the embedding of the structures in $\SO{5,5}$ puts bounds on the allowed number of multiplets. 
In order to have truncations with only tensor multiplets, the $\Gst$-structure must be a subgroup of $\SU{2} \times \Spin{1,5}$, the stabiliser of the triplet $J_\alpha$. 
Decomposing the bundles $N$ under 
$\SO{5,5}  \supset \SU{2}_R \times \SU{2} \times \Spin{1,5}$
\begin{equation}
\rep{10}  = (\rep{1},\rep{1},\rep{6}) \oplus (\rep{2},\rep{2},\rep{1}) \, ,
\end{equation}
and using the fact that the singlet tensors must be invariant under the R-symmetry,  we see that the maximum number of allowed  singlets is 6, which gives  $n_{\rm T} =5$ tensor multiplets. 

Similarly, for truncations with only hypermultiplets 
the structures must be subgroups of the stabiliser of the singlet tensor $Q$, namely $\Gst \subset \Spin{4,5}$. Recall 
that the hypermultiplets scalars are associated to the $\Gst$-singlets among the non-compact element of $\SO{5,5}$ which transform non-trivially under the $R$-symmetry. Thus in the 
decomposition of the 
adjoint bundle under 
$\SO{5,5}  \supset \SO{5} \times \SO{4} \subset  \Spin{4,5}$ the
relevant elements must be in the representation 
\begin{equation}
\rep{45} \supset  (\rep{5}, \rep{2}, \rep{2}) \, ,
   \end{equation}
which sets the bound $n_{\rm H}\leq 5$ for the number of hypermultiplets. 
The number of singlets among the generalised vectors follows from the specific $\Gst$ group, but cannot clearly exceed $n_{\rm V}$, the dimension of the generalised tangent bundle. 

As a consequence, none of the theories we find as consistent truncations satisfy the anomaly cancellation condition.

 We find only a few examples of truncations to gauged supergravity. The theories we find contain tensors and vector multiplets, but  no hypermultiplets. 

The truncations and corresponding $\Gst$-structures are listed in Table \ref{ntnv6dn1}. 
\begingroup
\renewcommand{\arraystretch}{1.5}
\begin{table}[H]
\begin{center}
\begin{tabular}{|c|c|c|c|c|c|}
\hline
$\Gst$ &  $n_{\rm T}$ & $n_{\rm V}$   &  
$G_{\rm sym}$ &  $\mathcal{R}_{\rm T}$ 
&  $\mathcal{R}_{\rm V}$  \\
\hline
$\SU{2} \times  \SU{2}_{\rm diag}$   &  1   & 1  & $\SO{1,1_{\rm T}}$  & $\rep{2}$ &  -1    \\
\hline
$\SU{2} \times  \U{1}$  &  1   & 2  & $\SO{1,1_{\rm T}} \times \U{1}$  & $\rep{2}$ &   $2 (-1)$  \\
\hline
$\SU{2}$   &  2    & 2  & $\SO{1,2_{\rm T}}$ & $\rep{3}$  &  $\rep{2}_{\rm real}$  \\
\hline
$\U{1}$   &  3    & 4  & $\SO{1,3_{\rm T}} \times \U{1}$   & $ (\rep{2},\rep{2})_0$  &  $(\rep{2},\rep{1})_{2} \oplus (\rep{1},\rep{2})_{-2}  $ \\
\hline
$\mathbb{Z}_2$   &  5    & 8  &$\SO{1,5_{\rm T}} \times \SU{2}$  &  $(\rep{6}, \rep{1}) $ &  $ (\rep{\bar 4} , \rep{2})  $ \\
\hline
\end{tabular}             
\end{center}
\caption{Truncations with tensor and vector multiplets. $\mathcal{R}_{\rm T}$  and  $\mathcal{R}_{\rm V}$ denote the representation of $G_{\rm sym}$ in which tensors and vectors transform. }
\label{ntnv6dn1}
\end{table}
\endgroup

The case $n_{\rm T} = n_{\rm V} =1$ reproduces 
the  field content of  the Salam-Sezgin model \cite{Salam:1984cj}. 

The last three entries in Table \ref{ntnv6dn1}  correspond to six-dimensional magical supergravities \cite{Gunaydin:1983rk,Gunaydin:1983bi,Gunaydin:2010fi}
with no hypermultiplets.\footnote{The magical supergravity with $n_{\rm T}=9$ and $n_{\rm V}=16$ cannot be obtained as a truncation since its symmetry group $\SO{1,9}$ does not embeds in $\SO{5,5}$.} In these theories the vectors carry spinorial representations of the tensor global isometry group and, when extra symmetry groups not acting  on the tensors are present, they are also charged under these latters. 

The property of having vectors fields 
with "spinorial" charges under the isometry group of the tensors, is also common to the theories in the first two lines in Table \ref{ntnv6dn1}.

The $\Gst$-structures of  Table \ref{ntnv6dn1} are 
defined by the singlets $\{Q_i,  K_I, J_\alpha \} $ satisfying a generalisation of the algebraic conditions \eqref{acN1d6} for any $i,j = 0 \dots , n_{\rm T}$, $I=1, \dots, n_{\rm V}$ and $\alpha = 1,2,3$. The singlets $J_\alpha$ in the adjoint generate the $\SU{2}_R$ symmetry 
\begin{equation}
[J_\alpha, J_\beta] = \epsilon_{\alpha \beta}{}^\gamma J_\gamma
\end{equation}
and are normalised to 
$\tr(J_\alpha, J_\beta) = \delta_{\alpha \beta}$.  Altogether the singlets
 $\{Q_i,  K_I, J_\alpha \} $ satisfy the compatibility conditions 
 \begin{equation}
    \begin{aligned}
       &  \eta(Q_i, Q_j) = \eta_{ij} \kappa^2 \\ 
       & (Q_i \times_\ad Q_j ) \cdot K_I= - \kappa^2 \tfrac14 (\gamma_{ij})^J{}_I   K_J \\
       & J_A \cdot Q_i = 0 
    \end{aligned}
 \label{N16dtcond}
\end{equation}
where  $\eta_{ij}$ is the invariant $\SO{1,n_{\rm T}}$ metric, $\tfrac14 (\gamma_{ij})^J{}_I$ are the generators of $\SO{1,n_{\rm T}}$, as well as the 
 differential conditions 
\begin{equation}
    \begin{aligned}
      \partial \times_E Q_i & = X_i^I K_I \, ,  \\  
      L_{K_I} Q_i & = X_{I i}{}^j  Q_j \, 
       \end{aligned}
       \qquad 
       \begin{aligned}
      L_{K_I} J_\alpha &  = p_{I\alpha}{}^\beta J_\beta  \, ,\\ 
      L_{K_I} K_J &= X_{IJ}{}^K K_K  \, , 
    \end{aligned}
 \label{N16it}
\end{equation}
with $X_{Ii}{}^j\propto X^J_i (\gamma^j)_{IJ} - X^{jJ} (\gamma_{i})_{IJ}$.
The tensors $X^i_J$, $p_{I \alpha}{}^\beta$ and $X_{IJ}{}^K$
are the component of the intrinsic torsion. 

In all the examples in Table \ref{ntnv6dn1} the component 
 $X_{IJ}{}^K$ vanishes. Morevover, since 
the vectors are in spinorial representations of the tensor isometry group,  $X_i^I$ transforms in a tensorial representation of $\SO{1, n_{\rm T}}$ of dimension $(1+n_{\rm T}) n_{\rm V}$, which has two irreducible components, a trace part and a traceless one
\begin{equation}
    X_I^{(tr)} =  (\gamma^i)_{IJ} X_i^J 
\qquad 
    X_i^{(0)I} = X_i^I + \frac{1}{4} \gamma^{IJ} X_J \, . 
\end{equation}

This is easily seen by looking at the $\Gst$-singlets in the intrinsic torsion in the table below, where, in each line,
the first element corresponds to $X_i^{(0)I}$, the second to $X_I^{(tr)}$ and the last one to $p_{I \alpha}{}^\beta$.

The requirement that the differential conditions \eqref{N16it} give a Leibniz algebra  sets to zero the trace component of the singlet intrinsic torsion.

Recall that the intrinsic torsion determines 
the embedding tensor of the truncated theory, where 
$X_{i}^I, X_{Ii}{}^j$ and $p_{I\alpha}{}^\beta$  provide the gauging of the tensor isometries and R-symmetry, while $X_{IJ}{}^K$  determine the structure constants of the tensor scalar isometry group and of possible extra symmetries that doe not act on the scalars. 
Since  the representation in which $X_{IJ}{}^K$ transforms never appears in the intrinsic torsion of Table \ref{ntnv6dn1t},   we find that only abelian gaugings 
of the tensor isometries and of the R-symmetry are possible. Thus we reproduce exactly  the  embedding tensor derived in \cite{Gunaydin:2010fi} for magical supergravities.

\begingroup
\renewcommand{\arraystretch}{1.5}
\begin{table}[H]
\begin{center}
\begin{tabular}{|c|c|c|c|c|}
\hline
 $n_{\rm T}$ & $n_{\rm V}$   &  
$G_{\rm sym}$ &  $W_{\rm int}$   \\
\hline
 1   & 1  & $\SO{1,1_{\rm T}}$  &  $\rep{1}_3  \oplus  \rep{1}_{-1} \oplus \rep{3}_{-1} $  \\
\hline
  1   & 2  & $\SO{1,1_{\rm T}} \times \U{1}$  & 
$ \rep{1}_{3, \pm 2} \oplus \rep{1}_{-1 , \pm 2} \oplus 
\rep{3}_{-1, \pm2  } $
\\
\hline
  2    & 2  & $\SO{1,2_{\rm T}}$ & $ (\rep{4}, \rep{1}) \oplus  (\rep{2}, \rep{1}) \oplus (\rep{2}, \rep{3})  $  \\
\hline
  3    & 4  & $\SO{1,3_{\rm T}} \times \U{1}$   & 
$ \big((\rep{3}, \rep{2} )_2 \oplus (\rep{2}, \rep{3} )_{-2} ,\rep{1}\big)
\oplus  \big(( \rep{1}, \rep{2} )_{-2} \oplus  (\rep{2}, \rep{1} )_{2},
  \rep{1} \big) 
 $
\\
& & & $\oplus \big(( \rep{1}, \rep{2} )_2 \oplus ( \rep{2}, \rep{1} )_{-2}, \rep{3} \big)$\\
\hline
  5    & 8  &$\SO{1,5_{\rm T}} \times \SU{2}$  & 
$ (\rep{20} , \rep{2}, \rep{1} ) \oplus (\rep{4} , \rep{2}, \rep{1} ) \oplus (\rep{\bar{4}} , \rep{2}, \rep{3} ) $
\\
\hline
\end{tabular}             
\end{center}
\caption{Singlet intrinsic torsion in representations of $ G_{\rm sym} \times \SU{2}_R$.}
\label{ntnv6dn1t}
\end{table}
\endgroup

\medskip

The other truncations we find give  ungauged supergravity. 

We find a family of truncations with only tensor multiplets, associated  to the structures 
\begingroup
\renewcommand{\arraystretch}{1.5}
\begin{table}[H]
\begin{center}
\begin{tabular}{|c|c|c|c|c|c|}
\hline
$n_{\rm T}$ & 1 & 2 & 3 & 4& 5 \\ 
\hline 
$\Gst$ &  $\SU{2} \times \SU{2}\times \SU{2}$ & 
$\SU{2}_{\rm diag} \times \SU{2} $ & 
$\SU{2} \times \U{1}_{\rm diag} $ & $\U{1}$ &  $\mathbb{Z}_2$ \\
\hline
\end{tabular}
\end{center}
\caption{Generalised structure groups for truncation with only tensor multiplets}
\label{6dN1onlyt}
\end{table}
\endgroup

The commutant of $\Gst$ in $\SO{5,5}$ gives the global isometry  group of the tensor multiplet scalars
\begin{equation}
    G_{\rm iso} = \SO{1, n_{\rm T}} = \Com{\Gst}{\SO{5,5}}\, ,
\end{equation}
and, from \eqref{scalarman}, we recover the  scalar manifold \eqref{6dN1tm}.

Since for any value of $n_{\rm T} \leq 5$ there are no $\Gst$-singlets in 
intrinsic torsion, all the truncations in this family give  ungauged supergravity. The only differential constraint is 
again $\partial \times_E Q_i =0$, with $i=0, \ldots, n_{\rm T}$.

\medskip

Finally we find two truncations 
with $n_{\rm T} =3$ tensor multiplets and $n_{\rm H} =1,2$ hypermutliplets, corresponding to the $\Gst= \U{1}$ and $\Gst= \mathbb{Z}_4$, respectively. The scalar of the truncated theories are 
\begin{equation}
  \mathcal{M} =   \frac{\SO{1,3}}{\SO{3}} \times \frac{\SU{n_{\rm H}, 2}}{\SU{n_{\rm H}}\times\SU{2} \times \U{1}}  \qquad n_{\rm H} = 1,2
\end{equation}
and, again, there are no singlets in the intrinsic torsion.

\subsection{Truncations to 7
dimensions}

Consistent truncations to 7 dimensional supergravity have been discussed in the context of Exceptional Generalised Geometry or Exceptional Field theory in \cite{Lee:2014mla,Malek:2016bpu,Cassani:2016ncu, Malek:2018zcz, Malek:2019ucd}. 
The relevant generalised geometry is 
$\Ex{4} =\SL{5,\mathbb{R}}$. The generalised tangent bundle transforms in the $\rep{10}$ of $\SL{5,\mathbb{R}}$, while the supersymmetry parameters transform as spinors of $\USp{4}$, the maximally compact subgroup of $\SL{5,\mathbb{R}}$ (see Table \ref{table:higherED}).
The allowed amount of superymmetry gives maximal and half-maximal supergravities. 

\subsubsection{Maximal supergravity}

Maximal supergravity in 7 dimensions has $\mathcal{N}=4$ supercharges, transforming  in the fundamental of the $\USp{4}$ R-symmetry group. The fields are organised in the gravity multiplet, consisting of the 
graviton, 4 gravitini, 10 vectors, 5 two-forms, 16 spin 1/2 fermions and 14 scalars.
The bosonic fields carry non trivial representations of the global symmetry group $\SL{5,\mathbb{R}}$ group, while the fermions are symplectic Majorana and transform under 
$\USp{4}$ \cite{Sezgin:1982gi,Samtleben:2005bp}. 
The 14 scalars parameterise the coset  
\begin{equation}
    \mathcal{M} = \frac{\SL{5, \mathbb{R}}}{\SO{5}} \, . 
\end{equation}
The gaugings of the global $\SL{5, \mathbb{R}}$ symmetry are given by 
\begin{equation}
    D_\mu =\nabla_\mu - g A_\mu^{[ij]} \Theta_{[ij], k}^l t^k{}_l \,  ,
\end{equation}
where  $i,j,k,l = 1, \dots, 5$ are $\SL{5, \bbR}$ indices, 
$A_\mu^{[ij]}$  are the 10 vectors and 
$t^k{}_l$ (with $t^k{}_k =0$) are the $\SL{5, \mathbb{R}}$ generators.
The embedding tensor  has two components \cite{Samtleben:2005bp}
\begin{equation}
\label{d7n4et}
    \Theta_{I}{}^{\alpha} = (Y_{(ij)}, Z^{[ij],k} ) \qquad \quad Z^{[ij,k]} =0
\end{equation}
transforming in the $\rep{15}$ and $\rep{40^\prime}$ of $\SL{5, \mathbb{R}}$.

\medskip

In generalised geometry, truncations to maximal supergravity correspond to a $\Gst = \id$ structure  defined by 10 generalised vectors $\{K_I\} = \{K_{[ij]}\}$  transforming in the $\rep{10}$ of $\SL{5, \mathbb{R}}$ ($I=1,\dots, 10$ and $i,j =1, \dots, 5$).  They realise a Leibniz parallelisation 
\begin{equation}
\label{genL7}
    L_{K_I} K_J = X_{IJ}{}^K K_K \,\qquad I,J,K = 1, \ldots, 10
\end{equation}
and are normalised to $G(K_I, K_J) = \delta_{IJ}$,  with $G$ the generalised metric.
They give the 10 vectors of the truncated theory. 

As in the six-dimensional case, the generalised singlet vectors also define a parallelisation of the bundle $N$ via the projection $E \times_N E$. The 5 tensors in $N$ give the 5 two-forms of the truncated theory. 
The scalars are again given by \eqref{scalarman}. 

The generalised Lie derivative among the  vectors $K_I$ \eqref{genL7}
determines the intrinsic torsion $X_{IJ}{}^K$, which transforms as the 
\begin{equation}
    W_{\rm int} = \rep{40^\prime} \oplus \rep{15}
\end{equation}
of the $\SL{5, \mathbb{R}}$ global isometry group.  $X_{IJ}{}^K$  reproduces the embedding tensor \eqref{d7n4et} and determines the gaugings of the truncated theory.

\medskip

Examples of truncations to maximal supegravity are the truncation of 11-dimensionsal supergravity on $S^4$ with gauge group $\SO{5}$ \cite{Nastase:1999kf, Nastase:1999cb} and 
of massless IIA theory on $S^3$ with gauge group $\rm{ISO}(4)$
\cite{Cvetic:2000ah}. 
In generalised geometry the truncations have been reproduced in \cite{Cassani:2016ncu}.
In both cases the intrinsic torsion is only in the $\rep{15}$ component \cite{Lee:2014mla}
\begin{equation}
    X_{[i i^\prime] , [j j^\prime]}{}^{[k k^\prime]} \sim - R^{-1}  (Y_{ij}  \delta_{i^\prime j^\prime}^{[k k^\prime]} - Y_{i j^\prime}  \delta_{i^\prime j}^{[k k^\prime]} - Y_{i^\prime j}  \delta_{i j^\prime}^{[k k^\prime]} + Y_{i^\prime j^\prime}  \delta_{i j}^{[k k^\prime]} ) \,  ,
\end{equation}
where $R$ is the radius of the internal manifold and $i,j = 1, \ldots, 5$ are again $\SL{5, \bbR}$ indices. 
For the M-theory truncation  on $S^4$  \eqref{genL7} reproduces the algebra of the $\SO{5}$ with 
\begin{equation}
    Y_{i i^\prime} \sim {\rm diag}(1,1,1,1,1) \, , 
\end{equation}
while for the truncation of type IIA on $S^3$  it  gives the algebra of the ${\rm ISO}(4)$
\begin{equation}
    Y_{i i^\prime} \sim  {\rm diag}(1,1,1,1,0) \, . 
\end{equation}

\subsubsection{Half-maximal supergravity}
\label{7dhalfmax}

Half maximal supergravity has $\mathcal{N}=2$ supersymmetry with $\SU{2}$ R-symmetry. Its  field content consists of  the gravity multiplet (the graviton, 2 gravitini, 3 vectors, 2 spin 1/2 fermions, a two-form and a scalar) and $n_{\rm V}$  vector multiplets, each containing a vector, 2 spin 1/2 fermions and 3 scalars \cite{Bergshoeff:2005pq}. The fermions are all symplectic Majorana and the scalars in the vector multiplets parameterise the coset
 \begin{equation}
 \label{7dhmaxsc}
     \mathcal{M} = \frac{\SO{3,n_{\rm V}}}{\SO{3}\times \SO{n_{\rm V}}}  \times \mathbb{R}^+\, ,
 \end{equation}
 where the $\mathbb{R}^+$ factor is parameterised by the scalar in the gravity multiplet. 

The full embedding tensor formalism for half-maximal supergravity in 7 dimensions has not been worked out yet. Using the results from Kac-Moody analysis \cite{Bergshoeff:2007vb}, two components of the embedding tensor
\begin{equation}
\label{embdthalf7d}
    \Theta_I{}^{\alpha} = (f_I{}^{JK} + \delta_I^{[J} \xi^{K]} , \xi_I) \, . 
\end{equation}
have been studied in \cite{Louis:2015mka}. The tensors $\xi^I$ and  $f_I{}^{JK}$  transform as the fundamental and the three-index anti-symmetric representations of $\SO{3,n_{\rm V}}$. They  give the 
gaugings
\begin{equation}
    D_\mu = \nabla_\mu - A^I_\mu( f_I{}^{JK} t_{JK} + \xi^J t_{IJ} + \xi_I t_0) \, ,
\end{equation}
where $t_{IJ}$ and $t_0$ are the generators of $\SO{3,n_{\rm V}}$ and of the $\mathbb{R}^+$ shifts.

\medskip

In generalised geometry, truncations to half-maximal supegravity were classified in \cite{Malek:2016bpu, Malek:2018zcz, Malek:2019ucd}. They correspond to the generalised structures 
\begin{equation}
   \Gst = \SO{3 - n_{\rm V}} \subset \USp{4} \qquad n_{\rm V} = 0,1,2,3
\end{equation}
with $\SO{0} =\mathbb{Z}_2$. The truncation with  $n_{\rm V}=2$ vector multiplets automatically enhances to the one with  $n_{\rm V}=3$. 

The $\Gst$-structures  are defined by $3+n_{\rm V}$ singlet generalised vectors $K_I$  and a singlet element  $Q$ of the bundle $N$ (see Table \ref{table:higherED} again) satisfying the compatibility conditions
\begin{equation}
    \begin{aligned}
& K_I  \times_N  K_J  - \frac{1}{4}  \eta_{IJ }  K_K  \times_N  K^K = 0 \, ,  \\
 &  \epsilon (K_I ,  K_J , Q) = \eta_{IJ}  \kappa^2 \, ,
 \end{aligned} \qquad  I,J,K =1, \ldots, 3 + n_{\rm V} \, ,
\end{equation}
where  $\epsilon(\cdot , \cdot, \cdot)$ is the $\SL{5} $ invariant tensor and $\eta_{IJ}$ is the $\SO{3,n_{\rm V}}$ invariant metric. The generalised vectors $K_I$ with $I=1,2,3$ and the tensor $Q$ give the 3 vectors and the one-form of the gravity multiplets, while the remaining vectors are associated to the vector multiplets. From  \eqref{scalarman}
one reproduces the scalar manifold \eqref{7dhmaxsc}. 

The differential conditions 
\begin{equation}
\begin{aligned}
&  L_{K_I} K_J =  X_{IJ}{}^K K_K \, ,  \\
&  L_{K_I} Q  =  \xi_I Q  \, , \\
& \dd Q =  \xi_0 (J_I \times_N J^I)  \, , 
  \end{aligned}
\end{equation}
with $X_{IJK} = X_{[IJK]}$ give the  singlet components of the intrinsic torsion 
\begin{equation}
    W_{\rm int} = \rep{X}_{IJK} \oplus \rep{N} \oplus \rep{1}
\end{equation}
transforming in the three-index antisymmetric, fundamental and singlet representation of the global isometry group $\SO{3, n_{\rm V}}$. With respect to \eqref{embdthalf7d} we find an extra singlet component, whose presence is also confirmed by a  Kac-Moody analysis.\footnote{We thank
G. Bossard for confirming this results.}. The components of $W_{\rm int}$ should then give the full content of the  embedding tensor of the truncated theory. 

\medskip

Minimal half-maximal supergravity with gauge group $\SU{2}$  \cite{Townsend:1983kk} was obtained in \cite{Lu:1999bc} by truncating 11-dimensional supergravity on $S^4$. 
The same theory can be obtained as a truncation of massive IIA on a deformation of $S^3$ \cite{Passias:2015gya}.

A complete classification of half-maximal truncations of massless and massive type IIA on $AdS_7 \times M_3$, where $M_3$ is an $S^2$ fibration over a segment were studied in 
\cite{Malek:2018zcz,Malek:2019ucd}. 
For massive IIA only the truncation to pure supergravity with 
$\Gst= \SO{3}$  is allowed and it corresponds to the truncation in  \cite{Passias:2015gya}.   In massless IIA, there exists a truncation with $\U{1}$ structure on $S^3$ giving  gauged supergravity with one-vector mutltiplet and gauge group $\SU{2} \times  \U{1}$. The same truncation can be obtained by dimensionally reducing and  keeping only the $\U{1}$ invariant modes of the 
maximally supersymmetric consistent truncation of 11-dimensional supergravity on $S^4$. 

The truncation with $n_{\rm V}=3$ and $\Gst=\mathbb{Z}_2$ corresponds to the theory in \cite{Salam:1983fa} and it can be obtained by further imposing a $\mathbb{Z}_2$
structure to the truncation of 11-dimensional on $S^4$ to maximal supergravity.

\section{Conclusions}
\label{sec:Conclusions}

In this article we pursued the programme of classifying  lower dimensional supergravities that can be obtained as consistent truncations of 11/10-dimensional supergravity, using  the formalism of Exceptional Geometry.
A consistent truncation is determined by its field content and gauge symmetries. 
In  Exceptional Geometry these properties are captured by an exceptional $\Gst$-structure with singlet, constant intrinsic torsion. 
The field content of the reduced theory, as well as its supersymmetry and bosonic symmetries are given by globally defined $\Gst$-invariant generalised tensors on the compactification manifold $M$.

\medskip

Our main result is the classification of the truncations to 4 dimensional supergravity. In this case, 
the exceptional structure group is $\Ex{7}$, and, since we want supersymmetric truncations, the 
possible $\Gst$-structures are subgroups of $\SU{8}$, the double cover of the maximal compact subgroup of $\Ex{7}$.

In the formalism of $\Gst$-structures the derivation of a consistent truncation consists of an algebraic problem and a differential one. 
In this paper we focused on the algebraic question, namely 
the classification  of  the subgroups $\Gst \subset \SU{8}$ and the derivation of the field content and symmetries of the truncated theory in terms of   $\Gst$  singlets.
We do not address the algebraic problem of checking that 
the intrinsic torsion of the $\Gst$-structure consists of $\Gst$-singlets only (or is zero). We simply assume that this  is the case. 

We first scan through  supersymmetry
and determine the largest $\Gst \subset \SU{8}$ compatible with the fixed the number  $2 \leq  \mathcal{N} \leq 8$ of supercharges. 
For $\mathcal{N} \geq 5$  the 4-dimensional theory is unique, due to the
large amount of supersymmetry. For $2 \leq \mathcal{N} \leq 4$ there is a $\Gst^{max} \subset \SU{8}$ structure group, corresponding to the truncation to minimal supergravity. 
Then for any fixed $2 \leq \mathcal{N} \leq 4$, we scanned for all continous $\Gst \subset \Gst^{max}$ leading to inequivalent truncations
allowing for extra matter multiplets. 
We find a limited number of possible truncations, which are listed in \eqref{N4-sm} and Tables \ref{N=3tr} - \ref{MainProcedure_Table-nv-nh}.  For any $2 \leq \mathcal{N} \leq 4$  there is a  truncation with maximal number of matter multiplets, which  corresponds to a discrete structure group. 
In Tables \ref{N=2TruncationsFromPureDiscreteStructures} and \ref{MainProcedure_TableN=2FromN=4} we list other examples of truncations 
 associated to discrete $\Gst$-structure groups. However, in this case our analysis is far from being complete and what we give are just  few examples. 

Even if our analysis is performed looking explicitly at the various embedding of the structure groups $\Gst$ into $\SU{8}$ and $\Ex{7}$, there are some general features that emerge. 
In particular we can use group theory to exclude some of the a priori allowed $\Gst$-structures. 
There are two types of embeddings of  $\Gst$ into $\SU{8}$, what we call regular embeddings, where the number of Cartan generators is preserved, and special ones, when the number is not preserved. 
We find that Schur's lemma excludes a good deal of the special branchings. It would be interesting to see if more rigourous group theoretical arguments can be used to justify this result.

A general prediction of generalised geometry is that  the scalar manifolds of the truncated theories must all be homogeneous and symmetric. In \cite{Koerber:2008rx} it was shown that coset manifolds $\frac{G}{H}$,  where $H$ does not contain non-trivial $G$-invariant subgroups, have $H$ as tructure group. 
Our analysis suggest that this result extends to generalised geometry: 
all coset manifold compactifications $\frac{G}{H}$ correspond to a generalised $H$-structure. 

\medskip

The same algebraic approach can be applied to truncations to any external dimension $D \geq 4$.  These supergravity theories play an important role in the gauge/gravity duality.  

For truncations to 5, 6 and 7 dimensions most of the results have already been obtained in the literature (see for instance \cite{Cassani:2019vcl, Lee:2014mla,Hohm:2014qga,Baguet:2015sma,Ciceri:2016dmd,Cassani:2016ncu,Malek:2016bpu,Malek:2017njj,Malek:2018zcz,Malek:2019ucd,Cassani:2020cod,Josse:2021put}) in Exceptional Generalised Geometry and/or Exceptional Field Theory. Section \ref{sec:higherD} is devoted to a review of these results, with  the goal of  presenting them in a uniform language. Along the way, we completed them with  some missing details.  In particular we derived all components of the embedding tensors for half-maximal supergravities in 6 and 7 dimensions and checked that the extra terms we find with respect to the  literature are indeed predicted by the Kac-Moody analysis of the reduced theory. We also complete the analysis of the allowed truncations to 6 dimensions with different supersymmetry and matter content. For the chiral theories  $\mathcal{N}=(2,1)$ and $\mathcal{N}=(1,0)$ we only find anomalous theories, since the limited number of allowed matter fields does not fulfill the anomaly cancellation conditions.

\medskip

With this paper we complete the classification of all supergravity theories that have an 11/10-dimensional origin for any amount  of supersymmetry in dimension larger than 4, and with $\mathcal{N} \geq 2$ in dimension 4. 
It is straightforward to perform the same analysis for $\calN=1$ truncations in 4 dimensions. However, in this case we find a large variety  of allowed theories and we did not find a nice and interesting way of presenting them. We leave this to a further publication.
The large structure groups of the generalised tangent bundle makes it possible to construct also truncations with no supersymmetry. This is an  interesting direction to explore in the future.

\medskip

A similar classification for truncations to 3 dimensions requires 
 $\Ex{8}$ generalised geometry. Its structure is more involved than those described in this article as it  requires additional 
covariantly-constrained fields to close the algebra of the generalised Lie derivative, which makes it harder to introduce the notion of $\Gst$-structures.  These issues were resolved in \cite{Galli:2022idq} where a classification of consistent truncations with $\mathcal{N} \geq 4$ was given.  It would be interesting to extend this analysis to consistent truncations with less supersymmetry, which might be relevant for the study of truncations around GK geometries \cite{Gauntlett:2007ts}, for instance.

\medskip 

Our classifications only provides a list of 4-dimensional theories that can a priori be obtained as consistent truncations.  In some cases the explicit truncations have been worked out in the  literature (we mention the examples known to us in the various sections), for others an explicit higher dimensional realisation is still missing.
This would imply finding manifolds with the right differential properties to give the required singlet intrinsic torsion. 
For maximal supersymmetry, it is possible to provide necessary and sufficient conditions  that  an embedding tensor has to fullfil to give rise to a generalised Leibniz parallelisation and hence to be associated to a consisent truncations 
 \cite{Inverso:2017lrz, Bugden:2021wxg, Bugden:2021nwl}. It would be interesting to see whether similar conditions can be found also for less supersymmetric truncations. These would allow to 
 identify lower-dimensional theories with
interesting features for the construction of black-holes or solutions relevant for the AdS/CFT correspondence, and then uplift them to full 11/10-dimensional solutions.

\medskip

Finally, it would be interesting to explore  the space of vacua, $AdS$ in particular, of the theories we find in our analysis. 
In 6 and 7 dimensions a complete classfication of the possible $AdS$ vacua is known \cite{Apruzzi:2013yva,Apruzzi:2014qva}. $AdS$ vacua for the $\mathcal{N}=2$ 5-dimensional supergravities found in \cite{Josse:2021put} were studied in \cite{Josse:2023lsp} based on the analysis of  the embedding tensor and the supersymmetry equation.  A similar approach can be extended to the other theories discussed in this article.

\section*{Acknowledgments}

We would like to thank Guillaume Bossard, Mattia Cesaro, Emanuel Malek, Henning Samtleben, Mario Trigiante, Óscar Varela, Dan Waldram and Alberto Zaffaroni for useful discussions. M. Pico gratefully thanks the LPTHE, and in particular M. Petrini, for their hospitality during the initial stages of this work. GJ is supported by the Deutsche Forschungsgemeinschaft (DFG, German Research Foundation) via the Emmy Noether program “Exploring the landscape of string theory flux vacua using exceptional field theory” (project number 426510644). M. Pico is supported by predoctoral award FPU22/02084 of  the Spanish Government and also (partially) by grants CEX2020-001007-S and PID2021-123017NB- I00, funded by MCIN/AEI/10.13039/501100011033 and by ERDF A way of making Europe.


\appendix

\section{Details on \texorpdfstring{$\mathbf{E}_{7(7)}$}{E7(7)}}
\label{PreliminariesE77_Mth}

The classification of the $\Gst \subset \Ex{7}$-structures  is determined by 
determining the number of 
$\Gst$-singlets in the generalised spinor, vector and ajoint bundles. This amounts at solving the equations
\begin{equation}
\begin{matrix}
g  \cdot \epsilon &= &\epsilon\,\\
g  \cdot R_{\SU{8}} &= &R_{\SU{8}} \, \\
g \cdot V &= &V\, \\
g  \cdot R &= &R \, ,
\label{app:Preliminaries_Singlets_SingletsEquationGroup}
\end{matrix}  
\end{equation}
for all $ g \in \Gst$, where $\epsilon$ and $R_{\SU{8}}$ are in the spinorial and adjoint representation of $\SU{8}$ and 
$V$ and $R$ are in $\rep{56}$ and $\rep{133}$ of $\Ex{7}$. 

Since the action of the groups $\Gst$ on the vector and tensors spaces we are interested is linear,  for continuous $\Gst$, i.e  Lie groups, we can use the fact that any element connected to the identity is the exponentiation of the Lie algebra elements,  $g= e^{\lambda t}$, where $t$ is a generator in the Lie algebra and $\lambda$ a parameter, to replace \eqref{Preliminaries_Singlets_SingletsEquationGroup} with its Lie algebra analog 
\begin{equation}
\begin{matrix}
t \cdot \epsilon &= &0\,\\
t \cdot R_{\SU{8}} &= &0 \, \\
t \cdot V &= &0\, \\
t \cdot R &= &0 \, ,
\label{Preliminaries_Singlets_SingletsEquation}
\end{matrix} 
\end{equation}
for all $ \forall \, t \in \mathfrak{g}_S$, where $\mathfrak{g}_S$ denotes the Lie algebra of $\Gst$. 

\medskip

In this appendix we give our conventions for the action of $\Ex{7}$ and $\SU{8}/\mathbb{Z}_2$, its maximally compact subgroup.  The main references for this appendix are \cite{Pacheco:2008ps, Trigiante:2016mnt}.

We are mostly interested in the fundamental and adjoint representations, $\rep{56}$ and $\rep{133}$ of $\Ex{7}$.  We will denote $\Ex{7}$ fundamental indices by $M,N,P =1,\ldots, 56$ and adjoint indices by $A, B, C =1 ,\dots 133$.

\medskip
There are two relevant decompositions of $\Ex{7}$, one according to $\SL{8}$ and the other according to $\SU{8}$. 
We will relate the two  via $\SO{8}$ representations, since 
\begin{equation}
    \SO{8}= \SL{8} \cap \SU{8}\,.
\end{equation}

The group 
 $\SO{8}$ has three inequivalent  representations of dimension 8:  $\rep{8}_v$, $\rep{8}_c$ and $\rep{8}_s$. The $\rep{8}_v$ is identified with the $\rep{8}$ of $\SL{8}$ under the branching $\SO{8}\subset \SL{8}$, whereas the $\rep{8}_s$ is identified with the $\rep{8}$ of $\SU{8}$ under the branching $\SO{8}\subset \SU{8}$. We denote $\rep{8}_v$ indices by $a$, $b$ , $\rep{8}_c$ indices  by $\dot \alpha$, $\dot \beta$ and $\rep{8}_s$ indices by $\alpha$, $\beta$, all of them running from 1 to 8. 
 These three representations are connected by triality.
$\SO{8}$ indices are raised and lowered with the three invariant tensors: $\delta^{ab}$, $C^{\alpha \beta}$ and $C^{\dot{\alpha}\dot{\beta}}$. In our conventions $C^{\alpha \beta}=\delta^{\alpha \beta}$ and $C^{\dot{\alpha}\dot{\beta}}=\delta^{\dot{\alpha}\dot{\beta}}$. 
 
 We are interested in the $\SO{8}$ generators in the $\rep{8}_s$ representation, which we will use as intertwiners between the $\SL{8}$ and $\SU{8}$ representations, 
\begin{equation}
\label{SOgensp}
    (\gamma_{ab})^{\alpha \beta} = (\gamma_{[a} \gamma_{b]})^{\alpha \beta} =  (\gamma_{[a})^{\alpha}{}_{\dot \gamma} (\gamma_{b]})^\beta{}_{\dot \delta} \, C^{\dot \gamma \dot \delta}  \, ,   
\end{equation}
 where the gamma matrices are 
  \begin{equation}
\begin{aligned}
& \gamma_1= \sigma_2 \otimes \id_2 \otimes \sigma_1 \\ 
& \gamma_2= \sigma_2 \otimes \sigma_3 \otimes \sigma_3 \\ 
& \gamma_3= -\sigma_2 \otimes \sigma_1 \otimes \sigma_3  \\
& \gamma_4= -\sigma_1 \otimes \sigma_2 \otimes \sigma_1  
\end{aligned}
\qquad \qquad 
\begin{aligned}
& \gamma_5= \sigma_3 \otimes \sigma_2 \otimes \sigma_1 \\
& \gamma_6= -\id_2 \otimes \sigma_2 \otimes \sigma_3 \\
& \gamma_7= -\id_2 \otimes \id_2 \otimes \sigma_2 \\
& \gamma_8= -i \id_2 \otimes \id_2 \otimes \id_2 
\end{aligned}
\end{equation}
with Pauli matrices 
 \begin{equation}
\sigma_1 = \left( \begin{matrix}
0 &1\\
1 &0
\end{matrix} \right)  , \; \;
\sigma_2 = \left( \begin{matrix}
0 &-i\\
i &0
\end{matrix} \right) , \; \;
\sigma_3 = \left( \begin{matrix}
1 &0\\
0 &-1
\end{matrix} \right) \, . 
\end{equation} 

 We will also need the four-gamma antisymmetric product
 \begin{equation}
     \gamma^{abcd}{}_{\alpha \beta}=  (\gamma^{[a }\gamma^{b} \gamma^{c} \gamma^{d]})_{\alpha \beta} = \gamma^{[a }{}_{\alpha \dot{\alpha}} \gamma^{b}{}_{\gamma \dot{\beta}} \gamma^{c}{}_{\delta \dot{\gamma}} \gamma^{d]}{}_{\beta \dot{\delta}} C^{\dot{\alpha}\dot{\beta}} C^{\gamma\delta} C^{\dot{\gamma}\dot{\delta}}\,.
 \end{equation}

\subsection{$\SL{8}$ decomposition }
\label{PreliminariesSL8}

Under $\SL{8} \subset \Ex{7}$ the fundamental representations of $\Ex{7}$ decompose as 
\begin{equation}
    \begin{aligned}
        \rep{56}  & = \rep{28} \oplus \rep{28}'  \\
          V&  = (V^{ab} , \tilde{V}_{ab} ) 
    \end{aligned}
     \label{SL8E7f} 
\end{equation}
where  $V^{ab}, \tilde{V}_{ab}$ are  two index antisymmetric tensors $V^{ab }=-V^{ba}$, while the adjoint gives
\begin{equation}
    \begin{aligned}
\rep{133} & =  \rep{63} \oplus \rep{70} \\
         \mu & = ( \mu^a{}_b, \mu^{abcd} ) \, 
    \end{aligned}
     \label{SL8E7ad} 
\end{equation}
with  $\mu^a{}_a = 0$ and $\mu^{abcd} = \mu^{[abcd]}$.

\noindent Contractions between two vectors are given by 
\begin{equation}
    V^M W_M = \tfrac{1}{2}V^{ab}W_{ab} + \tfrac{1}{2}V_{ab}W^{ab} \, . 
\end{equation}
The action of $\rep{133}$ on the $\rep{56}$ becomes
\begin{equation}
    \begin{aligned}
     (\mu \cdot V)^{ab} &  =  \mu^a{}_c V^{cb} +  \mu^b{}_c V^{ac} + \frac{1}{2} \mu^{abcd} \tilde{V}_{cd} \\
     (\mu \cdot \tilde{V})_{ab} &  =  - \mu^c{}_a \tilde{V}_{cb} - \mu^c{}_b \tilde{V}_{ac} +  \frac{1}{2} (\ast\mu)_{abcd} V^{cd} \, ,
    \end{aligned}
\end{equation}
where $ (\ast\mu)_{abcd} = \frac{1}{4!} \epsilon_{abcdefgh} \mu^{efgh}$, and the commutator of two adjoints reads 
\begin{equation}
    \begin{aligned}
     (\mu \cdot \mu^\prime)^a{}_b &  =  \mu^a{}_c \mu^{\prime c}{}_b - \mu^{\prime a}{}_c \mu^c{}_b + \tfrac{1}{3!} \mu^{acde} (\ast \mu^\prime)_{bcde}  \\
     (\mu \cdot \mu^\prime)^{abcd} &  =  -4 ( \mu^{[a}{}_e  \mu^{\prime bcd]e} - 
     \mu^{\prime[a}{}_e  \mu^{bcd]e} ) \, . 
    \end{aligned}
\end{equation}

\medskip

For computational reasons, 
it is more convenient to treat the elements of the $\rep{56}$ and the $\rep{133}$ as 56-dimensional vectors and $56 \times 56$ dimensional matrices. In this way  the action of the $\Ex{7}$ adjoint on the $\rep{56}$ becomes a matrix multiplication 
\begin{equation}
    V^M \rightarrow \mu^M{}_N V^N \, . 
\end{equation}
The idea is to flatten the antisymmetric tensors $V^{ab}$ and $\tilde{V}_{ab}$ into two 28-component vectors whose elements are ordered as $(V^{12}, V^{13}, \ldots, V^{78})$,  and then to construct a 56-dimensional vector 
\begin{equation}
    V^M=(V^{ab}, \tilde{V}_{ab}) \qquad a<b \, .
\end{equation}   
Similarly, a generic $\Ex{7}$ Lie algebra element can be written as a $56 \times 56$ matrix in terms of a basis of generators:
\begin{equation}
  \mu^M{}_N =   \mu^a{}_b (t_a{}^b)^M{}_N + \frac{1}{4!}\mu^{abcd} (t_{a b c d})^M{}_N\, ,
\label{PreliminariesE77_Mth_E77AlgebraElementSL8}
\end{equation}
where\footnote{
 When flattening matrix indices we must take into account that each contribution in the $\SL{8}$ basis appears twice, due to the antisymmetry in the $ab$ indices. More explicitly,  if we  fix the first index of an adjoint element $\mu^{ab}{}_N$, its action is given by 
\begin{equation}
\mu^{ab}{}_N V^N =  \frac{1}{2} \big( \sum_{c,d}  \mu^{a b}{}_{c d}V^{cd} + \sum_{c,d}  \mu^{abcd} V_{cd} \big) =\sum_{c < d} \mu^{ab}{}_{cd}V^{cd} + \sum_{d <c} \mu^{ab}{}_{c d}V^{cd}\,.
\end{equation}} 
\begin{equation}
\label{t-133}
\begin{array}{lll}
(t_a{}^b)^M{}_N  &= \begin{pmatrix}
 (t_a{}^b)^{cd}{}_{ef} &0 \\
0 & (t_a{}^b)_{cd}{}^{ef}
\end{pmatrix} &= \begin{pmatrix}
4 \delta^{[c}{}_{[e}(t_a{}^b)^{d]}{}_{f]} &0 \\
0 & -4 \delta^{[e}{}_{[c}(t_a{}^b)^{f]}{}_{d]}
\end{pmatrix} \,\\\\
(t_{abcd})^M{}_N &= \begin{pmatrix}
  0 & (t_{a{} b cd})^{efgh} \\
(t_{abcd})_{efgh} & 0
\end{pmatrix} &= \begin{pmatrix}
0 & 4! \delta^{efgh}_{a{} bcd} \\
\varepsilon_{abcdefgh} & 0
\end{pmatrix} \,,
\end{array}
\end{equation}
and where the generators $(t_a{}^b)^c{}_d$ of $\SL{8}$ in the fundamental are $8 \times 8$ matrices defined as
\begin{equation}
(t_a{}^b)^c{}_d= \delta_a^c \delta_d^b - \tfrac18 \delta_a^b \delta_d^c \, . 
\end{equation}
Thus, the  action of a generic $\Ex{7}$ generator on the fundamental is given by:
\begin{equation}
\mu^M{}_N  V^N =  \begin{pmatrix}
\mu^a{}_{a'} V^{a'b}+\mu^b{}_{b'} V^{ab'} + \tfrac12 \mu^{a bcd} V_{cd} \\
- \mu^{a'}{}_a V_{a'b} - \mu^{b'}{}_b V_{ab'} +\tfrac12 (\ast \mu)_{a b cd} V^{cd}
\end{pmatrix}\,.
\end{equation}
Note how a $\tfrac12$ factor has been introduced due to our contraction conventions, to avoid over-counting. 

Finally, denoting by $t_A$ the full set of $\Ex{7}$ generators \eqref{t-133} , the adjoint representation of $\Ex{7}$ can be obtained from the commutator as:
\begin{equation}
t_A\cdot t_B= [t_A, t_B]  = f_{AB}{}^{C}  t_C\,.
\end{equation}

\subsection{$\SU{8}/\mathbb{Z}_2$ decomposition }
\label{PreliminariesSU8}

Under $\SU{8}/\mathbb{Z}_2$ the fundamental of $\Ex{7}$ decomposes as
\begin{equation}
    \begin{aligned}
        \rep{56}  & = \rep{28} \oplus \rep{\overline{28}}  \\
          V&  = (V^{\alpha \beta} , \bar{V}_{\alpha \beta} ) 
    \end{aligned}
     \label{SU8E7f} 
\end{equation}
where  $\bar{V}^{\alpha \beta} = V^*_{\alpha \beta}$, and  the adjoint decomposes as 
\begin{equation}
    \begin{aligned}
\rep{133} & =  \rep{63} \oplus \rep{70} \\
         \mu & = ( \mu^\alpha{}_\beta, \mu^{\alpha \beta \gamma \delta}) \, 
    \end{aligned}
     \label{SU8E7ad} 
\end{equation}
with  $\mu^\alpha{}_\alpha = 0$ and $\mu^{\alpha \beta \gamma \delta} = \mu^{[\alpha \beta \gamma \delta]}$.

The action of the adjoint of $\Ex{7}$ on the $\rep{56}$ decomposes as
\begin{equation}
\label{PreliminariesSU8_E7ActionSU8Basis}
    \begin{aligned}
     (\mu \cdot V)^{\alpha \beta} &  =  \mu^\alpha{}_\gamma V^{\gamma \beta} +  \mu^\beta{}_\gamma V^{\alpha \gamma} + \tfrac12 \mu^{\alpha \beta \gamma \delta} \bar{V}_{\gamma \delta} \\
     (\mu \cdot \bar{V})_{\alpha \beta} &  =  - \mu^\gamma{}_\alpha \bar{V}_{\gamma \beta} - \mu^\gamma{}_\beta \bar{V}_{\alpha \gamma}  + \tfrac12 (\ast {\mu})_{\alpha \beta \gamma \delta} V^{ \gamma \delta} \, ,
    \end{aligned}
\end{equation}
 and the commutator of two adjoints reads 
\begin{equation}
    \begin{aligned}
     (\mu \cdot \mu^\prime)^\alpha{}_\beta &  =  \mu^\alpha{}_\gamma \mu^{\prime \gamma}{}_\beta - \mu^{\prime \alpha}{}_\gamma \mu^\gamma{}_\beta + \tfrac{1}{3!} \mu^{\alpha \gamma \delta \delta^\prime} (\ast \mu^\prime)_{\beta \gamma \delta \delta^\prime}  \\
     (\mu \cdot \mu^\prime)^{\alpha \beta \gamma \delta} &  =  -4 ( \mu^{[\alpha}{}_{\delta^\prime}  \mu^{\prime \beta \gamma \delta ] \delta^\prime} - 
     \mu^{\prime[\alpha}{}_{\delta^\prime}  \mu^{\beta \gamma \delta ] \delta^\prime} ) \, . 
    \end{aligned}
\end{equation}

\subsection{Relation between $\SL{8}$ and $\SU{8}/\mathbb{Z}_2$ decomposition }
\label{PreliminariesSU8SL8Relation}

The idea is to express the $\Ex{7}$ generators  in terms of $\SU{8}$ generators. This can be done by connecting the $\SL{8}$ and $\SU{8}/\mathbb{Z}_2$ basis. 

The relation between a vector $V$ in the $\SL{8}$ and $\SU{8}/\mathbb{Z}_2$ basis is obtained using the $\SO{8}$ generators in \eqref{SOgensp}
\begin{equation}
    \begin{aligned}
        V^{\alpha \beta} & = \frac{1}{4 \sqrt{2}} ( V^{ab} \gamma_{ab} + i \tilde{V}_{ab} \gamma^{ab} )^{\alpha \beta} \\
        \bar{V}_{\alpha \beta} & = \frac{1}{4 \sqrt{2}} ( V^{ab} \gamma_{ab} - i \tilde{V}_{ab} \gamma^{ab} )_{\alpha \beta}\, .
    \end{aligned}
\end{equation}
In matrix notation this is performed in terms of the unitary matrix 
\begin{equation}
S^{M}{}_{\underline{N}} = 
\frac{1}{\sqrt{2}}  
\left( \begin{matrix}
\tfrac12 \gamma^{ab}{}_{\alpha \beta} & \tfrac12 \gamma^{ab}{}^{\alpha \beta}\\
 - \frac{i}{2} \gamma_{ab}{}_{\alpha \beta} & \frac{i}{2} \gamma_{ab}{}^{\alpha \beta}
\end{matrix} \right)\,.
\end{equation}
Explicitly, the change of basis in $56$ flattened indices is obtained as
\begin{equation}
\label{PreliminariesSU8SL8Relation_RotationFundamentalE7}
\begin{pmatrix}
V^{\alpha \beta}\\
V_{\alpha \beta}
\end{pmatrix}=V^{\underline{M}}= (S^\dagger)^{\underline{M}}{}_M V^{M}= \frac{1}{2\sqrt{2}}\begin{pmatrix}
   \tfrac12 \gamma_{a b}{}^{\alpha \beta}V^{a b} +\frac{i}{2} \gamma^{a b}{}^{\alpha \beta}V_{a b}\\
 \tfrac12  \gamma_{a b}{}_{\alpha \beta}V^{a b} -\frac{i}{2} \gamma^{a b}{}_{\alpha \beta}V_{a b}
\end{pmatrix}\,.
\end{equation}

To connect the elements of the $\rep{133}$ of $\Ex{7}$
 in the $\SL{8}$ basis \eqref{SL8E7f} to those in the $\SU{8}$ basis \eqref{SU8E7f}, it is convenient to recall how those two split under the common $\SO{8}$ factor. 

Under $\SL{8} \supset \SO{8}$, the elements of the $\rep{63}$ decompose as
\begin{equation}
\begin{aligned}
    \rep{63} & = \rep{28} \oplus \rep{35}_v \\
    \mu^a{}_b & = \mu^{(a) a}{}_{b} + \mu^{(s)a}{}_{b} \,, 
\end{aligned}
\end{equation}
where $\mu^{(a)}_{ab} = \mu_{[ab]}$ and  $\mu^{(s)}_{ab} = \mu_{(ab)}$  are the antisymmetric and symmetric components. Similarly the elements of the 
$\rep{70}$ decompose as 
\begin{equation}
    \begin{aligned}
        \rep{70} &= \rep{35}_c \oplus \rep{35}_s \\
        \mu^{abcd} & = \mu^{[abcd]_+} + \mu^{[abcd]_- } \,,
    \end{aligned}
\end{equation}
where $\mu^{[abcd]_\pm}$ correspond to self-dual and anti-self dual totally antisymmetric rank four tensors
\begin{equation}
    \mu^{[abcd]_\pm} =  \pm (\ast\mu)^{[abcd]_\pm}  \, .
\end{equation}

On the other hand, under $\SU{8} \supset \SO{8}$ 
we have the following splitting
\begin{equation}
\begin{aligned}
    \rep{63} & = \rep{28} \oplus \rep{35}_s \\
    \mu^\alpha{}_\beta & = \mu^{(a)\alpha}{}_\beta + i  \mu^{(s)\alpha}{}_\beta \,, 
\end{aligned}
\end{equation}
where $\mu^{(a)}_{\alpha \beta} = \mu_{[\alpha \beta]}$ and  $\mu^{(s)}_{\alpha \beta} = \mu_{(\alpha \beta)}$  are again the (real) antisymmetric and symmetric components, and  
\begin{equation}
    \begin{aligned}
        \rep{70} &= \rep{35}_v \oplus \rep{35}_c \\
        \mu^{\alpha \beta \gamma \delta} & = \mu^{[\alpha \beta \gamma \delta]_+} + i \mu^{[\alpha \beta \gamma \delta]_- } 
    \end{aligned}
\end{equation}
with $\mu^{[\alpha \beta \gamma \delta]_\pm}= \pm (\ast\mu)^{[\alpha \beta \gamma \delta]_\pm}$  (real) self-dual and anti-self dual totally antisymmetric rank four tensors. 

We can use the $\SO{8}$ triality to connect the representations of $\SU{8}$, in $\rep{8}_s$ indices, to those of $\SL{8}$ in $\SO{8}$ vector indices
\begin{equation}
\renewcommand\arraystretch{1.5}
    \begin{array}{lcl}
    \rep{28}:  &  \quad \quad & \mu^{(a)a}{}_{b} = \tfrac14 (\gamma^{a}{}_{b} )_{\alpha \beta}  \mu^{(a)\alpha\beta}  \\
      \rep{35}_v : &  \quad \quad & \mu^{(s)a}{}_{b} = \tfrac14 (\gamma^{a c} \gamma_{bc})_{[\alpha \beta \gamma\delta]_+ }  \mu^{[\alpha \beta \gamma \delta]_+ }  \\
 \rep{35}_c : &  \quad \quad  & \mu^{[abcd]_+} =  \tfrac32 \tfrac{1}{4!} (\gamma^{[a b} \gamma^{c d]})_{[\alpha \beta \gamma\delta]_- }  \mu^{[\alpha \beta \gamma \delta]_- } 
 \\
 \rep{35}_s  : &  \quad \quad &  \mu^{[abcd]_-} = \tfrac{1}{4} \frac{1}{4!} (\gamma^{[a bcd]_-})_{\alpha \beta} \mu^{(s)\alpha \beta}\,.
    \end{array}
    \label{PreliminariesE77_Mth_SL8SU8relation}
\end{equation}
 By plugging \eqref{PreliminariesE77_Mth_SL8SU8relation} into \eqref{PreliminariesE77_Mth_E77AlgebraElementSL8} we can express an $\Ex{7}$ adjoint element acting on the $\rep{56}$ as\footnote{Note that the generators can be expressed either in the $\SL{8}$ basis \eqref{t-133} or the $\SU{8}$ basis, for which we have to rotate them according to \eqref{PreliminariesSU8SL8Relation_RotationFundamentalE7}, this is:
\begin{equation}
   (t_A)^{\underline{M}}{}_{\underline{N}}= (S^\dagger)^{\underline{M}}{}_M (t_A)^{M}{}_{N}S^{N}{}_{\underline{N}} \,.
\end{equation}} 
\begin{equation}
   \mu^{\underline{M}}{}_{\underline{N}} =   \mu^{\alpha\beta} (t_{\alpha \beta})^{\underline{M}}{}_{\underline{N}}  + \frac{1}{4!}\mu^{ \alpha \beta\gamma \delta }  (t_{ \alpha \beta \gamma \delta})^{\underline{M}}{}_{\underline{N}} \,,
\end{equation}
where $\mu^{\alpha \beta} \in \rep{63}$ and  $\mu^{\alpha \beta \gamma \delta} \in \rep{70}$ of $\SU{8}$ and, omitting fundamental $\Ex{7}$ indices for the sake of simplicity, 
\begin{equation}
\begin{split}
     t_{\alpha \beta} &=  \tfrac{1}{4} \gamma^{a}{}_{b\, \alpha \beta} \; t_a{}^b - i  \tfrac{1}{4} \tfrac{1}{4!}\gamma^{[a bcd]_-}{}_{\alpha \beta}\; t_{a bcd} \,\\
 t_{\alpha \beta \gamma \delta} &= \tfrac14\gamma^{a c}{}_{[\alpha \beta |}\gamma_{bc \, | \gamma\delta]_+} \; t_a{}^b -i  \tfrac32 \tfrac{1}{4!}\gamma^{[a b}{}_{[\alpha \beta} \gamma^{c d]_+}{}_{\gamma\delta]_-}\; t_{a b c d}\,.
\end{split}
\end{equation}  
The generators $t_a{}^b$ and $t_{a bcd}$ are given in \eqref{t-133}.
It is straightforward, albeit tedious, to check that the action in the $\SU{8}$ basis matches \eqref{PreliminariesSU8_E7ActionSU8Basis}:
\begin{equation}
\left(  \mu^{\gamma \delta} t_{\gamma \delta}  + \frac{1}{4!}\mu^{\gamma \delta \gamma' \delta'}  t_{\gamma \delta \gamma' \delta'} \right) \cdot V =  \begin{pmatrix}
\mu^\alpha{}_{\alpha'} V^{\alpha'\beta}+\mu^\beta{}_{\beta'} V^{\alpha\beta'} +\tfrac12 \mu^{\alpha \beta \alpha' \beta'} V_{\alpha' \beta'} \\
- \mu^{\alpha'}{}_\alpha V_{\alpha'\beta} - \mu^{\beta'}{}_\beta V_{\alpha\beta'} + \tfrac12 (*\mu)_{\alpha \beta \alpha' \beta'} V^{\alpha' \beta'}
\end{pmatrix}\,.
\end{equation}

\subsection{Embeddings for $\mathcal{N}=2$ truncations}

We conclude the appendix with some details about some of the embeddings relevant for the classifications of $\mathcal{N}=2$ truncations. 

Since $\mathcal{N}=2$ truncations are associated to $\Gst$-structures that are subgroups of $\SU{6}$, we are interested in the  branching
\begin{equation}
\label{eq:SU8toSU6}
\begin{aligned}
    \SU{8} & \supset \SU{6}\times \SU{2}_R \times \U{1}_R \\
    \rep{63} & = (\rep{35}, \rep{1})_0 \oplus 
    (\rep{1}, \rep{3})_0 \oplus (\rep{1}, \rep{1})_0 \oplus (\rep{2}, \rep{2})_{-4} \oplus  (\rep{\bar 6}, \rep{2})_4 
\end{aligned}
\end{equation}
The explicit embedding of the $\SU{6}$ and $\SU{2}_R$ generators is given by
\begin{equation}
   \mu_{\SU{6}\times \SU{2}_R} = \begin{pmatrix}
       \mu_{\SU{6}} &  0 \\ 
     0   & 0 \end{pmatrix} + 
    \begin{pmatrix}
        0  &  0  \\ 
     0   & \mu_{\SU{2}_R}
    \end{pmatrix} \,,
\end{equation}
while $\U{1}_R$ embeds as 
\begin{equation}
\label{app:U1R}
  \mu_{\U{1}_R} =   \begin{pmatrix}
       i \id_6 &  \\ 
        & -3i \id_2
  \end{pmatrix} \,.
\end{equation}

Accordingly, the  $\SU{8}$ fundamental indices split into $\alpha=(m,i)$,  where $i=1,2\in\SU{2}_R$ and $m=1,\dots, 6\in \SU{6}$. 
With this choice  a generalised vector \eqref{SU8E7f} splits under \eqref{eq:SU8toSU6} as 
\begin{equation}
\label{vecSU6spl}
    V^{\alpha\beta}=(V^{mn},V^{mi},V^{ij}) \,  , 
\end{equation}  
and similarly for its conjugate.

In studying truncations with only  vector multiplets we need to consider $\Gst$-structures 
\begin{equation}
    \Gst \subset \SU{6} \subset \SOs{12}
\end{equation}
where $\SOs{12}$ is the stabiliser of the triplet of adjoint singlets generating the $\SU{2}_R$ R-symmetry and 
\begin{equation}
\label{eq:discretev}
 \begin{aligned}
   \Ex{7} \supset \SOs{12} \times \SU{2}_R \supset \SU{6} \times \U{1}_R \times \SU{2}_R \, . 
 \end{aligned}
\end{equation}
The $\SOs{12}$ generators are given, in terms of the $\SU{6}\times\SU{2}_R$ indices $(\alpha,i)$, as: 
\begin{equation}
\begin{split}
   & \SU{6}=\rep{35}_0: (\mu_{\SU{6}})^{\alpha \beta}(t_{\alpha \beta})^{M}{}_{N}\, \\
    &\U{1}_R=\rep{1}_0: (\mu_{\U{1}_R})^{\alpha \beta}(t_{\alpha \beta})^{M}{}_{N} \, \\
    &\rep{\overline{15}}_{-4}: \tfrac14 \mu^{mnij} (t_{mnij})^{M}{}_{N}  \, \\
    &\rep{15}_{4}: \tfrac12 \mu^{mnpq}(t_{mnpq})^{M}{}_{N} \,.
    \end{split}
\end{equation}

Similarly, in studying truncations with only  hypermultiplets we need to consider $\Gst$-structures 
\begin{equation}
    \Gst \subset \SU{6} \subset \rm E_{6(2)}
\end{equation}
where $\rm E_{6(2)}$ is the stabiliser of the generalised vectors $K$ and $\hat{K}$, and 
\begin{equation}
    \Ex{7}\supset \mathrm{E}_{6(2)} \times \U{1}_R \,.
\end{equation}
The $\mathrm{E}_{6(2)}$ generators are given, in terms of the $\SU{6}\times\SU{2}_R$ indices $(\alpha,i)$, as: 
\begin{equation}
\begin{split}
   & \SU{6}=(\rep{35},\rep{1}): (\mu_{\SU{6}})^{\alpha \beta}(t_{\alpha \beta})^{M}{}_{N} \\
    &\SU{2}_R=(\rep{1},\rep{3}): (\mu_{\SU{2}_R})^{\alpha \beta}(t_{\alpha \beta})^{M}{}_{N} \\
    &(\rep{20},\rep{2}): \mu^{imnp} (t_{imnp})^M{}_N\,.
    \end{split}
\end{equation}

\section{Example: $\mathcal{N}=4$ truncations}
\label{explicit_ex_subsec}
In this appendix we give the details of the derivation of the truncations with $\mathcal{N}=4$ supersymmetry of Section \ref{sec:N4}.

To classify the possible $\mathcal{N}=4$ truncations we have to look at  $\Gst$-structures  that are subgroups 
\begin{equation}
    \Gst \subseteq \SU{4} \, , 
    \label{explicit_ex_subsec_GSinsideSU4}
\end{equation}
where $\SU{4}$ is the commutant in $\SU{8}$ of the $\SU{4}_R$ symmetry
\begin{equation}
    \SU{8} \supset \SU{4} \times \SU{4}_R \, , 
    \label{explicit_ex_subsec_SU4SEmbedding}
\end{equation}
and only preserve 4 singlets in the $\rep{8}$ of $\SU{8}$. 

\medskip

The idea is to proceed from the largest to the smallest subgroup  $\Gst \subset \SU{8}$, with the property \eqref{explicit_ex_subsec_GSinsideSU4}. The largest  subgroup of  $\SU{8}$ leading to a $\calN=4$ truncation is $\SU{4}_S\cong \Spin{6}$ \cite{Cassani:2019vcl}. The $\SU{4}_S$ generators are embbeded in $\SU{8}$ as anti-hermitean matrices of the form 
\begin{equation}
\label{app:SU4semb}
  \mu_{\SU{4}_S} \sim   \left(\begin{array}{cc}
        \su(4)_4 &  \\ 
         & 0_{4}
    \end{array} \right)\, . 
\end{equation}
This defines the embedding of the $\rep{4}$ of $\SU{4}_S$ in the $\rep{8}$ of $\SU{8}$ according to \eqref{explicit_ex_subsec_SU4SEmbedding}. 
Using the expressions of  Appendix \ref{PreliminariesE77_Mth} we also build the embedding of the $\SU{4}_S$ generators in all relevant representations of $\Ex{7}$: $\rep{56}$ and $\rep{133}$.

Then the singlets in the $\rep{8}$ and $\rep{63}$ of $\SU{8}$ and the $\rep{56}$ and $\rep{133}$ of $\Ex{7}$  are given by the solutions to the equations
\begin{equation}
\label{sing-cond-bis}
\begin{matrix}
\mu_{\SU{4}_S} \cdot \epsilon & = & 0\, \\
\mu_{\SU{4}_S} \cdot V &= &0\, \\
\mu_{\SU{4}_S} \cdot R &= &0 \, \\
\mu_{\SU{4}_S} \cdot R_{\SU{8}} &= &0 \, . 
\end{matrix}
\end{equation}

We find 12 singlets in the $\rep{56}$, 18 in the $\rep{133}$ and 16 in the $\rep{63}$, in agreement with the branchings:
\begin{equation}
    \begin{array}{lll}
       \Ex{7}  &\rightarrow & \SU{4}_S\times\SU{4}\times \U{1}  \\ 
          \rep{56}  &\rightarrow & (\rep{1},\rep{6})_{-2} \oplus (\rep{6},\rep{1})_2 \oplus    (\rep{4},\rep{4})_0 \oplus \text{c.c.}
          \\
          \rep{133}  &\rightarrow &  (\rep{1},\rep{1})_4 \oplus (\rep{1},\rep{1})_{-4} \oplus (\rep{1},\rep{1})_0 \oplus (\rep{1},\rep{15})_0 \oplus (\rep{15},\rep{1})_0  \oplus    (\rep{4},\rep{\overline{4}})_{2} \oplus (\rep{\overline{4}},\rep{4})_{-2}  \\
          & & \quad \oplus   (\rep{4},\rep{\overline{4}})_{-2} \oplus  (\rep{\overline{4}},\rep{4})_{2} \oplus (\rep{6},\rep{6})_0 \, , 
    \end{array}
    \label{Section_Algorithm_E7repsBranchedUnderSU4}
\end{equation}
and
\begin{equation}
    \begin{array}{lll}
       \SU{8}  &\rightarrow & \SU{4}_S\times\SU{4}\times \U{1}  \\ 
        \rep{8}  &\rightarrow & (\rep{4},\rep{1})_{1} \oplus (\rep{1},\rep{4})_{-1}
          \\
          \rep{63}  &\rightarrow & (\rep{1},\rep{1})_0 \oplus (\rep{1},\rep{15})_0 \oplus (\rep{15},\rep{1})_0 \oplus   (\rep{4},\rep{\overline{4}})_{2} \oplus (\rep{\overline{4}},\rep{4})_{-2}   \,.
    \end{array}
\end{equation}
The 12 singlets in the $\rep{56}$ correspond to 6 vectors plus their magnetic duals, whereas the 2 singlets in $\Ex{7}/\SU{8}$ correspond to 2 scalars parametrizing the manifold $\frac{\SL{2,\mathbb{R}}}{\U{1}}$. Together with the spin 2 degrees of freedom, these form the bosonic content of the gravity multiplet in a duality covariant form. The truncation associated to this structure is pure supergravity.

\medskip

In order to obtain an $\calN = 4$ theory with matter coupled to gravity, the largest  subgroup of $\SU{4}_S$ is $\USp{4}_S\cong \Spin{5}_S$, whose generators are the subset of  $\SU{4}_S$ generators  in \eqref{app:SU4semb} preserving  the symplectic form 
\begin{equation}
\omega_{\alpha \beta}=
\begin{pmatrix}
&\id_2  & \\
- \id_2 & & \\
& & 0_{4}
\end{pmatrix} 
\end{equation}
As for the $\SU{4}_S$-structure, we can embed the $\USp{4}_S$ generators 
in the relevant representations and determine the 
singlets in the $\rep{56}$ and $\rep{133}$ of $\Ex{7}$ and in the $\rep{63}$ of $\SU{8}$ imposing \eqref{sing-cond-bis} with $\mu_{\USp{4}_S}$.

Beside the singlets corresponding to the gravity multiplet, we get one extra vector (with its magnetic dual) and 6 scalars parameterising the manifold $\SO{6,1}/\SO{6}_R$. All together they constitute the bosonic field content of an $\calN = 4$ vector multiplet. 

These results are in agreement with the further breaking $\SU{4}  \supset  \USp{4}$ in \eqref{Section_Algorithm_E7repsBranchedUnderSU4}
\begin{equation}
    \begin{array}{lll}
       \SU{8}  &\rightarrow & \USp{4}_S\times\SU{4}\times \U{1}  \\ 
        \rep{8}  &\rightarrow & (\rep{4},\rep{1})_{1} \oplus (\rep{1},\rep{4})_{-1} \\ 
          \rep{63}  &\rightarrow & (\rep{1},\rep{1})_0 \oplus (\rep{1},\rep{15})_0 \oplus (\rep{5},\rep{1})_0 \oplus (\rep{10},\rep{1})_0 \oplus   (\rep{4},\rep{\overline{4}})_{2} \oplus (\rep{4},\rep{4})_{-2}   \,,
    \end{array}
\end{equation}
and
\begin{equation}
    \begin{array}{lll}
       \Ex{7}  &\rightarrow & \USp{4}_S\times\SU{4}\times \U{1}  \\ 
          \rep{56}  &\rightarrow & (\rep{1},\rep{6})_{-2} \oplus (\rep{1},\rep{1})_2 \oplus  (\rep{5},\rep{1})_2 \oplus  (\rep{4},\rep{4})_0 \oplus \text{c.c.}
          \\
          \rep{133}  &\rightarrow &  (\rep{1},\rep{1})_4 \oplus (\rep{1},\rep{1})_{-4} \oplus (\rep{1},\rep{6})_0 \oplus (\rep{1},\rep{1})_0 \oplus (\rep{1},\rep{15})_0  \oplus (\rep{5},\rep{1})_0  \oplus (\rep{10},\rep{1})_0   \\
          & & \quad \oplus    (\rep{4},\rep{\overline{4}})_{2} \oplus (\rep{4},\rep{4})_{-2} \oplus   (\rep{4},\rep{\overline{4}})_{-2} \oplus  (\rep{4},\rep{4})_{2} \oplus (\rep{5},\rep{6})_0  \,.
    \end{array}
    \label{app:Section_Algorithm_E7repsBranchedUnderUSp4}
\end{equation}

\medskip

The next $\Gst \subset \SU{4}$ preserving $\mathcal{N}=4$ supersymmetry and providing extra matter is $\Spin{4}_S\cong \SU{2}_S\times \SU{2}_S \subset \USp{4}_S$. Schematically, its Lie algebra reads\footnote{In our explicit realization, this $\SU{2}\times\SU{2}$ arises upon a permutation of the spinor coordinates $\epsilon_2 \leftrightarrow \epsilon_3$. As far as the structure is concerned, this permutation is not needed. We include it for readability.} 
\begin{equation}
\mu_{\Spin{4}_S} \sim \begin{pmatrix}
        \su(2)_2 &  &  \\ 
          & \su(2)_2 & \\
&  &  0_{4}
\end{pmatrix}  \, . 
    \label{Section_Algorithm_SO4embedding}
\end{equation}
This corresponds to the  branching:
\begin{equation}
    \begin{array}{lll}
       \SU{8}  &\rightarrow & \SU{2}_S \times \SU{2}_S\times\SU{4}\times \U{1}  \\
        \rep{8}  &\rightarrow & (\rep{2},\rep{1}, \rep{1})_{1} \oplus (\rep{1},\rep{2} , \rep{1})_{1} \oplus (\rep{1},\rep{1} ,\rep{4})_{-1} \\
          \rep{63}  &\rightarrow & (\rep{1},\rep{1},\rep{1})_0 \oplus (\rep{1},\rep{1},\rep{15})_0 \oplus (\rep{1},\rep{1},\rep{1})_0 \oplus \dots  \,,
    \end{array}
\end{equation}
and 
\begin{equation}
    \begin{array}{lll}
       \Ex{7}  &\rightarrow & \SU{2}_S \times \SU{2}_S\times\SU{4}\times \U{1}  \\
          \rep{56}  &\rightarrow & (\rep{1},\rep{1},\rep{6})_{-2} \oplus (\rep{1},\rep{1},\rep{1})_2 \oplus  (\rep{1},\rep{1},\rep{1})_2 \oplus \dots \oplus  \text{c.c.}
          \\
          \rep{133}  &\rightarrow &  (\rep{1},\rep{1},\rep{1})_4 \oplus (\rep{1},\rep{1},\rep{1})_{-4} \oplus (\rep{1},\rep{1},\rep{6})_0 \oplus    (\rep{1},\rep{1},\rep{6})_0       \\
          & & \quad  \oplus (\rep{1},\rep{1},\rep{1})_0 \oplus (\rep{1},\rep{1},\rep{15})_0 \oplus (\rep{1},\rep{1},\rep{1})_0 \oplus \dots \, ,
    \end{array}
    \label{app:Section_Algorithm_E7repsBranchedUnderSO4}
\end{equation}
where $\dots$ denote non-singlets representations of $\Spin{4}_S$, 
which we have ommited for simplicity of notation. 
The numbers of vector and scalar singlets corresponds to a truncation with two vector multiplets, 
whose 12 scalars parametrize the manifold $\frac{\SO{6,2}}{\SO{6}_R\times \SO{2}}$.

\medskip

From \eqref{Section_Algorithm_SO4embedding}, one immediately realises that the diagonal combination, $\SU{2}_{S,\text{diag}}$, of the two  $\SU{2}$'s also preserves the  same amount of supersymmetries. This diagonal $\SU{2}_{S,\text{diag}}$  corresponds to the branching
\begin{equation}
    \begin{array}{lll}
       \SU{8}  &\rightarrow & \SU{2}_{S,\text{diag}}\times\SU{4}\times \U{1}  \\
        \rep{8}  &\rightarrow & (\rep{2},\rep{1})_{1} \oplus (\rep{2},\rep{1})_1 \oplus  (\rep{1},\rep{4})_{-1} 
       \\
          \rep{63}  &\rightarrow & (\rep{1},\rep{1})_0 \oplus (\rep{1},\rep{15})_0 \oplus (\rep{1},\rep{1})_0 \oplus (\rep{1},\rep{1})_0 \oplus (\rep{1},\rep{1})_0 \oplus \dots  \, ,
    \end{array}
\end{equation}
and
\begin{equation}
    \begin{array}{lll}
       \Ex{7}  &\rightarrow & \SU{2}_{S,\text{diag}}\times\SU{4}\times \U{1}  \\
          \rep{56}  &\rightarrow & (\rep{1},\rep{6})_{-2} \oplus (\rep{1},\rep{1})_2 \oplus  (\rep{1},\rep{1})_2 \oplus  (\rep{1},\rep{1})_2 \oplus \dots \oplus  \text{c.c.}
          \\
          \rep{133}  &\rightarrow &  (\rep{1},\rep{1})_4 \oplus (\rep{1},\rep{1})_{-4} \oplus (\rep{1},\rep{6})_0 \oplus    (\rep{1},\rep{6})_0 \oplus    (\rep{1},\rep{6})_0       \\
          & & \quad  \oplus (\rep{1},\rep{1})_0 \oplus (\rep{1},\rep{15})_0 \oplus (\rep{1},\rep{1})_0 \oplus (\rep{1},\rep{1})_0 \oplus (\rep{1},\rep{1})_0 \oplus \dots  \, .
    \end{array}
    \label{Section_Algorithm_E7repsBranchedUnderSU2}
\end{equation}
and gives three vector multiplets, whose 18 scalars parametrize the coset $\frac{\SO{6,3}}{\SO{6}_R\times \SO{3}}$.

\medskip

Finally, the last $\Gst \subset \SU{4}_S$ group preserving $\mathcal{N}=4$ consists of  the  $\U{1}_S \subset \SU{2}_{S,\text{diag}}$. In spinor indices, it is given by\footnote{Here we keep the permutation $\epsilon_2 \leftrightarrow \epsilon_3$ introduced before.}:
\begin{equation}
    \mu_{\U{1}_S}=\begin{pmatrix}
        i & & & & \\
         & -i  & &  & \\
         &   & i & &  \\
           &   & & -i &  \\
           &   & &  &  0_4
    \end{pmatrix} \, , 
    \label{Section_Algorithm_SO2embedding}
\end{equation}
 and it  corresponds to the branching
\begin{equation}
\begin{array}{lll}
       \SU{8}  &\rightarrow & \U{1}_S \times\SU{4}\times \U{1}  \\ 
          \rep{8}  &\rightarrow &  2 \times [ (\rep{1} , \rep{1})_{1,1}
            \oplus (\rep{1} , \rep{1})_{-1,1} ] \oplus (\rep{1} , \rep{4})_{0,-1} \\
             \rep{63}  &\rightarrow &    \rep{1}_{(0,0)} \oplus \rep{15}_{(0,0)} \oplus \rep{1}_{(0,0)} \oplus \rep{1}_{(0,0)} \oplus \rep{1}_{(0,0)} \oplus \rep{1}_{(0,0)}  \\
          & & \quad \oplus \rep{1}_{(0,0)} \oplus \rep{1}_{(0,0)} \oplus \rep{1}_{(0,0)} \oplus \dots  \,, 
    \end{array}
\end{equation}
and 
\begin{equation}
    \begin{array}{lll}
       \Ex{7}  &\rightarrow & \U{1}_S \times\SU{4}\times \U{1}  \\
          \rep{56}  &\rightarrow & \rep{6}_{(0,-2)} \oplus \rep{1}_{(0,2)} \oplus  \rep{1}_{(0,2)} \oplus  \rep{1}_{(0,2)} \oplus \dots \oplus  \text{c.c.}
          \\
          \rep{133}  &\rightarrow &  \rep{1}_{(0,4)} \oplus \rep{1}_{(0,-4)} \oplus \rep{6}_{(0,0)} \oplus    \rep{6}_{(0,0)} \oplus    \rep{6}_{(0,0)} \oplus    \rep{6}_{(0,0)}       \\
          & & \quad  \oplus \rep{1}_{(0,0)} \oplus \rep{15}_{(0,0)} \oplus \rep{1}_{(0,0)} \oplus \rep{1}_{(0,0)} \oplus \rep{1}_{(0,0)} \oplus \rep{1}_{(0,0)}  \\
          & & \quad \oplus \rep{1}_{(0,0)} \oplus \rep{1}_{(0,0)} \oplus \rep{1}_{(0,0)} \oplus \dots 
    \end{array}
    \label{Section_Algorithm_E7repsBranchedUnderU1}
\end{equation}
From the singlets in the above decomposition, one sees that the $\U{1}_S$ structure preserves 4 vector multiplets on top of the gravity multiplet.
The 24 scalars in the vector multiplets parametrize the manifod $\frac{\SO{6,4}}{\SO{6}_R\times \SO{4}}$.

\medskip

These are all the unequivalent continuous $\Gst$ structures one can find with $\mathcal{N}=4$ supersymmetries. Since the field content of each truncation is fixed by the singlets of the corresponding structure, every structure can be systematically studied by solving \eqref{Preliminaries_Singlets_SingletsEquation}. The number of independent solutions to this equation provides the number of singlets in each representation and, therefore, the field content for each truncation.

\bibliographystyle{JHEP}

\bibliography{Bibliography}

\providecommand{\href}[2]{#2}\begingroup\raggedright\begin{thebibliography}{100}

\bibitem{Cassani:2010uw}
D.~Cassani, G.~Dall'Agata and A.~F. Faedo, \emph{{Type IIB Supergravity on
  Squashed Sasaki-Einstein Manifolds}},
  \href{http://dx.doi.org/10.1007/JHEP05(2010)094}{\emph{JHEP} {\bf 05} (2010)
  094}, [\href{http://arxiv.org/abs/1003.4283}{{\tt 1003.4283}}].

\bibitem{Gauntlett:2010vu}
J.~P. Gauntlett and O.~Varela, \emph{{Universal Kaluza-Klein Reductions of Type
  IIB to ${\mathcal{N}}\!=4$ Supergravity in Five Dimensions}},
  \href{http://dx.doi.org/10.1007/JHEP06(2010)081}{\emph{JHEP} {\bf 06} (2010)
  081}, [\href{http://arxiv.org/abs/1003.5642}{{\tt 1003.5642}}].

\bibitem{Gauntlett:2009zw}
J.~P. Gauntlett, S.~Kim, O.~Varela and D.~Waldram, \emph{{Consistent
  Supersymmetric Kaluza-Klein Truncations with Massive Modes}},
  \href{http://dx.doi.org/10.1088/1126-6708/2009/04/102}{\emph{JHEP} {\bf 04}
  (2009) 102}, [\href{http://arxiv.org/abs/0901.0676}{{\tt 0901.0676}}].

\bibitem{KashaniPoor:2007tr}
A.-K. Kashani-Poor, \emph{{Nearly K\"ahler Reduction}},
  \href{http://dx.doi.org/10.1088/1126-6708/2007/11/026}{\emph{JHEP} {\bf 11}
  (2007) 026}, [\href{http://arxiv.org/abs/0709.4482}{{\tt 0709.4482}}].

\bibitem{Cassani:2009ck}
D.~Cassani and A.-K. Kashani-Poor, \emph{{Exploiting ${\mathcal{N}}\!=2$ in
  Consistent Coset Reductions of Type IIA}},
  \href{http://dx.doi.org/10.1016/j.nuclphysb.2009.03.011}{\emph{Nucl. Phys.}
  {\bf B817} (2009) 25--57}, [\href{http://arxiv.org/abs/0901.4251}{{\tt
  0901.4251}}].

\bibitem{Cassani:2011fu}
D.~Cassani and P.~Koerber, \emph{{Tri-Sasakian Consistent Reduction}},
  \href{http://dx.doi.org/10.1007/JHEP01(2012)086}{\emph{JHEP} {\bf 01} (2012)
  086}, [\href{http://arxiv.org/abs/1110.5327}{{\tt 1110.5327}}].

\bibitem{Cassani:2012pj}
D.~Cassani, P.~Koerber and O.~Varela, \emph{{All Homogeneous
  ${\mathcal{N}}\!=2$ M-theory Truncations with Supersymmetric AdS4 Vacua}},
  \href{http://dx.doi.org/10.1007/JHEP11(2012)173}{\emph{JHEP} {\bf 11} (2012)
  173}, [\href{http://arxiv.org/abs/1208.1262}{{\tt 1208.1262}}].

\bibitem{Cassani:2019vcl}
D.~Cassani, G.~Josse, M.~Petrini and D.~Waldram, \emph{{Systematics of
  consistent truncations from generalised geometry}},
  \href{http://dx.doi.org/10.1007/JHEP11(2019)017}{\emph{JHEP} {\bf 11} (2019)
  017}, [\href{http://arxiv.org/abs/1907.06730}{{\tt 1907.06730}}].

\bibitem{Lee:2014mla}
K.~Lee, C.~Strickland-Constable and D.~Waldram, \emph{{Spheres, generalised
  parallelisability and consistent truncations}},
  \href{http://dx.doi.org/10.1002/prop.201700048}{\emph{Fortsch. Phys.} {\bf
  65} (2017) 1700048}, [\href{http://arxiv.org/abs/1401.3360}{{\tt
  1401.3360}}].

\bibitem{Hohm:2014qga}
O.~Hohm and H.~Samtleben, \emph{{Consistent Kaluza-Klein Truncations via
  Exceptional Field Theory}},
  \href{http://dx.doi.org/10.1007/JHEP01(2015)131}{\emph{JHEP} {\bf 01} (2015)
  131}, [\href{http://arxiv.org/abs/1410.8145}{{\tt 1410.8145}}].

\bibitem{Baguet:2015sma}
A.~Baguet, O.~Hohm and H.~Samtleben, \emph{{Consistent Type IIB Reductions to
  Maximal 5D Supergravity}},
  \href{http://dx.doi.org/10.1103/PhysRevD.92.065004}{\emph{Phys. Rev.} {\bf
  D92} (2015) 065004}, [\href{http://arxiv.org/abs/1506.01385}{{\tt
  1506.01385}}].

\bibitem{Ciceri:2016dmd}
F.~Ciceri, A.~Guarino and G.~Inverso, \emph{{The exceptional story of massive
  IIA supergravity}},
  \href{http://dx.doi.org/10.1007/JHEP08(2016)154}{\emph{JHEP} {\bf 08} (2016)
  154}, [\href{http://arxiv.org/abs/1604.08602}{{\tt 1604.08602}}].

\bibitem{Cassani:2016ncu}
D.~Cassani, O.~de~Felice, M.~Petrini, C.~Strickland-Constable and D.~Waldram,
  \emph{{Exceptional Generalised Geometry for Massive IIA and Consistent
  Reductions}}, \href{http://dx.doi.org/10.1007/JHEP08(2016)074}{\emph{JHEP}
  {\bf 08} (2016) 074}, [\href{http://arxiv.org/abs/1605.00563}{{\tt
  1605.00563}}].

\bibitem{Malek:2017njj}
E.~Malek, \emph{{Half-Maximal Supersymmetry from Exceptional Field Theory}},
  \href{http://dx.doi.org/10.1002/prop.201700061}{\emph{Fortsch. Phys.} {\bf
  65} (2017) 1700061}, [\href{http://arxiv.org/abs/1707.00714}{{\tt
  1707.00714}}].

\bibitem{Malek:2018zcz}
E.~Malek, H.~Samtleben and V.~Vall~Camell, \emph{{Supersymmetric AdS$_{7}$ and
  AdS$_6$ vacua and their minimal consistent truncations from exceptional field
  theory}}, \href{http://dx.doi.org/10.1016/j.physletb.2018.09.037}{\emph{Phys.
  Lett.} {\bf B786} (2018) 171--179},
  [\href{http://arxiv.org/abs/1808.05597}{{\tt 1808.05597}}].

\bibitem{Malek:2019ucd}
E.~Malek, H.~Samtleben and V.~Vall~Camell, \emph{{Supersymmetric AdS$_7$ and
  AdS$_6$ vacua and their consistent truncations with vector multiplets}},
  \href{http://dx.doi.org/10.1007/JHEP04(2019)088}{\emph{JHEP} {\bf 04} (2019)
  088}, [\href{http://arxiv.org/abs/1901.11039}{{\tt 1901.11039}}].

\bibitem{Cassani:2020cod}
D.~Cassani, G.~Josse, M.~Petrini and D.~Waldram, \emph{{$\mathcal{N} $ = 2
  consistent truncations from wrapped M5-branes}},
  \href{http://dx.doi.org/10.1007/JHEP02(2021)232}{\emph{JHEP} {\bf 02} (2021)
  232}, [\href{http://arxiv.org/abs/2011.04775}{{\tt 2011.04775}}].

\bibitem{Josse:2021put}
G.~Josse, E.~Malek, M.~Petrini and D.~Waldram, \emph{{The higher-dimensional
  origin of five-dimensional $ \mathcal{N} $ = 2 gauged supergravities}},
  \href{http://dx.doi.org/10.1007/JHEP06(2022)003}{\emph{JHEP} {\bf 06} (2022)
  003}, [\href{http://arxiv.org/abs/2112.03931}{{\tt 2112.03931}}].

\bibitem{Malek:2016bpu}
E.~Malek, \emph{{7-dimensional ${\cal N}=2$ Consistent Truncations using
  $\mathrm{SL}(5)$ Exceptional Field Theory}},
  \href{http://dx.doi.org/10.1007/JHEP06(2017)026}{\emph{JHEP} {\bf 06} (2017)
  026}, [\href{http://arxiv.org/abs/1612.01692}{{\tt 1612.01692}}].

\bibitem{Ashmore:2015joa}
A.~Ashmore and D.~Waldram, \emph{{Exceptional Calabi-Yau spaces: the geometry
  of $\mathcal{N}=2$ backgrounds with flux}},
  \href{http://dx.doi.org/10.1002/prop.201600109}{\emph{Fortsch. Phys.} {\bf
  65} (2017) 1600109}, [\href{http://arxiv.org/abs/1510.00022}{{\tt
  1510.00022}}].

\bibitem{Hull:2007zu}
C.~M. Hull, \emph{{Generalised Geometry for M-theory}},
  \href{http://dx.doi.org/10.1088/1126-6708/2007/07/079}{\emph{JHEP} {\bf 07}
  (2007) 079}, [\href{http://arxiv.org/abs/hep-th/0701203}{{\tt
  hep-th/0701203}}].

\bibitem{Pacheco:2008ps}
P.~Pires~Pacheco and D.~Waldram, \emph{{M-theory, Exceptional Generalised
  Geometry and Superpotentials}},
  \href{http://dx.doi.org/10.1088/1126-6708/2008/09/123}{\emph{JHEP} {\bf 09}
  (2008) 123}, [\href{http://arxiv.org/abs/0804.1362}{{\tt 0804.1362}}].

\bibitem{Coimbra:2011ky}
A.~Coimbra, C.~Strickland-Constable and D.~Waldram, \emph{{$E_{d(d)} \times
  \mathbb{R}^+$ generalised geometry, connections and M theory}},
  \href{http://dx.doi.org/10.1007/JHEP02(2014)054}{\emph{JHEP} {\bf 02} (2014)
  054}, [\href{http://arxiv.org/abs/1112.3989}{{\tt 1112.3989}}].

\bibitem{Coimbra:2014uxa}
A.~Coimbra, C.~Strickland-Constable and D.~Waldram, \emph{{Supersymmetric
  Backgrounds and Generalised Special Holonomy}},
  \href{http://dx.doi.org/10.1088/0264-9381/33/12/125026}{\emph{Class. Quant.
  Grav.} {\bf 33} (2016) 125026}, [\href{http://arxiv.org/abs/1411.5721}{{\tt
  1411.5721}}].

\bibitem{Samtleben:2008pe}
H.~Samtleben, \emph{{Lectures on Gauged Supergravity and Flux
  Compactifications}},
  \href{http://dx.doi.org/10.1088/0264-9381/25/21/214002}{\emph{Class. Quant.
  Grav.} {\bf 25} (2008) 214002}, [\href{http://arxiv.org/abs/0808.4076}{{\tt
  0808.4076}}].

\bibitem{Trigiante:2016mnt}
M.~Trigiante, \emph{{Gauged Supergravities}},
  \href{http://dx.doi.org/10.1016/j.physrep.2017.03.001}{\emph{Phys. Rept.}
  {\bf 680} (2017) 1--175}, [\href{http://arxiv.org/abs/1609.09745}{{\tt
  1609.09745}}].

\bibitem{Lee2014mla}
K.~Lee, C.~Strickland-Constable and D.~Waldram, \emph{{Spheres, generalised
  parallelisability and consistent truncations}},
  \href{http://dx.doi.org/10.1002/prop.201700048}{\emph{Fortsch. Phys.} {\bf
  65} (2017) 1700048}, [\href{http://arxiv.org/abs/1401.3360}{{\tt
  1401.3360}}].

\bibitem{Cremmer:1978ds}
E.~Cremmer and B.~Julia, \emph{{The N=8 Supergravity Theory. 1. The
  Lagrangian}},
  \href{http://dx.doi.org/10.1016/0370-2693(78)90303-9}{\emph{Phys. Lett. B}
  {\bf 80} (1978) 48}.

\bibitem{deWit:2002vt}
B.~de~Wit, H.~Samtleben and M.~Trigiante, \emph{{On Lagrangians and gaugings of
  maximal supergravities}},
  \href{http://dx.doi.org/10.1016/S0550-3213(03)00059-2}{\emph{Nucl. Phys. B}
  {\bf 655} (2003) 93--126}, [\href{http://arxiv.org/abs/hep-th/0212239}{{\tt
  hep-th/0212239}}].

\bibitem{deWit:2007kvg}
B.~de~Wit, H.~Samtleben and M.~Trigiante, \emph{{The Maximal D=4
  supergravities}},
  \href{http://dx.doi.org/10.1088/1126-6708/2007/06/049}{\emph{JHEP} {\bf 06}
  (2007) 049}, [\href{http://arxiv.org/abs/0705.2101}{{\tt 0705.2101}}].

\bibitem{Baron:2014yua}
W.~H. Baron, \emph{{Gaugings from $E_{7(7)}$ extended geometries}},
  \href{http://dx.doi.org/10.1103/PhysRevD.91.024008}{\emph{Phys. Rev.} {\bf
  D91} (2015) 024008}, [\href{http://arxiv.org/abs/1404.7750}{{\tt
  1404.7750}}].

\bibitem{Inverso:2017lrz}
G.~Inverso, \emph{{Generalised Scherk-Schwarz reductions from gauged
  supergravity}}, \href{http://dx.doi.org/10.1007/JHEP12(2017)124}{\emph{JHEP}
  {\bf 12} (2017) 124}, [\href{http://arxiv.org/abs/1708.02589}{{\tt
  1708.02589}}].

\bibitem{deWit:1982bul}
B.~de~Wit and H.~Nicolai, \emph{{N=8 Supergravity}},
  \href{http://dx.doi.org/10.1016/0550-3213(82)90120-1}{\emph{Nucl. Phys. B}
  {\bf 208} (1982) 323}.

\bibitem{deWit:1986oxb}
B.~de~Wit and H.~Nicolai, \emph{{The Consistency of the S**7 Truncation in D=11
  Supergravity}},
  \href{http://dx.doi.org/10.1016/0550-3213(87)90253-7}{\emph{Nucl. Phys. B}
  {\bf 281} (1987) 211--240}.

\bibitem{Guarino:2015vca}
A.~Guarino and O.~Varela, \emph{{Consistent $ \mathcal{N}=8 $ truncation of
  massive IIA on S$^{6}$}},
  \href{http://dx.doi.org/10.1007/JHEP12(2015)020}{\emph{JHEP} {\bf 12} (2015)
  020}, [\href{http://arxiv.org/abs/1509.02526}{{\tt 1509.02526}}].

\bibitem{Nilsson:1984bj}
B.~E.~W. Nilsson and C.~N. Pope, \emph{{Hopf Fibration of Eleven-dimensional
  Supergravity}},
  \href{http://dx.doi.org/10.1088/0264-9381/1/5/005}{\emph{Class. Quant. Grav.}
  {\bf 1} (1984) 499}.

\bibitem{deWit:1981yv}
B.~de~Wit and H.~Nicolai, \emph{{Extended Supergravity With Local SO(5)
  Invariance}},
  \href{http://dx.doi.org/10.1016/0550-3213(81)90107-3}{\emph{Nucl. Phys. B}
  {\bf 188} (1981) 98--108}.

\bibitem{Schon:2006kz}
J.~Schon and M.~Weidner, \emph{{Gauged N=4 supergravities}},
  \href{http://dx.doi.org/10.1088/1126-6708/2006/05/034}{\emph{JHEP} {\bf 05}
  (2006) 034}, [\href{http://arxiv.org/abs/hep-th/0602024}{{\tt
  hep-th/0602024}}].

\bibitem{Das:1977uy}
A.~K. Das, \emph{{SO(4) Invariant Extended Supergravity}},
  \href{http://dx.doi.org/10.1103/PhysRevD.15.2805}{\emph{Phys. Rev. D} {\bf
  15} (1977) 2805}.

\bibitem{Cremmer:1977tc}
E.~Cremmer and J.~Scherk, \emph{{Algebraic Simplifications in Supergravity
  Theories}}, \href{http://dx.doi.org/10.1016/0550-3213(77)90214-0}{\emph{Nucl.
  Phys. B} {\bf 127} (1977) 259--268}.

\bibitem{Cremmer:1977tt}
E.~Cremmer, J.~Scherk and S.~Ferrara, \emph{{SU(4) Invariant Supergravity
  Theory}}, \href{http://dx.doi.org/10.1016/0370-2693(78)90060-6}{\emph{Phys.
  Lett. B} {\bf 74} (1978) 61--64}.

\bibitem{Freedman:1978ra}
D.~Z. Freedman and J.~H. Schwarz, \emph{{N=4 Supergravity Theory with Local
  SU(2) x SU(2) Invariance}},
  \href{http://dx.doi.org/10.1016/0550-3213(78)90526-6}{\emph{Nucl. Phys. B}
  {\bf 137} (1978) 333--339}.

\bibitem{deRoo:1984zyh}
M.~de~Roo, \emph{{Matter Coupling in N=4 Supergravity}},
  \href{http://dx.doi.org/10.1016/0550-3213(85)90151-8}{\emph{Nucl. Phys. B}
  {\bf 255} (1985) 515--531}.

\bibitem{Bergshoeff:1985ms}
E.~Bergshoeff, I.~G. Koh and E.~Sezgin, \emph{{Coupling of Yang-Mills to N=4,
  D=4 Supergravity}},
  \href{http://dx.doi.org/10.1016/0370-2693(85)91034-2}{\emph{Phys. Lett. B}
  {\bf 155} (1985) 71}.

\bibitem{deRoo:1985jh}
M.~de~Roo and P.~Wagemans, \emph{{Gauge Matter Coupling in $N=4$
  Supergravity}},
  \href{http://dx.doi.org/10.1016/0550-3213(85)90509-7}{\emph{Nucl. Phys. B}
  {\bf 262} (1985) 644}.

\bibitem{DallAgata:2023ahj}
G.~Dall'Agata, N.~Liatsos, R.~Noris and M.~Trigiante, \emph{{Gauged D = 4 $
  \mathcal{N} $ = 4 supergravity}},
  \href{http://dx.doi.org/10.1007/JHEP09(2023)071}{\emph{JHEP} {\bf 09} (2023)
  071}, [\href{http://arxiv.org/abs/2305.04015}{{\tt 2305.04015}}].

\bibitem{Cvetic:1999au}
M.~Cvetic, H.~Lu and C.~N. Pope, \emph{{Four-dimensional N=4, SO(4) gauged
  supergravity from D = 11}},
  \href{http://dx.doi.org/10.1016/S0550-3213(99)00828-7}{\emph{Nucl. Phys. B}
  {\bf 574} (2000) 761--781}, [\href{http://arxiv.org/abs/hep-th/9910252}{{\tt
  hep-th/9910252}}].

\bibitem{Guarino:2024gke}
A.~Guarino, C.~Sterckx and M.~Trigiante, \emph{{Consistent N=4, D=4 truncation
  of type IIB supergravity on S1\texttimes{}S5}},
  \href{http://dx.doi.org/10.1103/PhysRevD.111.046019}{\emph{Phys. Rev. D} {\bf
  111} (2025) 046019}, [\href{http://arxiv.org/abs/2410.23149}{{\tt
  2410.23149}}].

\bibitem{Duboeuf:2023dmq}
B.~Duboeuf, M.~Galli, E.~Malek and H.~Samtleben, \emph{{Holographic RG flow
  from the squashed to the round S7}},
  \href{http://dx.doi.org/10.1103/PhysRevD.108.086002}{\emph{Phys. Rev. D} {\bf
  108} (2023) 086002}, [\href{http://arxiv.org/abs/2306.11789}{{\tt
  2306.11789}}].

\bibitem{Castellani:1985ka}
L.~Castellani, A.~Ceresole, S.~Ferrara, R.~D'Auria, P.~Fre and E.~Maina,
  \emph{{The Complete $N=3$ Matter Coupled Supergravity}},
  \href{http://dx.doi.org/10.1016/0550-3213(86)90157-4}{\emph{Nucl. Phys. B}
  {\bf 268} (1986) 317--348}.

\bibitem{Karndumri:2016miq}
P.~Karndumri and K.~Upathambhakul, \emph{{Gaugings of four-dimensional N=3
  supergravity and AdS4/CFT3 holography}},
  \href{http://dx.doi.org/10.1103/PhysRevD.93.125017}{\emph{Phys. Rev. D} {\bf
  93} (2016) 125017}, [\href{http://arxiv.org/abs/1602.02254}{{\tt
  1602.02254}}].

\bibitem{Duff:1999gh}
M.~J. Duff and J.~T. Liu, \emph{{Anti-de Sitter black holes in gauged N = 8
  supergravity}},
  \href{http://dx.doi.org/10.1016/S0550-3213(99)00299-0}{\emph{Nucl. Phys. B}
  {\bf 554} (1999) 237--253}, [\href{http://arxiv.org/abs/hep-th/9901149}{{\tt
  hep-th/9901149}}].

\bibitem{Andrianopoli:1997wi}
L.~Andrianopoli, R.~D'Auria, S.~Ferrara, P.~Fre and M.~Trigiante,
  \emph{{E(7)(7) duality, BPS black hole evolution and fixed scalars}},
  \href{http://dx.doi.org/10.1016/S0550-3213(97)00675-5}{\emph{Nucl. Phys. B}
  {\bf 509} (1998) 463--518}, [\href{http://arxiv.org/abs/hep-th/9707087}{{\tt
  hep-th/9707087}}].

\bibitem{Bertolini:1999uz}
M.~Bertolini and M.~Trigiante, \emph{{Regular RR and NS NS BPS black holes}},
  \href{http://dx.doi.org/10.1142/S0217751X00002078}{\emph{Int. J. Mod. Phys.
  A} {\bf 15} (2000) 5017}, [\href{http://arxiv.org/abs/hep-th/9910237}{{\tt
  hep-th/9910237}}].

\bibitem{Azizi:2016noi}
A.~Azizi, H.~Godazgar, M.~Godazgar and C.~N. Pope, \emph{{Embedding of gauged
  STU supergravity in eleven dimensions}},
  \href{http://dx.doi.org/10.1103/PhysRevD.94.066003}{\emph{Phys. Rev. D} {\bf
  94} (2016) 066003}, [\href{http://arxiv.org/abs/1606.06954}{{\tt
  1606.06954}}].

\bibitem{Gauntlett:2009bh}
J.~P. Gauntlett, J.~Sonner and T.~Wiseman, \emph{{Quantum Criticality and
  Holographic Superconductors in M-theory}},
  \href{http://dx.doi.org/10.1007/JHEP02(2010)060}{\emph{JHEP} {\bf 02} (2010)
  060}, [\href{http://arxiv.org/abs/0912.0512}{{\tt 0912.0512}}].

\bibitem{Blair:2024ofc}
C.~D.~A. Blair, M.~Pico and O.~Varela, \emph{{Infinite and finite consistent
  truncations on deformed generalised parallelisations}},
  \href{http://dx.doi.org/10.1007/JHEP09(2024)065}{\emph{JHEP} {\bf 09} (2024)
  065}, [\href{http://arxiv.org/abs/2407.01298}{{\tt 2407.01298}}].

\bibitem{Donos:2010ax}
A.~Donos, J.~P. Gauntlett, N.~Kim and O.~Varela, \emph{{Wrapped M5-branes,
  consistent truncations and AdS/CMT}},
  \href{http://dx.doi.org/10.1007/JHEP12(2010)003}{\emph{JHEP} {\bf 12} (2010)
  003}, [\href{http://arxiv.org/abs/1009.3805}{{\tt 1009.3805}}].

\bibitem{Slansky:1981yr}
R.~Slansky, \emph{{Group Theory for Unified Model Building}},
  \href{http://dx.doi.org/10.1016/0370-1573(81)90092-2}{\emph{Phys. Rept.} {\bf
  79} (1981) 1--128}.

\bibitem{Yamatsu:2015npn}
N.~Yamatsu, \emph{{Finite-Dimensional Lie Algebras and Their Representations
  for Unified Model Building}},  \href{http://arxiv.org/abs/1511.08771}{{\tt
  1511.08771}}.

\bibitem{Cremmer:1980gs}
E.~Cremmer, \emph{{Supergravities in 5 Dimensions}},  8, 1980.

\bibitem{Gunaydin:1984qu}
M.~Gunaydin, L.~J. Romans and N.~P. Warner, \emph{{Gauged N=8 Supergravity in
  Five-Dimensions}},
  \href{http://dx.doi.org/10.1016/0370-2693(85)90361-2}{\emph{Phys. Lett. B}
  {\bf 154} (1985) 268--274}.

\bibitem{Gunaydin:1985cu}
M.~Gunaydin, L.~J. Romans and N.~P. Warner, \emph{{Compact and Noncompact
  Gauged Supergravity Theories in Five-Dimensions}},
  \href{http://dx.doi.org/10.1016/0550-3213(86)90237-3}{\emph{Nucl. Phys. B}
  {\bf 272} (1986) 598--646}.

\bibitem{Pernici:1985ju}
M.~Pernici, K.~Pilch and P.~van Nieuwenhuizen, \emph{{Gauged N=8 D=5
  Supergravity}},
  \href{http://dx.doi.org/10.1016/0550-3213(85)90645-5}{\emph{Nucl. Phys. B}
  {\bf 259} (1985) 460}.

\bibitem{Ferrara:1998zt}
S.~Ferrara, M.~Porrati and A.~Zaffaroni, \emph{{N=6 supergravity on AdS(5) and
  the SU(2,2/3) superconformal correspondence}},
  \href{http://dx.doi.org/10.1023/A:1007592711262}{\emph{Lett. Math. Phys.}
  {\bf 47} (1999) 255--263}, [\href{http://arxiv.org/abs/hep-th/9810063}{{\tt
  hep-th/9810063}}].

\bibitem{Liu:2019cea}
J.~T. Liu and B.~McPeak, \emph{{Gauged Supergravity from the Lunin-Maldacena
  background}},  \href{http://arxiv.org/abs/1905.06861}{{\tt 1905.06861}}.

\bibitem{Maldacena:2000mw}
J.~M. Maldacena and C.~Nunez, \emph{{Supergravity description of field theories
  on curved manifolds and a no go theorem}},
  \href{http://dx.doi.org/10.1142/S0217751X01003935,
  10.1142/S0217751X01003937}{\emph{Int. J. Mod. Phys.} {\bf A16} (2001)
  822--855}, [\href{http://arxiv.org/abs/hep-th/0007018}{{\tt
  hep-th/0007018}}].

\bibitem{Cheung:2019pge}
K.~C.~M. Cheung, J.~P. Gauntlett and C.~Rosen, \emph{{Consistent KK truncations
  for M5-branes wrapped on Riemann surfaces}},
  \href{http://arxiv.org/abs/1906.08900}{{\tt 1906.08900}}.

\bibitem{Bhattacharya:2024tjw}
R.~Bhattacharya, A.~Katyal and O.~Varela, \emph{{Class S Superconformal Indices
  from Maximal Supergravity}},
  \href{http://dx.doi.org/10.1103/PhysRevLett.134.181601}{\emph{Phys. Rev.
  Lett.} {\bf 134} (2025) 181601}, [\href{http://arxiv.org/abs/2411.16837}{{\tt
  2411.16837}}].

\bibitem{Varela:2025xeb}
O.~Varela, \emph{{Trombone gaugings of five-dimensional maximal supergravity}},
   \href{http://arxiv.org/abs/2509.12391}{{\tt 2509.12391}}.

\bibitem{Gunaydin:1999zx}
M.~Gunaydin and M.~Zagermann, \emph{{The Gauging of five-dimensional, N=2
  Maxwell-Einstein supergravity theories coupled to tensor multiplets}},
  \href{http://dx.doi.org/10.1016/S0550-3213(99)00801-9}{\emph{Nucl. Phys. B}
  {\bf 572} (2000) 131--150}, [\href{http://arxiv.org/abs/hep-th/9912027}{{\tt
  hep-th/9912027}}].

\bibitem{Ceresole:2000jd}
A.~Ceresole and G.~Dall'Agata, \emph{{General matter coupled N=2, D = 5 gauged
  supergravity}},
  \href{http://dx.doi.org/10.1016/S0550-3213(00)00339-4}{\emph{Nucl. Phys. B}
  {\bf 585} (2000) 143--170}, [\href{http://arxiv.org/abs/hep-th/0004111}{{\tt
  hep-th/0004111}}].

\bibitem{Bergshoeff:2002qk}
E.~Bergshoeff, S.~Cucu, T.~De~Wit, J.~Gheerardyn, R.~Halbersma, S.~Vandoren
  et~al., \emph{{Superconformal N=2, D = 5 matter with and without actions}},
  \href{http://dx.doi.org/10.1088/1126-6708/2002/10/045}{\emph{JHEP} {\bf 10}
  (2002) 045}, [\href{http://arxiv.org/abs/hep-th/0205230}{{\tt
  hep-th/0205230}}].

\bibitem{Faedo:2019cvr}
A.~F. Faedo, C.~Nunez and C.~Rosen, \emph{{Consistent truncations of
  supergravity and $\frac{1}{2}$-BPS RG flows in $4d$ SCFTs}},
  \href{http://dx.doi.org/10.1007/JHEP03(2020)080}{\emph{JHEP} {\bf 03} (2020)
  080}, [\href{http://arxiv.org/abs/1912.13516}{{\tt 1912.13516}}].

\bibitem{Bah:2012dg}
I.~Bah, C.~Beem, N.~Bobev and B.~Wecht, \emph{{Four-Dimensional SCFTs from
  M5-Branes}}, \href{http://dx.doi.org/10.1007/JHEP06(2012)005}{\emph{JHEP}
  {\bf 06} (2012) 005}, [\href{http://arxiv.org/abs/1203.0303}{{\tt
  1203.0303}}].

\bibitem{Szepietowski:2012tb}
P.~Szepietowski, \emph{{Comments on a-maximization from gauged supergravity}},
  \href{http://dx.doi.org/10.1007/JHEP12(2012)018}{\emph{JHEP} {\bf 12} (2012)
  018}, [\href{http://arxiv.org/abs/1209.3025}{{\tt 1209.3025}}].

\bibitem{Tanii:1998px}
Y.~Tanii, \emph{{Introduction to Supergravities in Diverse Dimensions}},  in
  \emph{{Yitp Workshop on Supersymmetry Kyoto, Japan, March 27-30, 1996}},
  1998.
\newblock \href{http://arxiv.org/abs/hep-th/9802138}{{\tt hep-th/9802138}}.

\bibitem{Bergshoeff:2007ef}
E.~Bergshoeff, H.~Samtleben and E.~Sezgin, \emph{{The Gaugings of Maximal D=6
  Supergravity}},
  \href{http://dx.doi.org/10.1088/1126-6708/2008/03/068}{\emph{JHEP} {\bf 03}
  (2008) 068}, [\href{http://arxiv.org/abs/0712.4277}{{\tt 0712.4277}}].

\bibitem{Cowdall:1998rs}
P.~M. Cowdall, \emph{{On gauged maximal supergravity in six-dimensions}},
  \href{http://dx.doi.org/10.1088/1126-6708/1999/06/018}{\emph{JHEP} {\bf 06}
  (1999) 018}, [\href{http://arxiv.org/abs/hep-th/9810041}{{\tt
  hep-th/9810041}}].

\bibitem{Hull:2000zn}
C.~M. Hull, \emph{{Strongly coupled gravity and duality}},
  \href{http://dx.doi.org/10.1016/S0550-3213(00)00323-0}{\emph{Nucl. Phys. B}
  {\bf 583} (2000) 237--259}, [\href{http://arxiv.org/abs/hep-th/0004195}{{\tt
  hep-th/0004195}}].

\bibitem{Hull:2000rr}
C.~M. Hull, \emph{{Symmetries and compactifications of (4,0) conformal
  gravity}}, \href{http://dx.doi.org/10.1088/1126-6708/2000/12/007}{\emph{JHEP}
  {\bf 12} (2000) 007}, [\href{http://arxiv.org/abs/hep-th/0011215}{{\tt
  hep-th/0011215}}].

\bibitem{Bertrand:2022pyi}
Y.~Bertrand, S.~Hohenegger, O.~Hohm and H.~Samtleben, \emph{{Supersymmetric
  action for 6D (4, 0) supergravity}},
  \href{http://dx.doi.org/10.1007/JHEP08(2022)255}{\emph{JHEP} {\bf 08} (2022)
  255}, [\href{http://arxiv.org/abs/2206.04100}{{\tt 2206.04100}}].

\bibitem{Bertrand:2020nob}
Y.~Bertrand, S.~Hohenegger, O.~Hohm and H.~Samtleben, \emph{{Toward exotic 6D
  supergravities}},
  \href{http://dx.doi.org/10.1103/PhysRevD.103.046002}{\emph{Phys. Rev. D} {\bf
  103} (2021) 046002}, [\href{http://arxiv.org/abs/2007.11644}{{\tt
  2007.11644}}].

\bibitem{Cvetic:2000ah}
M.~Cvetic, H.~Lu, C.~N. Pope, A.~Sadrzadeh and T.~A. Tran, \emph{{S**3 and S**4
  reductions of type IIA supergravity}},
  \href{http://dx.doi.org/10.1016/S0550-3213(00)00466-1}{\emph{Nucl. Phys. B}
  {\bf 590} (2000) 233--251}, [\href{http://arxiv.org/abs/hep-th/0005137}{{\tt
  hep-th/0005137}}].

\bibitem{Julia81}
B.~Julia in \emph{Superspace and Supergravity} (S.~W. Hawking and M.~Rocek,
  eds.), p.~331, Cambridge, 1981.

\bibitem{DAuria:1997caz}
R.~D'Auria, S.~Ferrara and C.~Kounnas, \emph{{N = (4,2) chiral supergravity in
  six-dimensions and solvable Lie algebras}},
  \href{http://dx.doi.org/10.1016/S0370-2693(97)01508-6}{\emph{Phys. Lett. B}
  {\bf 420} (1998) 289--299}, [\href{http://arxiv.org/abs/hep-th/9711048}{{\tt
  hep-th/9711048}}].

\bibitem{Bergshoeff:1986hv}
E.~Bergshoeff, T.~W. Kephart, A.~Salam and E.~Sezgin, \emph{{Global Anomalies
  in Six-dimensions}},
  \href{http://dx.doi.org/10.1142/S021773238600035X}{\emph{Mod. Phys. Lett. A}
  {\bf 1} (1986) 267}.

\bibitem{DAuria:2000afl}
R.~D'Auria, S.~Ferrara and S.~Vaula, \emph{{Matter coupled F(4) supergravity
  and the AdS(6) / CFT(5) correspondence}},
  \href{http://dx.doi.org/10.1088/1126-6708/2000/10/013}{\emph{JHEP} {\bf 10}
  (2000) 013}, [\href{http://arxiv.org/abs/hep-th/0006107}{{\tt
  hep-th/0006107}}].

\bibitem{Giani:1984dw}
F.~Giani, M.~Pernici and P.~van Nieuwenhuizen, \emph{{GAUGED N=4 d = 6
  SUPERGRAVITY}}, \href{http://dx.doi.org/10.1103/PhysRevD.30.1680}{\emph{Phys.
  Rev. D} {\bf 30} (1984) 1680}.

\bibitem{Romans:1985tw}
L.~J. Romans, \emph{{The F(4) Gauged Supergravity in Six-dimensions}},
  \href{http://dx.doi.org/10.1016/0550-3213(86)90517-1}{\emph{Nucl. Phys. B}
  {\bf 269} (1986) 691}.

\bibitem{Karndumri:2016ruc}
P.~Karndumri and J.~Louis, \emph{{Supersymmetric $AdS_6$ vacua in
  six-dimensional $N=(1,1)$ gauged supergravity}},
  \href{http://dx.doi.org/10.1007/JHEP01(2017)069}{\emph{JHEP} {\bf 01} (2017)
  069}, [\href{http://arxiv.org/abs/1612.00301}{{\tt 1612.00301}}].

\bibitem{Bergshoeff:2007vb}
E.~A. Bergshoeff, J.~Gomis, T.~A. Nutma and D.~Roest, \emph{{Kac-Moody Spectrum
  of (Half-)Maximal Supergravities}},
  \href{http://dx.doi.org/10.1088/1126-6708/2008/02/069}{\emph{JHEP} {\bf 02}
  (2008) 069}, [\href{http://arxiv.org/abs/0711.2035}{{\tt 0711.2035}}].

\bibitem{Townsend:1983xt}
P.~K. Townsend, \emph{{A New Anomaly Free Chiral Supergravity Theory From
  Compactification on K3}},
  \href{http://dx.doi.org/10.1016/0370-2693(84)91081-5}{\emph{Phys. Lett. B}
  {\bf 139} (1984) 283--287}.

\bibitem{Romans:1986er}
L.~J. Romans, \emph{{Selfduality for Interacting Fields: Covariant Field
  Equations for Six-dimensional Chiral Supergravities}},
  \href{http://dx.doi.org/10.1016/0550-3213(86)90016-7}{\emph{Nucl. Phys. B}
  {\bf 276} (1986) 71}.

\bibitem{Riccioni:1997np}
F.~Riccioni, \emph{{Tensor multiplets in six-dimensional (2,0) supergravity}},
  \href{http://dx.doi.org/10.1016/S0370-2693(98)00070-7}{\emph{Phys. Lett. B}
  {\bf 422} (1998) 126--134}, [\href{http://arxiv.org/abs/hep-th/9712176}{{\tt
  hep-th/9712176}}].

\bibitem{Cvetic:1999un}
M.~Cvetic, H.~Lu and C.~N. Pope, \emph{{Gauged six-dimensional supergravity
  from massive type IIA}},
  \href{http://dx.doi.org/10.1103/PhysRevLett.83.5226}{\emph{Phys. Rev. Lett.}
  {\bf 83} (1999) 5226--5229}, [\href{http://arxiv.org/abs/hep-th/9906221}{{\tt
  hep-th/9906221}}].

\bibitem{Hong:2018amk}
J.~Hong, J.~T. Liu and D.~R. Mayerson, \emph{{Gauged Six-Dimensional
  Supergravity from Warped IIB Reductions}},
  \href{http://dx.doi.org/10.1007/JHEP09(2018)140}{\emph{JHEP} {\bf 09} (2018)
  140}, [\href{http://arxiv.org/abs/1808.04301}{{\tt 1808.04301}}].

\bibitem{Jeong:2013jfc}
J.~Jeong, O.~Kelekci and E.~O~Colgain, \emph{{An alternative IIB embedding of
  F(4) gauged supergravity}},
  \href{http://dx.doi.org/10.1007/JHEP05(2013)079}{\emph{JHEP} {\bf 05} (2013)
  079}, [\href{http://arxiv.org/abs/1302.2105}{{\tt 1302.2105}}].

\bibitem{Ale75}
D.~Alekseevskii, \emph{{Classification of quaternionic spaces with a~transitive
  solvable group of motions}}, {\emph{Math. USSR-Izv.} {\bf 9} (1975) 297}.

\bibitem{deWit:1991nm}
B.~de~Wit and A.~Van~Proeyen, \emph{{Special geometry, cubic polynomials and
  homogeneous quaternionic spaces}},
  \href{http://dx.doi.org/10.1007/BF02097627}{\emph{Commun. Math. Phys.} {\bf
  149} (1992) 307--334}, [\href{http://arxiv.org/abs/hep-th/9112027}{{\tt
  hep-th/9112027}}].

\bibitem{Nishino:1984gk}
H.~Nishino and E.~Sezgin, \emph{{Matter and Gauge Couplings of N=2 Supergravity
  in Six-Dimensions}},
  \href{http://dx.doi.org/10.1016/0370-2693(84)91800-8}{\emph{Phys. Lett. B}
  {\bf 144} (1984) 187--192}.

\bibitem{Nishino:1986dc}
H.~Nishino and E.~Sezgin, \emph{{The Complete $N=2$, $d=6$ Supergravity With
  Matter and {Yang-Mills} Couplings}},
  \href{http://dx.doi.org/10.1016/0550-3213(86)90218-X}{\emph{Nucl. Phys. B}
  {\bf 278} (1986) 353--379}.

\bibitem{Nishino:1997ff}
H.~Nishino and E.~Sezgin, \emph{{New couplings of six-dimensional
  supergravity}},
  \href{http://dx.doi.org/10.1016/S0550-3213(97)00357-X}{\emph{Nucl. Phys. B}
  {\bf 505} (1997) 497--516}, [\href{http://arxiv.org/abs/hep-th/9703075}{{\tt
  hep-th/9703075}}].

\bibitem{Ferrara:1997gh}
S.~Ferrara, F.~Riccioni and A.~Sagnotti, \emph{{Tensor and vector multiplets in
  six-dimensional supergravity}},
  \href{http://dx.doi.org/10.1016/S0550-3213(97)00837-7}{\emph{Nucl. Phys. B}
  {\bf 519} (1998) 115--140}, [\href{http://arxiv.org/abs/hep-th/9711059}{{\tt
  hep-th/9711059}}].

\bibitem{Riccioni:2001bg}
F.~Riccioni, \emph{{All couplings of minimal six-dimensional supergravity}},
  \href{http://dx.doi.org/10.1016/S0550-3213(01)00199-7}{\emph{Nucl. Phys. B}
  {\bf 605} (2001) 245--265}, [\href{http://arxiv.org/abs/hep-th/0101074}{{\tt
  hep-th/0101074}}].

\bibitem{Gunaydin:2010fi}
M.~Gunaydin, H.~Samtleben and E.~Sezgin, \emph{{On the Magical Supergravities
  in Six Dimensions}},
  \href{http://dx.doi.org/10.1016/j.nuclphysb.2011.02.010}{\emph{Nucl. Phys. B}
  {\bf 848} (2011) 62--89}, [\href{http://arxiv.org/abs/1012.1818}{{\tt
  1012.1818}}].

\bibitem{Randjbar-Daemi:1985tdc}
S.~Randjbar-Daemi, A.~Salam, E.~Sezgin and J.~A. Strathdee, \emph{{An Anomaly
  Free Model in Six-Dimensions}},
  \href{http://dx.doi.org/10.1016/0370-2693(85)91653-3}{\emph{Phys. Lett. B}
  {\bf 151} (1985) 351--356}.

\bibitem{Avramis:2005qt}
S.~D. Avramis, A.~Kehagias and S.~Randjbar-Daemi, \emph{{A New anomaly-free
  gauged supergravity in six dimensions}},
  \href{http://dx.doi.org/10.1088/1126-6708/2005/05/057}{\emph{JHEP} {\bf 05}
  (2005) 057}, [\href{http://arxiv.org/abs/hep-th/0504033}{{\tt
  hep-th/0504033}}].

\bibitem{Avramis:2005hc}
S.~D. Avramis and A.~Kehagias, \emph{{A Systematic search for anomaly-free
  supergravities in six dimensions}},
  \href{http://dx.doi.org/10.1088/1126-6708/2005/10/052}{\emph{JHEP} {\bf 10}
  (2005) 052}, [\href{http://arxiv.org/abs/hep-th/0508172}{{\tt
  hep-th/0508172}}].

\bibitem{Suzuki:2005vu}
R.~Suzuki and Y.~Tachikawa, \emph{{More anomaly-free models of six-dimensional
  gauged supergravity}}, \href{http://dx.doi.org/10.1063/1.2209767}{\emph{J.
  Math. Phys.} {\bf 47} (2006) 062302},
  [\href{http://arxiv.org/abs/hep-th/0512019}{{\tt hep-th/0512019}}].

\bibitem{Kumar:2009ae}
V.~Kumar and W.~Taylor, \emph{{A Bound on 6D N=1 supergravities}},
  \href{http://dx.doi.org/10.1088/1126-6708/2009/12/050}{\emph{JHEP} {\bf 12}
  (2009) 050}, [\href{http://arxiv.org/abs/0910.1586}{{\tt 0910.1586}}].

\bibitem{Taylor:2010wm}
W.~Taylor, \emph{{Anomaly constraints and string/F-theory geometry in 6D
  quantum gravity}},  in \emph{{Perspectives in Mathematics and Physics (Singer
  Conference 2009): in Celebration of I.M. Singer's 85th Birthday and His
  Legendary Contributions to Mathematics and Physics}}, 9, 2010.
\newblock \href{http://arxiv.org/abs/1009.1246}{{\tt 1009.1246}}.

\bibitem{Kumar:2010am}
V.~Kumar, D.~S. Park and W.~Taylor, \emph{{6D supergravity without tensor
  multiplets}}, \href{http://dx.doi.org/10.1007/JHEP04(2011)080}{\emph{JHEP}
  {\bf 04} (2011) 080}, [\href{http://arxiv.org/abs/1011.0726}{{\tt
  1011.0726}}].

\bibitem{Bonetti:2011mw}
F.~Bonetti and T.~W. Grimm, \emph{{Six-dimensional (1,0) effective action of
  F-theory via M-theory on Calabi-Yau threefolds}},
  \href{http://dx.doi.org/10.1007/JHEP05(2012)019}{\emph{JHEP} {\bf 05} (2012)
  019}, [\href{http://arxiv.org/abs/1112.1082}{{\tt 1112.1082}}].

\bibitem{Kim:2024eoa}
H.-C. Kim, C.~Vafa and K.~Xu, \emph{{Finite Landscape of 6d N=(1,0)
  Supergravity}},  \href{http://arxiv.org/abs/2411.19155}{{\tt 2411.19155}}.

\bibitem{Salam:1984cj}
A.~Salam and E.~Sezgin, \emph{{Chiral Compactification on Minkowski x S**2 of
  N=2 Einstein-Maxwell Supergravity in Six-Dimensions}},
  \href{http://dx.doi.org/10.1016/0370-2693(84)90589-6}{\emph{Phys. Lett. B}
  {\bf 147} (1984) 47}.

\bibitem{Gunaydin:1983rk}
M.~Gunaydin, G.~Sierra and P.~K. Townsend, \emph{{Exceptional Supergravity
  Theories and the MAGIC Square}},
  \href{http://dx.doi.org/10.1016/0370-2693(83)90108-9}{\emph{Phys. Lett. B}
  {\bf 133} (1983) 72--76}.

\bibitem{Gunaydin:1983bi}
M.~Gunaydin, G.~Sierra and P.~K. Townsend, \emph{{The Geometry of N=2
  Maxwell-Einstein Supergravity and Jordan Algebras}},
  \href{http://dx.doi.org/10.1016/0550-3213(84)90142-1}{\emph{Nucl. Phys. B}
  {\bf 242} (1984) 244--268}.

\bibitem{Sezgin:1982gi}
E.~Sezgin and A.~Salam, \emph{{Maximal Extended Supergravity Theory in
  Seven-dimensions}},
  \href{http://dx.doi.org/10.1016/0370-2693(82)90204-0}{\emph{Phys. Lett. B}
  {\bf 118} (1982) 359}.

\bibitem{Samtleben:2005bp}
H.~Samtleben and M.~Weidner, \emph{{The Maximal D=7 supergravities}},
  \href{http://dx.doi.org/10.1016/j.nuclphysb.2005.07.028}{\emph{Nucl. Phys. B}
  {\bf 725} (2005) 383--419}, [\href{http://arxiv.org/abs/hep-th/0506237}{{\tt
  hep-th/0506237}}].

\bibitem{Nastase:1999kf}
H.~Nastase, D.~Vaman and P.~van Nieuwenhuizen, \emph{{Consistency of the
  Ad$S^7$ $\times$ $S^4$ Reduction and the Origin of Selfduality in Odd
  Dimensions}},
  \href{http://dx.doi.org/10.1016/S0550-3213(00)00193-0}{\emph{Nucl. Phys.}
  {\bf B581} (2000) 179--239}, [\href{http://arxiv.org/abs/hep-th/9911238}{{\tt
  hep-th/9911238}}].

\bibitem{Nastase:1999cb}
H.~Nastase, D.~Vaman and P.~van Nieuwenhuizen, \emph{{Consistent nonlinear K K
  reduction of 11-d supergravity on AdS(7) x S(4) and selfduality in odd
  dimensions}},
  \href{http://dx.doi.org/10.1016/S0370-2693(99)01266-6}{\emph{Phys. Lett. B}
  {\bf 469} (1999) 96--102}, [\href{http://arxiv.org/abs/hep-th/9905075}{{\tt
  hep-th/9905075}}].

\bibitem{Bergshoeff:2005pq}
E.~Bergshoeff, D.~C. Jong and E.~Sezgin, \emph{{Noncompact gaugings, chiral
  reduction and dual sigma models in supergravity}},
  \href{http://dx.doi.org/10.1088/0264-9381/23/9/003}{\emph{Class. Quant.
  Grav.} {\bf 23} (2006) 2803--2832},
  [\href{http://arxiv.org/abs/hep-th/0509203}{{\tt hep-th/0509203}}].

\bibitem{Louis:2015mka}
J.~Louis and S.~L{\"u}st, \emph{{Supersymmetric AdS$_{7}$ backgrounds in
  half-maximal supergravity and marginal operators of (1, 0) SCFTs}},
  \href{http://dx.doi.org/10.1007/JHEP10(2015)120}{\emph{JHEP} {\bf 10} (2015)
  120}, [\href{http://arxiv.org/abs/1506.08040}{{\tt 1506.08040}}].

\bibitem{Townsend:1983kk}
P.~K. Townsend and P.~van Nieuwenhuizen, \emph{{GAUGED SEVEN-DIMENSIONAL
  SUPERGRAVITY}},
  \href{http://dx.doi.org/10.1016/0370-2693(83)91230-3}{\emph{Phys. Lett. B}
  {\bf 125} (1983) 41--46}.

\bibitem{Lu:1999bc}
H.~Lu and C.~N. Pope, \emph{{Exact embedding of N=1, D = 7 gauged supergravity
  in D = 11}},
  \href{http://dx.doi.org/10.1016/S0370-2693(99)01170-3}{\emph{Phys. Lett. B}
  {\bf 467} (1999) 67--72}, [\href{http://arxiv.org/abs/hep-th/9906168}{{\tt
  hep-th/9906168}}].

\bibitem{Passias:2015gya}
A.~Passias, A.~Rota and A.~Tomasiello, \emph{{Universal consistent truncation
  for 6d/7d gauge/gravity duals}},
  \href{http://dx.doi.org/10.1007/JHEP10(2015)187}{\emph{JHEP} {\bf 10} (2015)
  187}, [\href{http://arxiv.org/abs/1506.05462}{{\tt 1506.05462}}].

\bibitem{Salam:1983fa}
A.~Salam and E.~Sezgin, \emph{{SO(4) Gauging of $N=2$ Supergravity in
  Seven-dimensions}},
  \href{http://dx.doi.org/10.1016/0370-2693(83)90167-3}{\emph{Phys. Lett. B}
  {\bf 126} (1983) 295--300}.

\bibitem{Koerber:2008rx}
P.~Koerber, D.~Lust and D.~Tsimpis, \emph{{Type IIA AdS(4) compactifications on
  cosets, interpolations and domain walls}},
  \href{http://dx.doi.org/10.1088/1126-6708/2008/07/017}{\emph{JHEP} {\bf 07}
  (2008) 017}, [\href{http://arxiv.org/abs/0804.0614}{{\tt 0804.0614}}].

\bibitem{Galli:2022idq}
M.~Galli and E.~Malek, \emph{{Consistent truncations to 3-dimensional
  supergravity}}, \href{http://dx.doi.org/10.1007/JHEP09(2022)014}{\emph{JHEP}
  {\bf 09} (2022) 014}, [\href{http://arxiv.org/abs/2206.03507}{{\tt
  2206.03507}}].

\bibitem{Gauntlett:2007ts}
J.~P. Gauntlett and N.~Kim, \emph{{Geometries with Killing Spinors and
  Supersymmetric AdS Solutions}},
  \href{http://dx.doi.org/10.1007/s00220-008-0575-5}{\emph{Commun. Math. Phys.}
  {\bf 284} (2008) 897--918}, [\href{http://arxiv.org/abs/0710.2590}{{\tt
  0710.2590}}].

\bibitem{Bugden:2021wxg}
M.~Bugden, O.~Hulik, F.~Valach and D.~Waldram, \emph{{G-Algebroids: A Unified
  Framework for Exceptional and Generalised Geometry, and
  Poisson\textendash{}Lie Duality}},
  \href{http://dx.doi.org/10.1002/prop.202100028}{\emph{Fortsch. Phys.} {\bf
  69} (2021) 2100028}, [\href{http://arxiv.org/abs/2103.01139}{{\tt
  2103.01139}}].

\bibitem{Bugden:2021nwl}
M.~Bugden, O.~Hulik, F.~Valach and D.~Waldram, \emph{{Exceptional algebroids
  and type IIB superstrings}},  \href{http://arxiv.org/abs/2107.00091}{{\tt
  2107.00091}}.

\bibitem{Apruzzi:2013yva}
F.~Apruzzi, M.~Fazzi, D.~Rosa and A.~Tomasiello, \emph{{All AdS$_7$ solutions
  of type II supergravity}},
  \href{http://dx.doi.org/10.1007/JHEP04(2014)064}{\emph{JHEP} {\bf 04} (2014)
  064}, [\href{http://arxiv.org/abs/1309.2949}{{\tt 1309.2949}}].

\bibitem{Apruzzi:2014qva}
F.~Apruzzi, M.~Fazzi, A.~Passias, D.~Rosa and A.~Tomasiello, \emph{{AdS$_{6}$
  solutions of type II supergravity}},
  \href{http://dx.doi.org/10.1007/JHEP11(2014)099}{\emph{JHEP} {\bf 11} (2014)
  099}, [\href{http://arxiv.org/abs/1406.0852}{{\tt 1406.0852}}].

\bibitem{Josse:2023lsp}
G.~Josse and F.~Merenda, \emph{{Vacua scan of 5d, N=2 consistent truncations}},
  \href{http://dx.doi.org/10.1016/j.physletb.2024.138670}{\emph{Phys. Lett. B}
  {\bf 853} (2024) 138670}, [\href{http://arxiv.org/abs/2310.15272}{{\tt
  2310.15272}}].

\end{thebibliography}\endgroup

\end{document}